\documentclass[hidelinks,12pt]{article}

\usepackage{amssymb,amsmath,amsfonts,eurosym,geometry,ulem,graphicx,caption,color,setspace,sectsty,comment,footmisc,caption,natbib,pdflscape,subfigure,array,hyperref,booktabs,tabularx,changepage,lscape,appendix}
\usepackage[subfigure]{tocloft}

\usepackage{titlesec}
\titleformat*{\section}{\large\bfseries}
\titleformat*{\subsection}{\normalsize\bfseries}
\titleformat*{\subsubsection}{\small\bfseries}

\newcolumntype{L}[1]{>{\raggedright\let\newline\\arraybackslash\hspace{0pt}}m{#1}}
\newcolumntype{C}[1]{>{\centering\let\newline\\arraybackslash\hspace{0pt}}m{#1}}
\newcolumntype{R}[1]{>{\raggedleft\let\newline\\arraybackslash\hspace{0pt}}m{#1}}

\usepackage{lscape}

\usepackage{booktabs}
\usepackage[flushleft]{threeparttable}
\usepackage{multirow}
\usepackage{graphicx}

\usepackage{xurl}
\usepackage{multibib}
\newcites{app}{References}

\usepackage{capt-of}
\usepackage{tocloft}
\usepackage{xpatch}

\usepackage{hyperref}

\newcommand{\listofappendixfiguresname}{\large{List of Figures}}
\newlistof{appendixfigures}{apf}{\listofappendixfiguresname}
\newcommand{\listofappendixtablesname}{\large{List of Tables}}
\newlistof{appendixtables}{apt}{\listofappendixtablesname}

\makeatletter
\xapptocmd{\appendix}{%
  \write\@auxout{%
    \string\let\string\latex@tf@lof\string\tf@lof
    \string\let\string\tf@lof\string\tf@apf%
    \string\let\string\latex@tf@lof\string\tf@lot
    \string\let\string\tf@lot\string\tf@apt%
  }%
}{}{}
\makeatother

\renewcommand{\appendixtocname}{\large{Contents}}

\makeatletter
\let\oldappendix\appendices

\renewcommand{\appendices}{%
  \clearpage
  \renewcommand{\thesection}{\Roman{section}}
  \let\tf@toc\tf@app
  \addtocontents{app}{\protect\setcounter{tocdepth}{1}}
  \immediate\write\@auxout{%
    \string\let\string\tf@toc\string\tf@app^^J
  }
  \oldappendix
}%

\newcommand{\listofappendices}{%
  \begingroup
  \renewcommand{\contentsname}{\appendixtocname}
  \let\@oldstarttoc\@starttoc
  \def\@starttoc##1{\@oldstarttoc{app}}
  \tableofcontents
  \endgroup
}

\makeatother

\begin{document}

\begin{titlepage}
\title{\vspace{-0.4cm} Unintended Consequences of Sanitation Investment: Negative Externalities on Water Quality and Health \\ in India}
\author{Kazuki Motohashi\thanks{~Graduate School of Economics, Hitotsubashi University. E-mail: kazuki.motohashi@r.hit-u.ac.jp. 
I am grateful for the guidance and advice of Ujjayant Chakravorty, Shinsuke Tanaka, and Marc Jeuland. 
For helpful comments, I thank Khushboo Aggarwal, Jenny Aker, Antonella Bancalari, Yvonne Jie Chen, Yuxian Chen, Jishnu Das, Kyle Emerick, Jun Goto, Raymond Guiteras, Cynthia Kinnan, Motohiro Kumagai, Takashi Kurosaki, Daniele Lantagne, Leslie Martin, Nobuhiko Nakazawa, Patricia Ritter, Adam Storeygard, Haruka Takayama, Takanao Tanaka, Kensuke Teshima, Duc Anh Tran, Shunsuke Tsuda, Zachary Wagner, Junichi Yamasaki, Akio Yamazaki, and Hidefumi Yokoo; the seminar participants at Hitotsubashi University, Tufts University, Waseda University, JADE/GRIPS, SWELL, and Osaka University; the conference participants at 2020 Young JADE Conference, 2021 SEEPS Camp, 2021 AERE Summer Conference, 2021 NEUDC Conference, 2022 JEA Spring Meeting, Camp Resources XXVIII, 2022 AMES Tokyo, AMIE 2nd Workshop in Applied Microeconomics, Fourth SANEM-World Bank North America Discussion Forum, and 4th Asian Workshop on Econometrics and Health Economics. This study is supported by Grants-in-Aid for Scientific Research from the Japan Society for the Promotion of Science (grant number JP24K16353). The previous version of this paper was awarded the 2022 Moriguchi Prize by the Institute of Social and Economic Research at Osaka University.}}
\date{\vspace{0.5cm} April 1, 2026}

\maketitle

\vspace{-0.6cm}
\onehalfspacing
\begin{abstract}
\noindent Developing countries have increased sanitation investment to improve child health.
However, scaling up latrine construction can cause water pollution externalities owing to insufficient infrastructure for the treatment of fecal sludge, offsetting the direct health benefits.
I estimate the negative externalities of an Indian sanitation policy that subsidized the construction of over 100 million latrines.
Exploiting geographical variations in soil characteristics that affect the feasibility of latrine construction, I find that this policy increases fecal contamination of rivers by 72\%.
Although the policy reduces diarrheal child mortality overall, this positive health effect is eliminated when upstream areas lack adequate wastewater infrastructure. \\

\vspace{-0.4cm}


\bigskip
\end{abstract}
\setcounter{page}{1}

\end{titlepage}
\pagebreak
\onehalfspacing
\setcounter{page}{2}
\section{Introduction} \label{sec:introduction}

Policymakers and researchers have recognized the importance of sanitation investments in improving child health in developing countries. 
Poor access to sanitation facilities and the associated practice of open defecation increase the risk of diarrheal diseases and mortality, adversely affecting child health. 
Worldwide, according to WHO/UNICEF data, 688 million people practiced open defecation in 2016, leading to 432,000 deaths \citep{pruss2019burden}. 
Consequently, developing countries such as India and China have adopted nationwide policies that subsidize the construction of latrines (toilets).\footnote{~Like the Indian government's Swachh Bharat Mission examined in this paper, under the ``Toilet Revolution,'' the Chinese government built and upgraded over 10 million rural toilets in 2018.}
These policies are intended to reduce open defecation and exposure to fecal matter near human habitats, thereby improving child health. 
These direct health benefits have been well-documented as local impacts at the village level \citep[e.g.,][]{geruso2018neighborhood,cameron2021sanitation,cameron2022dirty}.

However, the negative externalities of scaling up latrine construction as a nationwide policy remain unclear and can offset its direct health benefits. 
The constructed latrines accumulate a large volume of fecal sludge, which requires periodic emptying.
Subsequently, the emptied fecal sludge should be treated by infrastructure, specifically wastewater treatment plants, to destroy the remaining active pathogens.
However, owing to insufficient infrastructure, a growing volume of fecal sludge, exacerbated by the intensive increase in latrine construction under the policy, may be dumped into rivers, thus polluting them.
These water pollution externalities can decrease the overall effectiveness of latrine construction in improving child health. 
In extreme cases, latrine construction can worsen health outcomes if the water pollution externalities exceed the direct positive effects of reduced open defecation. 

Therefore, I examine the negative externalities of latrine construction on water quality and health in the context of India's nationwide sanitation policy, the Swachh Bharat Mission (SBM). 
Since its inception in 2014, the SBM has allocated approximately USD 6 billion to subsidize the construction of over 100 million latrines at the household level in rural India.\footnote{~The types of latrines most commonly used in rural India are pit latrines and latrines with septic tanks. Because these are not connected to sewer pipes, they accumulate fecal sludge. The disposal of fecal sludge from these latrines can result in water pollution externalities.}
The impact of the SBM deserves careful examination as it is the largest sanitation policy in the world.
I use administrative panel datasets on the district-level number of household latrines from 2012 to 2019 under the SBM and the water quality of 1,189 monitoring stations along rivers in 337 districts from 2007 to 2019 to examine the negative externality of water quality.
I combine these data with district-level panel data on diarrheal child mortality rates to examine the effects on health.

To identify the causal effects of latrine construction on water quality and health, I adopt an instrumental variable (IV) design that exploits geographical variations in soil characteristics that affect the feasibility of latrine construction under the SBM.
Specifically, I use Available Water Capacity (AWC), a proxy for the soil infiltration rate, interacted with a post-SBM indicator, as an instrument for the number of latrines.
This IV design is conceptually similar to the difference-in-differences (DiD) design, in which the reduced-form regression uses AWC as a treatment variable.
Higher infiltration rates (lower AWC) increase the risk of groundwater contamination in wells from fecal sludge accumulated in the latrines. 
To address this risk, an official technical guideline \citep{cpheeo2013manual}, which has become effective since the SBM's inception in 2014, requires either greater distances between latrines and wells or the addition of impervious materials inside latrines in areas with high infiltration rates. 
Therefore, a lower AWC increased the difficulty and cost of latrine construction after the SBM started in 2014. 
Indeed, a lower AWC was associated with a smaller increase in latrines during the post-SBM period in the first stage.

I also adopt an upstream--downstream specification that examines the effects of upstream latrine construction on downstream water quality and health.
Dumped fecal sludge from latrines can flow downstream along rivers, causing water pollution externalities in downstream areas.
I test these spillover effects in the modified IV design, where I use the upstream AWC as an instrument for upstream latrine construction and examine its impact on downstream outcomes. 
This upstream--downstream specification addresses the concerns about the exclusion restriction in the baseline IV design, especially for health outcomes.
One potential violation could be that AWC, which measures soil quality, affects health outcomes in the same area by influencing agricultural output and income, which, in turn, affects the level of health investments.
However, upstream AWC is not expected to affect downstream health outcomes through this income channel, because upstream AWC is unlikely to be associated with downstream agricultural output and income.
Indeed, in the reduced-form event study regressions, I find that upstream AWC did not have differential effects on water quality and health prior to the SBM policy (before the official technical guideline was published), supporting the validity of the exclusion restriction.
Accordingly, I treat the upstream--downstream specification as the preferred design for health outcomes, and all subsequent health results rely on this specification.\footnote{~For the water quality outcome, I do not find differential effects of AWC prior to the SBM policy in either the baseline or the upstream--downstream specification. Therefore, I present results for both specifications for the water quality outcome.}

My results show that latrine construction under the SBM degrades river water quality while improving overall health. 
I find that one additional latrine per square kilometer increases fecal coliform in rivers by 3\%. 
The total effect of the SBM is estimated to be a 72\% increase in river pollution.
Moreover, the upstream--downstream specification shows that water pollution externalities spill over to downstream areas, especially several years after the SBM started.
However, additional upstream latrine construction per square kilometer reduces the downstream diarrheal post-neonatal mortality rate per 1,000 children by 0.011, which is a 0.4\% reduction from the pre-SBM period. 
The total health effect of the SBM is estimated to be a 10\% decrease in diarrheal post-neonatal mortality rate. 
This overall positive health effect suggests that the direct positive health effect of decreased open defecation outweighs the negative externality on health due to increased water pollution.

To explore the mechanisms behind these negative externalities, I examine whether the effects on water quality vary by the level of complementary infrastructure for fecal sludge treatment.
Sufficient infrastructure for treating wastewater can prevent the dumping of fecal sludge from latrines, which is the main mechanism of negative externalities. 
The most common infrastructure in India is sewage treatment plants (STPs), which co-treat urban sewage and rural fecal sludge.
Therefore, I compare the effects between areas with higher and lower treatment capacities of STPs than the median.
In both the baseline and upstream--downstream specifications, the negative externality of water quality is eliminated in areas with higher treatment capacities.
Conversely, in areas with lower treatment capacities, the negative externality of water quality is significant and spills over downstream, suggesting that dumping of untreated fecal sludge is the mechanism.

The same heterogeneity analysis by the treatment capacity of fecal sludge for the health outcome suggests that water pollution externalities offset the direct positive health effects.
The total effect of the SBM is a 15\% (42\%) decrease in diarrheal post-neonatal mortality rate when upstream districts (states) have higher treatment capacities, coupled with insignificant water pollution externalities.
However, the positive health effect is eliminated in cases with lower upstream treatment capacities when water pollution externalities are significant.
These heterogeneous health effects suggest that water pollution externalities offset the direct positive health effects, thereby reducing the overall effectiveness of latrine construction in improving child health in rural India.

A variety of tests corroborate my findings on the negative externalities of latrine constructions. 
First, falsification tests show no effects on unrelated water quality and health outcomes, strengthening the validity of the exclusion restriction.
Second, the results are robust to an alternative DiD design, consideration of spillovers from neighboring districts and urban areas, and adoption of a balanced panel and alternative mortality dataset.

My findings suggest that wastewater treatment infrastructure is a crucial complementary investment to latrine construction for mitigating negative externalities.
The back-of-the-envelope cost--benefit analysis at the district level shows that the net mortality benefit (USD 5.6 million) is approximately one-third of the subsidy cost of the SBM policy (USD 16.9 million).
However, complementing latrine construction with higher treatment capacities to mitigate negative externalities would substantially increase the mortality benefit (by USD 7.4 million) with lower additional construction and operating costs for more STPs (USD 4.5 million).
More generally, these findings highlight the importance of incentivizing private goods together with complementary public goods to prevent potential negative externalities.

This paper makes three main contributions to the literature. 
First, it contributes to the literature on the effects of sanitation interventions by revealing the negative externalities of toilet construction on the environment and health. 
Previous studies have thus far focused on the direct positive effects of sanitation interventions on child health and mortality \citep{duflo2015toilets,hammer2016village,coffey2018sanitation,geruso2018neighborhood,spears2018exposure,alsan2019watersheds,cameron2019scaling,cameron2021sanitation,cameron2022dirty,flynn2023watershed}.\footnote{~Past literature also confirms the positive effects of such interventions on educational outcomes \citep{spears2016effects,adukia2017sanitation}, labor supply \citep{wang2022sanitation}, and violence against women \citep{hossain2022access}. Another strand of literature has examined the constraints to latrine adoption, including financial constraints, inadequate information concerning the benefits of latrines and costs of open defecation \citep{pattanayak2009shame,guiteras2015encouraging,yishay2017microcredit,lipscomb2018subsidies}, and religious and caste beliefs that discourage latrine use \citep{spears2019puzzle,adukia2021religion}.}
I complement these findings by showing that toilet construction causes unintended water pollution externalities that offset the direct positive health effects.
My analysis leveraging policy variation across hundreds of districts in India allows me to examine negative externalities that can extend beyond villages, which have not been captured in most previous studies that relied on village-level field experiments.
Moreover, this paper provides new evidence on the impacts of the SBM policy, which is the world's largest toilet construction program.

Second, this paper contributes to the literature on the causes and effects of water pollution by providing the first causal estimate of the effect of toilet construction on river water quality.\footnote{~Public health literature has examined the association between pit latrines and groundwater quality based on a limited sample of a few hundred latrines \citep{graham2013pit}. This paper estimates the causal effect of latrines on river water quality based on nationwide administrative data.} 
Previous studies examined how water quality is affected by regulations \citep{greenstone2014environmental,keiser2019consequences}, political boundaries \citep{lipscomb2016decentralization,motohashi2024}, and court rulings \citep{do2018can,bhupatiraju2024environmental}.\footnote{~\cite{greenstone2014environmental} find limited impacts of water pollution regulations on water quality in India. My findings point to one potential explanation: contemporaneous toilet construction policies may have increased fecal contamination, potentially attenuating the measured effects of those regulations.}
Another set of studies investigated the effects of industrial and agricultural wastewater on health outcomes, including digestive cancer \citep{ebenstein2012consequences}, infant mortality \citep{brainerd2014seasonal,mettetal2019irrigation}, and birth outcomes \citep{dias2023down}. 
This paper shows that toilet construction can also substantially increase river pollution (by 72\% under the SBM, which is a large effect), and that this increase in domestic wastewater can offset the positive health effects.
A particularly relevant concurrent paper, \cite{lepault2024urban}, shows that the introduction of sewage treatment plants reduces water pollution in the urban context of India, with positive effects on child health. 
By contrast, this paper focuses on rural toilet expansion and provides evidence that large-scale toilet construction, in the absence of adequate fecal sludge management, can generate water pollution externalities with adverse health consequences. 
It further shows, in a way consistent with \cite{lepault2024urban}, that sewage treatment plants play a critical complementary role in mitigating these externalities.

Third, and more broadly, this paper advances the literature on the unintended negative effects of health policies in developing countries by showing that the negative effects of the displacement of pollution sources can be minimized with sufficient complementary infrastructure.
Previous literature has documented how health policies can worsen health outcomes due to reduced complementary health behaviors \citep{bennett2012does}, switching to alternative unsafe health behaviors \citep{buchmann2019throwing}, abandonment and delays in project completion \citep{bancalari2024unintended}.
This paper shows that unintended negative effects can also be caused by the displacement of pollution sources from open defecation sites to rivers where emptied fecal sludge is dumped. 
I then demonstrate that these effects can be mitigated by complementary infrastructure for pollution control.\footnote{~Relatedly, my findings underscore the importance of public capacity to cope with the increased consumer demand under demand-side incentives, as also observed in other contexts like healthcare \citep{andrew2024incentivizing}.}

The remainder of this paper is organized as follows. 
Section \ref{sec:background} describes the SBM and its potential effects on water quality and health.
Sections \ref{sec:data} and \ref{sec:model} describe my data and empirical strategies, respectively. 
Section \ref{sec:result} presents the baseline results of the effects on water quality and health.
Section \ref{sec:result_het_treatment} presents the heterogeneous effects of latrine construction.
Finally, Section \ref{sec:conclusion} concludes the paper.

\section{Background} \label{sec:background}

\subsection{Latrine Construction under Swachh Bharat Mission in India} \label{sec:sbm}

To eliminate open defecation, the Indian government subsidized the construction of over 100 million latrines in rural India under the SBM, the largest sanitation policy in the world.

In India, many people have historically practiced open defecation, which increases the risk of diarrheal diseases and mortality, adversely affecting child health.
About 470 million people in India practiced open defecation in 2013, according to the WHO/UNICEF Joint Monitoring Programme.
As such, India had the highest number of people practicing open defecation in the world, more than ten times that of the country with the second-highest number, Nigeria, in 2013 (Appendix Figure \ref{fig:od_india}).

To eliminate open defecation, the Indian government implemented a nationwide sanitation policy, the SBM, that subsidized household latrine construction in rural areas.\footnote{~This paper focuses on examining the impacts of household-level latrine construction in rural areas, as promoted under the SBM-Gramin, in contrast to the construction of school-level latrines in rural areas \citep{adukia2017sanitation} or public toilets that are more prevalent in urban areas.}
Since its inception in 2014, the SBM has set the ambitious goal of achieving universal latrine coverage by October 2, 2019, the 150th anniversary of Mahatma Gandhi's birth.
To achieve this goal, SBM substantially increased the subsidy to a maximum of INR 12,000 (approximately USD 140) per household, covering most of the initial costs of basic latrines in rural India.\footnote{~SBM is the most recent policy out of four consecutive sanitation policies at the central government level. Although state governments have primary responsibility for sanitation, these central policies were meant to influence the state-level sanitation efforts through policy guidance and budget allocation.}

\begin{figure}[p]
	\begin{center}
		\includegraphics[width=0.6\textwidth]{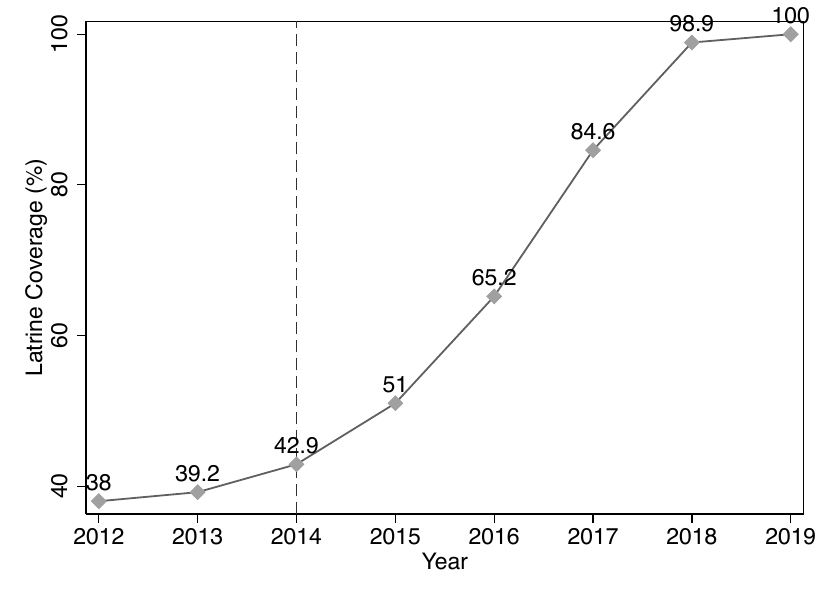}
		\caption{\label{fig:latrine_india} Latrine Coverage in Rural India}
		\medskip 
		\begin{minipage}{1\textwidth} 
			{\small 
				Notes: This figure documents the proportion of households that have latrines in rural India between 2012 and 2019, based on the administrative database of the SBM.
				A vertical dashed line shows the starting year of the SBM. \par}
		\end{minipage}
	\end{center}
\end{figure}

With this big push to construct latrines, SBM has become the world's largest sanitation policy, with central government expenditures totaling nearly USD 6 billion from 2014 to 2019, during which the number of rural households with latrines increased by 100 million.\footnote{~The annual budgets of the Indian government show that the central government spent USD 5.96 billion (INR 497 billion) from 2014 to 2019. The data source of the number of latrines built is the SBM website at \url{https://swachhbharatmission.gov.in/SBMCMS/about-us.htm}.}
According to the administrative database of the SBM, latrine coverage dramatically increased from 39.2\% in 2013 to almost 100\% in 2019 (Figure \ref{fig:latrine_india}).
Although the latrine coverage calculated from the administrative database is likely overestimated, recent independent surveys corroborate a substantial improvement in latrine coverage.
For example, the National Annual Rural Sanitation Survey, conducted by an independent verification agency with support from the World Bank, found that 85\% of the rural population used latrines in 2019--2020 \citep{narss2020}.\footnote{~The National Family Health Survey 5 also reported that 74.1\% of households in rural India used toilets or latrines without practicing open defecation in 2019--2021.}

\subsection{Negative Externality of Latrine Construction on Water Quality} \label{sec:consequence_wq}

Scaled-up latrine construction under the SBM may cause an unintended negative externality in river water quality owing to the insufficient treatment of fecal sludge.

Fecal sludge from household latrines requires periodic emptying by private vacuum truck operators or manual emptying services and subsequent transport to wastewater treatment plants for treatment.\footnote{~Although the fecal sludge contained in pits degrades to some degree with time, pathogens can be present even after long-term storage. The primary objective of pit latrines is fecal containment rather than pathogen reduction \citep{orner2018pit}.}
However, given insufficient treatment infrastructure, the growing volume of fecal sludge generated by scaled-up latrine construction may instead be dumped untreated into rivers, thereby causing pollution.\footnote{~The practice of dumping fecal sludge is highlighted in an ethnographic study on 32 truck operators who desludge latrines, although this study focuses on urban areas in Bangalore, Karnataka \citep{prasad2019pits}. Several news media reports have also highlighted the dumping of fecal sludge and the associated water pollution owing to insufficient wastewater infrastructure (e.g., DownToEarth (\url{https://www.downtoearth.org.in/news/waste/pollution-time-bomb-ticking-for-ganga-despite-odf-63790})). Moreover, emptied fecal sludge may also be dumped on fields rather than directly into rivers, but it can still reach surface water via drainage channels and rainfall runoff, thereby increasing fecal contamination in rivers.}

My analysis captures the differential effects of latrine construction relative to open defecation, which can also cause water pollution externalities.\footnote{~In my setting, the relevant counterfactual for private latrine construction is primarily open defecation rather than using community toilets or sharing a latrine with neighbors. Consistent with this, data from the National Family Health Survey 5 (2019--2021) show that 99.5\% of rural households either own a private latrine or practice open defecation; only 0.5\% use community toilets or shared latrines.}
Open defecation, which is practiced before latrine construction, generates small amounts of stool across a wide range of locations. 
In rural India, it is often practiced in open fields or bushy areas on the periphery of villages to avoid being seen by others \citep{routray2015socio}. 
As a result, fecal matter from open defecation is more spatially dispersed.
Moreover, because it is carried into rivers mainly during rainfall, only a fraction is likely to reach them.
By contrast, latrines accumulate a large volume of fecal sludge in a single location, so when emptied, the fecal sludge may be dumped into rivers in a concentrated manner.
Thus, the volume of fecal sludge that reaches rivers may increase after latrine construction, causing increased water pollution.

I argue that the water pollution externality of latrine construction is unintended, as evidenced by the absence of policy targets for the treatment of fecal sludge.
According to SBM guidelines \citep{sbm2018guideline}, the open-defecation-free status of a village is declared and verified based on a checklist of multiple indicators, including access to toilet facilities, 100\% usage, fly-proofing, and safe septage disposal.
In the safe septage disposal section, although the checklist stipulates that toilets should be connected to pits or septic tanks, it lacks specific guidance on how emptied fecal sludge should be treated properly.

\subsection{Negative Externality of Latrine Construction on Health} \label{sec:consequence_health}

Latrine construction under SBM may also result in a negative externality to health through exposure to increased river pollution.
This water pollution externality may offset the positive health effects of reduced open defecation.

My analysis investigates the overall health effect determined by the magnitudes of both the direct positive health effects and the indirect negative externality of latrine construction.
On the one hand, latrine construction has direct positive health effects by reducing open defecation and exposure to fecal matter near human habitats, leading to a reduction in diarrheal diseases and mortality among children.
Conversely, latrine construction can indirectly harm health by causing water pollution externalities.
Exposure to polluted water from activities such as drinking river water and bathing in rivers can increase the risk of diarrheal diseases and mortality among children \citep{moe1991bacterial,garg2018not,buchmann2019throwing}. 
Thus, the water pollution externalities of latrine construction may offset the direct positive health effects of reduced open defecation.
This tradeoff is formally presented in the conceptual framework in Appendix \ref{sec:conceptual_framework}.

\subsection{Complementary Infrastructure for Treatment of Fecal Sludge} \label{sec:complementary_investment}

The magnitude of the negative externalities of latrine construction is expected to vary by the level of complementary infrastructure for fecal sludge treatment. 
Adequate infrastructure for fecal sludge treatment can prevent the dumping of emptied fecal sludge, which is the main mechanism behind the negative externalities of latrine construction.

In India, local governments are tasked with developing wastewater infrastructure, such as STPs and fecal sludge treatment plants (FSTPs), to treat fecal sludge emptied from latrines.\footnote{~Wastewater infrastructure is used to treat fecal sludge from both pit latrines and latrines with septic tanks in rural areas, whereas in urban areas, it is used to treat sewage from sewer networks.}
STPs are large-scale infrastructure that has been available in India for many years. 
India had approximately 500 operating STPs in 2015 \citep{cpcb2015}. 
STPs are typically designed to treat urban sewage, but they are also increasingly used to co-treat fecal sludge owing to the underutilization of STP capacities in India.\footnote{~Although the data on the actual prevalence of co-treatment is not available, case studies are available for STPs in Panaji (Goa), Kanpur (Uttar Pradesh), and Chennai (Tamil Nadu). Also, policies and guidelines stipulating the co-treatments at STPs have been formulated by the central government and multiple states, including Punjab, Madhya Pradesh, Jharkhand, and Rajasthan \citep{cotreatment2018}.}
FSTPs are newly developed, small-scale facilities for the treatment of fecal sludge. 
FSTPs began operating in 2014, and approximately 30 FSTPs were in operation at the end of 2019 \citep{rao2020business}.

I use geographical variations in STP capacity in the pre-SBM period to examine the heterogeneous effects on water quality and health.\footnote{~I do not consider FSTP capacities because there were no FSTPs in the pre-SBM period.}
The negative externality of water quality is expected to be substantial in areas with lower treatment capacity. 
In these areas, the negative externality of health through pollution exposure is expected to be greater, suggesting a smaller overall positive health effect.
Conversely, I expect to find smaller water pollution externalities in areas with high treatment capacities, leading to larger positive health effects.
I develop a conceptual framework in Appendix \ref{sec:conceptual_framework} to derive these predictions for heterogeneous effects, which are tested in Section \ref{sec:result_het_treatment}.

\section{Data} \label{sec:data}

I combine administrative datasets on river water quality and household latrines across India to examine the negative externality of latrine construction on water quality.
I use diarrheal child mortality estimates as an additional outcome to examine the negative externality to health.
I also use AWC as an instrument for latrine construction.
These data are spatially matched based on 2011 district boundary data.
Additional details on the data are provided in Appendix \ref{sec:appendix_data}.

\subsection{Water Quality} \label{sec:data_wq}

\begin{figure}
	\begin{center}
		\includegraphics[width=0.5\textwidth]{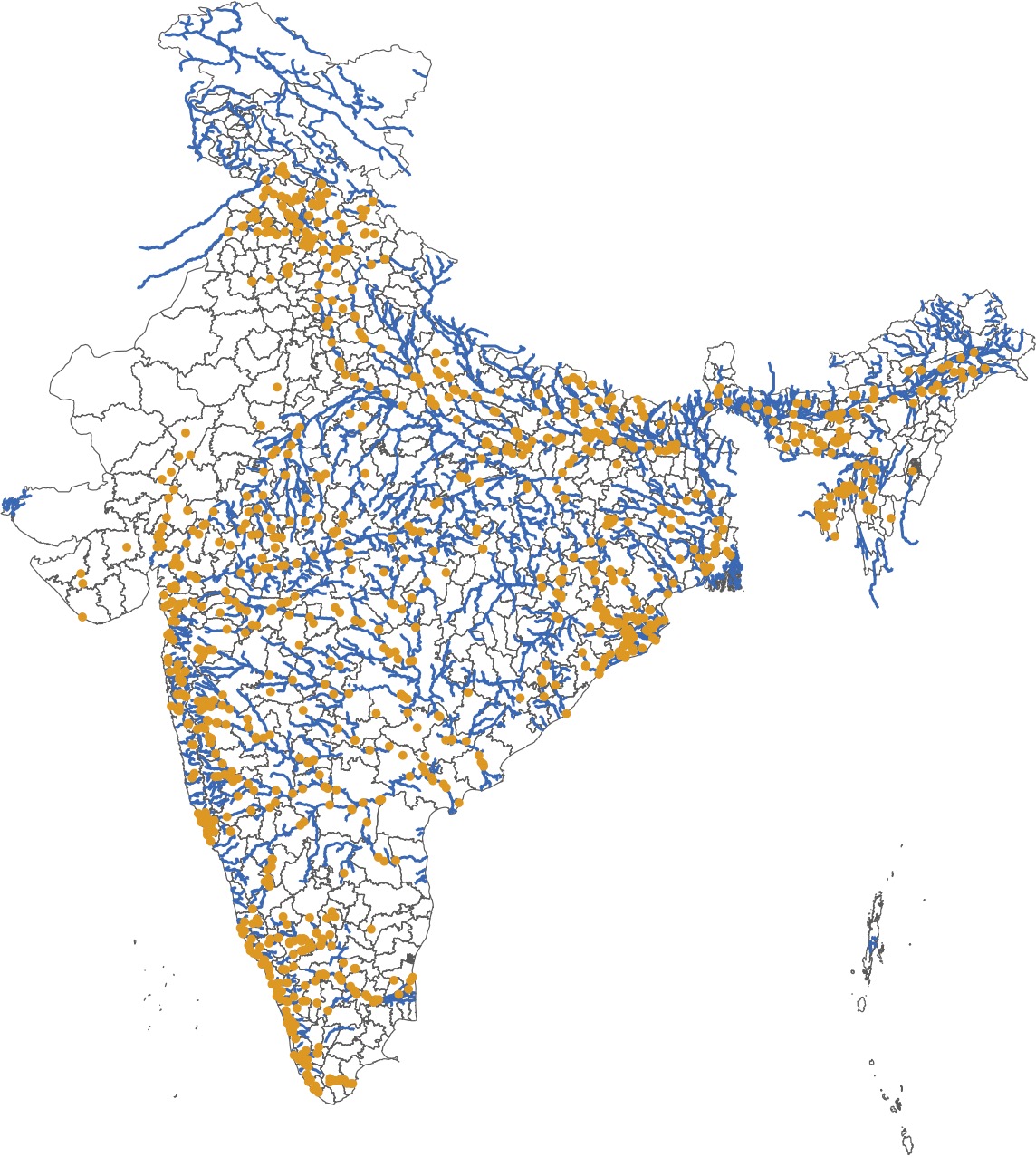}	
		\caption{\label{fig:stations} Distribution of Water Quality Monitoring Stations in India}
		\medskip 
		\begin{minipage}{1\textwidth} 
			{\small 
				Notes: This figure shows water quality monitoring stations in orange dots, district boundaries in black lines, and rivers in blue lines. The data source of river lines is \cite{allen2018global}. \par}
		\end{minipage}
	\end{center}
\end{figure}

I adopt two outcome variables in this paper: water quality and health. 
First, I use yearly water quality data from 1,189 monitoring stations along the rivers in India from 2007 to 2019 (Figure \ref{fig:stations}). 
The yearly data are provided based on the monthly or quarterly monitoring of water quality as part of the National Water Quality Monitoring Programme (NWMP) managed by the Central Pollution Control Board (CPCB).

Among multiple water quality indicators, I use fecal coliform as the main indicator because it is a direct measurement of fecal contamination caused by fecal sludge emptied from latrines.\footnote{~While fecal coliform can also originate from animal waste, the significant effects on fecal coliform observed only after the SBM started, as shown in Panels A and B of Figure \ref{fig:event_result_awc}, suggest that these effects are primarily driven by human waste.}
A higher number of fecal coliforms indicates a higher level of fecal contamination.
The baseline analysis uses the average of the yearly maximum and minimum values of fecal coliform, while I also use the maximum values for the robustness check, because peak pollution events can arise when fecal sludge is dumped during desludging, and such spikes may be more relevant for health risks.\footnote{~The average values of fecal coliform are used because the actual yearly mean values are only recorded up to 2014. The correlation between average and mean values of fecal coliform is 0.997, which suggests that average values are good proxies for mean values.}
Since the distribution of fecal coliform is right-skewed and approximately log-normal, I use the logarithm of fecal coliform as a water quality outcome in the analysis.\footnote{~There are only 28 observations with an average fecal coliform value of 0 out of approximately 7,200 observations in my sample, which are excluded when transforming the fecal coliform values into logarithms.}
Moreover, because fecal coliform values can be extremely high, as shown in Table \ref{tab:sumstats}, I conduct a robustness check by running the analysis after winsorizing fecal coliform values at the 99th, 95th, 90th, and 75th percentiles.\footnote{~The 99th, 95th, 90th, and 75th percentiles of fecal coliform values in the final sample of the water quality analysis are 0.674, 0.056, 0.018, and 0.003 million MPN/100 mL, respectively.}

\subsection{Health} \label{sec:data_health}

Another outcome variable is health, specifically focusing on diarrheal child mortality, owing to its close relationship with poor sanitation. 
I use diarrheal mortality rate estimates (per 1,000 children) from 2000 to 2019, provided as 5 km raster data by the Institute for Health Metrics and Evaluation \citep{ihme2020}.
This dataset includes estimates of diarrheal mortality rate in five age groups: early-neonatal (0--6 days), late-neonatal (7--27 days), post-neonatal (28 days--1 year), ages 1--4 years, and under 5 years.
These estimates are constructed based on geocoded datasets from multiple household surveys, including the India National Family Health Survey, the India District Level Household Survey, and the India Human Development Survey.\footnote{~\cite{ihme2020} applies a Bayesian model-based geostatistical framework to 15 geocoded variables and 3 national-level time-varying variables to predict the posterior distributions of diarrheal mortality. One of the national-level time-varying variables is the percentage of the population with access to improved latrines, but my analysis controls for these by including year fixed effects.}
For the analysis, the district-level mean of these estimates is computed based on raster data and district boundary data.

\subsection{Latrines} \label{sec:data_latrines}

The treatment variable is the number of household latrines. 
I use administrative data on the district-level number of household latrines in rural India from 2012 to 2019, scraped from the database that records toilet construction under the SBM policy. 
Based on this dataset, the number of latrines per square kilometer is computed as a normalized measure.\footnote{~My analysis does not consider whether constructed latrines are used owing to a lack of district-level panel data on latrine usage. Because non-use of constructed latrines has been documented in India \citep{coffey2014open}, my estimates represent a lower bound and could be even larger if the number of used latrines (or households using latrines) were used as the treatment indicator instead.}

One concern with this dataset is that the number of latrines may have been overreported because the data were collected by the Indian government under the SBM policy with the aim of achieving universal latrine coverage.
Such overreporting could lead to a downward bias in the magnitude of the estimates because the actual number of latrines may be lower than those reported in the administrative data.
In other words, the reported effects on water quality and health in this paper represent lower-bound estimates and could be even larger in reality.
As a partial solution to this concern, district (or monitoring station) and year fixed effects in the empirical analyses control for level differences across districts and nationwide trends in overreporting.
Moreover, the heterogeneity in effects by fecal sludge treatment capacity remains unaffected if overreporting impacts both high- and low-capacity areas similarly and is differenced out.

\subsection{Available Water Capacity} \label{sec:data_awc}

For the IV design, I use AWC as an instrument for latrine construction. 
AWC is the amount of water that a soil can store that is available for use by plants.
AWC represents the soil infiltration rate, that is, the velocity or speed at which water enters the soil.
Higher AWC is associated with a lower soil infiltration rate.
The AWC data are available in the Harmonized World Soil Database v1.2, provided by the Food and Agriculture Organization of the United Nations. 
This database provides 30 arc-second raster data for AWC across the globe.
I compute the district-level mean AWC for the analysis based on the raster data and district boundary data because AWC is mostly distributed contiguously within districts, as shown in Appendix Figure \ref{fig:soil_awc_raw}. 

\subsection{Other District Characteristics}

I supplement the above information with additional data to account for district characteristics that might affect latrine construction, water quality, and health.
Specifically, I use 0.25-degree raster data of precipitation from 2007 to 2019, provided by the India Meteorological Department \citep{pai2014development}. 
I further aggregate the daily raw data into annual data, and then construct the district-level mean precipitation based on raster data and district boundary data. 

\subsection{Data Matching and Sample Construction}

To match the water quality data with other data, I first use the 2011 district boundary data of the ML Infomap and the GPS coordinates of the monitoring stations to identify the district where each monitoring station is located. 
This process results in the unique assignment of each station to a specific district.
I then match the water quality data with the latrine data based on the district names.\footnote{~I deal with the changes in the district boundary by ensuring that all data are organized according to the 2011 boundary. Latrine data based on the 2019 boundary are aggregated to follow the 2011 boundary by considering the district splits from 2011 to 2019.}
All other data are similarly matched to the water quality and latrine data following the 2011 district boundary.

After data matching, I construct an unbalanced panel data of 1,189 water quality monitoring stations in 337 districts between 2012 and 2019 for the baseline water quality analysis.\footnote{~My analysis focuses only on districts that have monitoring stations. The average number of stations per district is 3.5, with a standard deviation of 3.4.} 
For the health analysis, I construct a balanced panel of the same 337 districts from 2012 to 2019.
In the reduced-form event study analysis in Section \ref{sec:validity_exclusion_restriction}, I use a longer panel of water quality and health data from 2007 to 2019.
In the upstream--downstream specification, I focus on a subset of monitoring stations and districts along major rivers, resulting in a sample of 365 stations in 154 districts for the water quality analysis and a sample of 103 districts for the health analysis, as explained in Section \ref{sec:iv_updown}.

Table \ref{tab:sumstats} presents summary statistics for the variables used in the main analysis.\footnote{~Appendix Table \ref{tab:sumstats_appendix} shows the summary statistics of variables used for robustness checks.}
To evaluate the representativeness of my samples relative to all districts in India (640 districts in 2011), I conduct balance tests by comparing pre-SBM means of key variables.
Appendix Table \ref{tab:balance_samples} shows that the treatment-related variable, latrine coverage, is almost balanced between my sample districts and the remaining districts (Columns 5-7).
Moreover, the health outcome, diarrheal mortality rate, is balanced between the upstream--downstream sample districts, which are used for health analysis, and the remaining districts (Column 7).\footnote{~A balance test on water quality outcomes is not conducted because the remaining districts outside my sample districts, by design, do not have water quality monitoring stations along rivers.}
Balance tests on other variables show that my samples are more populated and less developed (or more rural), where the SBM is likely to be intensively targeted and implemented.

\section{Empirical Strategy} \label{sec:model}

I empirically examine the effects of latrine construction under the SBM on river water quality and health. 
Ordinary least squares (OLS) estimates may be biased due to reverse causality and time-varying omitted variables affecting both latrine construction and outcomes.
For example, an increase in diarrheal mortality rate may encourage latrine construction to address this health issue, leading to reverse causality.
Moreover, unobserved practices of open defecation may discourage latrine construction while also increasing water pollution. 

To identify the causal effects of latrine construction, I adopt an IV design that exploits the geographical variation in soil characteristics that affect the feasibility of latrine construction. 

\subsection{Instrumental Variable Design} \label{sec:iv_baseline}

In the IV design, I use AWC, a proxy for soil infiltration rate, interacted with a post-SBM indicator as an instrument for latrine construction to examine the effects of latrine construction on water quality and health.

Higher soil infiltration rates (lower AWC) increase the risk of groundwater contamination from fecal sludge accumulated in pit latrines, which are widely adopted in rural India.
Pit latrines consist of a hole called a pit that accumulates fecal sludge without a completely sealed wall. 
Therefore, pathogens inside fecal sludge can percolate into soils, potentially causing fecal contamination of groundwater sources such as tube and dug wells.\footnote{~This groundwater contamination is different from river pollution caused by the dumping of fecal sludge emptied from latrines. The former (related to AWC) is considered to motivate the IV design, while the latter (related to STPs) is the effect investigated in this paper.}
The degree of fecal contamination depends on the soil infiltration rate.

To address the risk of groundwater contamination, an official technical guideline \citep{cpheeo2013manual}, which has been effective since the SBM's inception in 2014, requires additional precautionary measures for latrine construction in areas with high infiltration rates (lower AWC).
Specifically, if the effective size (ES) of the soil is 0.2 mm or less, that is, a lower infiltration rate (higher AWC), pits can be located at a minimum distance of 3 m from water sources.\footnote{~This requirement applies to dry pits under unsaturated soil conditions, that is, where the height between the bottom of the pit and the maximum groundwater level throughout the year is 2 m or more. In the other case of wet pits under saturated soil conditions, the minimum distance is increased to 10 m.}
However, for coarser soils with an ES greater than 0.2 mm, that is, a higher infiltration rate (lower AWC), the 3 m minimum distance is insufficient, requiring a greater separation with increased minimum distances.
Alternatively, households can maintain the same 3 m minimum distance by making additional investments in latrine construction.
Specifically, the bottom of the pits must be sealed with impervious materials such as puddle clay and plastic sheeting, and a 500 mm thick envelope of fine sand of 0.2 mm ES must surround the pit.\footnote{~Noncompliance with the requirements in the technical guideline can weaken the first-stage relationship. However, I find that the F-statistics of the first-stage regressions are not too low and show the confidence interval of the \cite{anderson1949estimation} test, which is robust to the weak instrument in Section \ref{sec:result}.} 
In short, higher infiltration rates (lower AWC) make it more difficult to find space for latrines and increase construction costs because of the need for additional investments after the SBM started in 2014.\footnote{~This consideration of AWC applies not only to pit latrines but also to latrines with septic tanks, which are usually equipped with soak pits that treat septic tank effluent. Soak pits are subject to similar requirements that depend on soil infiltration rates to prevent groundwater contamination \citep{cpheeo2013manual}.}
These additional requirements are expected to be particularly substantial in rural India, where houses are typically closely spaced, and financially constrained households are unlikely to afford additional investments beyond the fixed subsidy amount, which does not depend on AWC levels.

\begin{figure}[p]
	\begin{center}
		\includegraphics[width=0.5\textwidth]{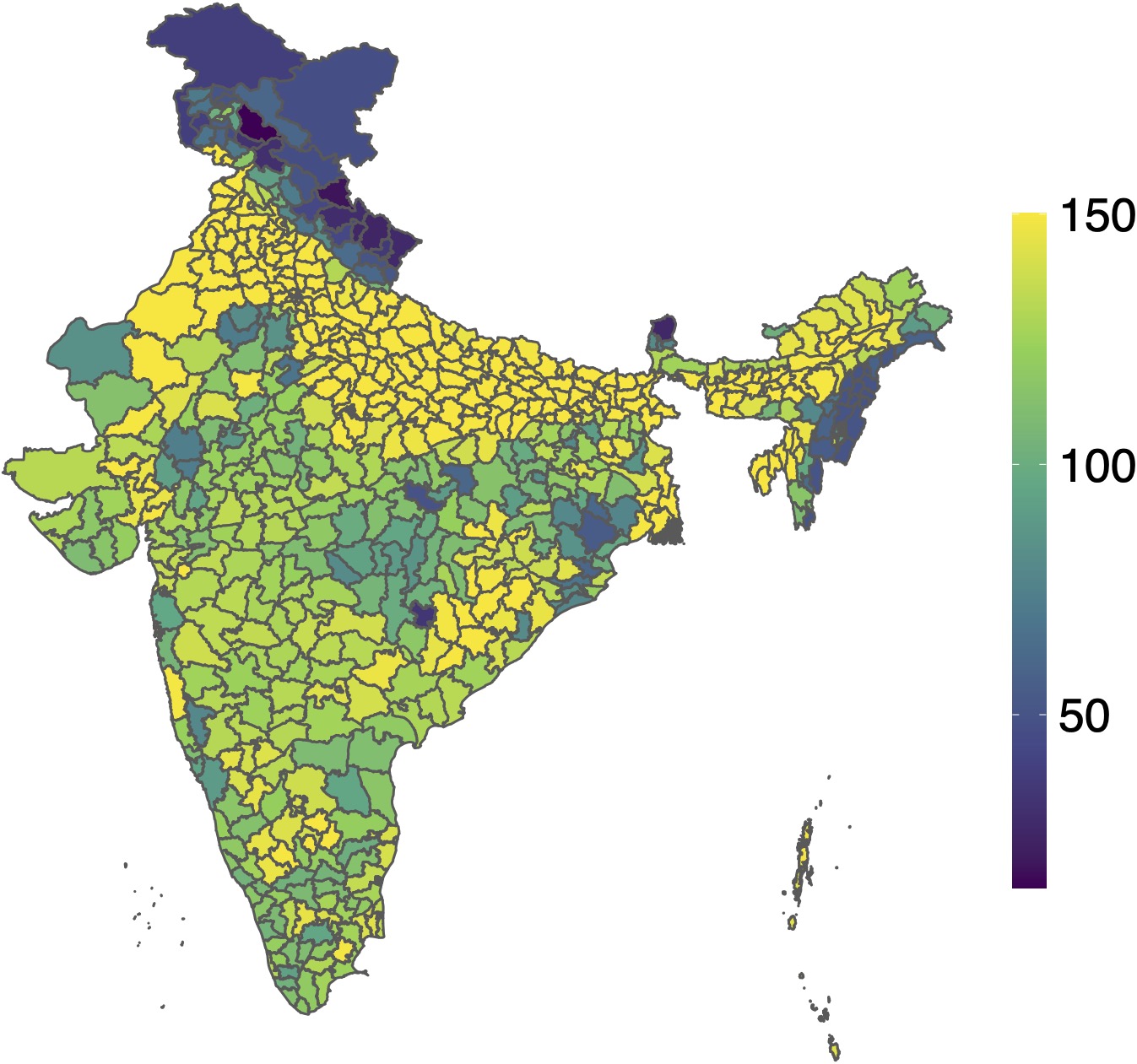}
		\caption{\label{fig:soil_awc} District-level Mean Available Water Capacity (mm/m)}
	\end{center}
\end{figure}

Therefore, in the first stage of the IV design, areas with a lower AWC are expected to experience a smaller increase in the number of latrines post-SBM. 
As expected, I find that a 1 mm/m decrease in AWC is associated with a smaller increase in the number of latrines per square kilometer by approximately 0.3 (Column 2 of Tables \ref{tab:result_water} and \ref{tab:result_health}). 
The F-statistics of the first-stage regressions are 30--50 for the water quality analysis and 79 for the health analysis.
Relatedly, Figure \ref{fig:soil_awc} shows the substantial variation in AWC across districts in India.

In the water quality analysis, I adopt the following two-stage least squares regressions, where regressions \ref{iv_st} and \ref{iv_ft} are the second- and first-stage regressions, respectively.
This IV design is conceptually similar to the DiD design, in which the reduced-form regression in this IV design uses AWC as a treatment variable.

\vspace*{-0.8cm}

\begin{equation}\label{iv_st}
	\begin{split}
		Y_{i,d,t} =  \delta_i + \theta_t + \beta_{IV} Latrine_{d,t} + \gamma_1 Precip_{d,t} + \varepsilon_{i, t} 
	\end{split}
\end{equation}

\vspace*{-0.8cm}

\begin{equation}\label{iv_ft}
	Latrine_{d,t} =  \delta_i + \theta_t + \pi_1 AWC_{d} \cdot Post_{t} + \pi_2 Precip_{d,t} + \nu_{d, t} 
\end{equation}

\vspace*{0.2cm}

\noindent where $Y_{i,d,t}$ is a water quality indicator, represented by the logarithm of fecal coliform, at monitoring station $i$ located in district $d$ in year $t$. 
$Latrine_{d, t}$ is the number of household latrines per square kilometer in district $d$ in year $t$. 
$Precip_{d,t}$ is the precipitation in district $d$ in year $t$, which is added to control for rainfall and associated floods, which may affect both water quality and latrine construction. 
I construct a time-variant instrument for the panel data analysis by interacting the time-invariant AWC in district $d$ with a post-SBM indicator that takes the value of one after 2014, when the SBM started. 
Monitoring station fixed effects ($\delta_{i}$) are included to control for the time-invariant characteristics of each monitoring station (and, more broadly, of each district), including the positions of stations along rivers and cross-sectional socioeconomic disparities across districts.
Year fixed effects ($\theta_{t}$) are also included to account for secular trends in water quality over time, which may be influenced by changes in water quality regulations.
Standard errors are clustered at the district level because variation in the number of latrines is observed at this level. 
Lastly, the coefficient of interest is $\beta_{IV}$ and is expected to be positive in the water quality analysis.

The IV design builds on the key assumption of exclusion restrictions: the instrument ($AWC_{d} \cdot Post_{t}$) must affect outcomes only through the channel of latrine construction after controlling for precipitation, monitoring station fixed effects, and year fixed effects.
Conditioning on monitoring station fixed effects implies that any violation of the exclusion restriction would arise from time-varying monitoring station (or district) characteristics that are correlated with AWC and change post-SBM.

To address the concern of exclusion restrictions, I choose fecal coliform as a water quality outcome. 
One potential concern is that AWC, which measures soil quality, affects the agricultural yield of crops. 
This, in turn, can affect the volume of agricultural runoff, leading to changes in water quality. 
Therefore, as a water quality outcome, I use fecal coliform, which is primarily affected by fecal sludge from latrines and is unrelated to crop production.

In the health analysis, the exclusion restriction is a more legitimate concern, which motivates me to adopt the upstream--downstream specification described in the following section.
For instance, AWC might affect the agricultural yield of crops, which, in turn, could determine household income.
This change in income could affect the level of health investment, leading to changes in health conditions.\footnote{~Although district fixed effects control for time-invariant agricultural productivity across districts, differential growth in agricultural yield and income caused by different levels of AWC may still be present.}
To address this concern, I adopt the upstream--downstream specification, using upstream AWC as an instrument for upstream latrine construction, and examine its effect on downstream outcomes.
Further tests to check the validity of the exclusion restriction, including parallel pre-trends and falsification tests, are presented in Section \ref{sec:validity_exclusion_restriction}.

My IV estimates, both in the baseline and upstream--downstream specifications, identify the local average treatment effect (LATE) of latrine construction. 
This LATE is likely to be primarily driven by backward districts, where latrine construction is more constrained by higher costs or construction difficulties when AWC is lower.
In these complier districts, lower education levels are likely to translate into lower awareness of pollution risks and weaker waste-management practices, making fecal sludge dumping more prevalent. 
As a result, the IV estimate of the water pollution effect may be larger than the average treatment effect (ATE) across districts in India.
At the same time, backward districts tend to have higher baseline diarrheal mortality, leaving more room for health improvements under the SBM. 
Thus, the IV estimate of the health effect may also be larger than the ATE.
However, these backward districts are also likely to be the main beneficiaries of the SBM subsidy program, so the IV estimates are policy-relevant and informative about the effects of latrine construction under the SBM program.

\subsection{Upstream--Downstream Specification} \label{sec:iv_updown}

The negative externalities of water and health may spill over to downstream areas because dumped fecal sludge can flow downstream along rivers. 
Thus, I adopt an additional upstream--downstream specification to examine the effects of upstream latrine construction on downstream water quality and health. 

The upstream--downstream specification addresses the concern of exclusion restrictions in the baseline IV specification.
The baseline specification is modified using upstream AWC as an instrument for upstream latrine construction to examine its effects on downstream outcomes while controlling for downstream AWC. 
Upstream AWC may affect upstream agricultural output and income, which, in turn, could affect health outcomes in the same area.
However, I do not expect upstream AWC to affect downstream health outcomes through changes in downstream income, as upstream AWC is unlikely to affect downstream agricultural output after controlling for downstream AWC.
Therefore, by adopting upstream AWC, which is unrelated to downstream agricultural output and income, as an instrument and focusing on downstream health as an outcome, this specification addresses the concern of the exclusion restriction.
In other words, the upstream--downstream specification, which relies on the instrument and outcomes in different locations along the rivers, enhances the validity of the exclusion restriction.\footnote{~This approach of using upstream--downstream relationship in the IV design aligns with the methodology adopted in \cite{dias2023down}.}

\begin{figure}[p]
	\begin{center}
		\includegraphics[width=0.5\textwidth]{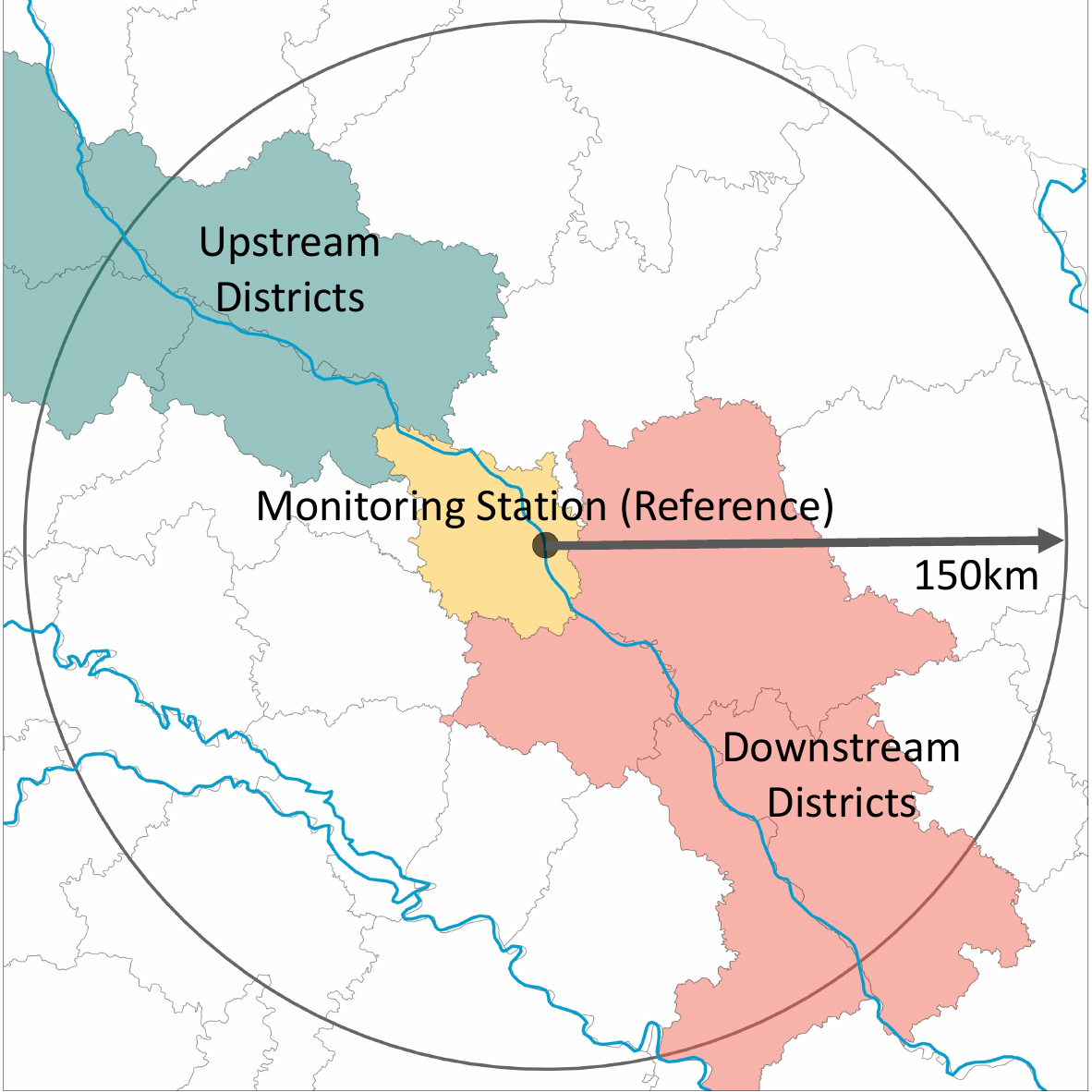}
		\caption{\label{fig:updown_map} Illustration of Upstream--Downstream Analysis}
		\medskip 
		\begin{minipage}{1\textwidth} 
			{\small
				Notes: This figure illustrates the upstream--downstream analysis, which analyzes the effect of upstream latrine construction on water quality in a reference monitoring station (or health in a reference district). Upstream districts are selected as districts that (i) intersect with river segments whose elevations are higher than the elevation of the reference station (district) and (ii) fall within a range of [0, 150] km from the reference station (district) in the baseline specification. This figure shows district boundaries in grey lines and rivers in blue lines. It highlights the upstream districts in green, the reference district in yellow, and the downstream districts in red.
				\par}
		\end{minipage}
	\end{center}
\end{figure}

I identify the upstream--downstream relationships between monitoring stations and districts using elevation data along 43 major rivers.\footnote{~I focus on major rivers included in the Version 4.1.0 GIS polygons of rivers provided by the Natural Earth. Upstream--downstream relationships along major rivers are less susceptible to measurement errors because the river systems are simpler than those that include hundreds of rivers. I also use 90 m raster elevation data, called the Shuttle Radar Topography Mission data Version 4.1 \citep{reuter2007evaluation}.} 
The analysis focuses on a subset of monitoring stations and districts located along these major rivers, with districts situated further upstream.\footnote{~This focus results in a sample of 365 stations in 154 districts for the water quality analysis and a sample of 103 districts for the health analysis. Here, I also drop districts where more than one major river flows, owing to the complexity of determining the upstream--downstream relationships.}
As shown in Figure \ref{fig:updown_map}, the upstream districts of a given district (station) are selected as those that intersect with river segments whose elevations are higher than that of the given district (station).\footnote{~As this process is repeated for all districts (stations) along major rivers, nearly all districts (except for the most downstream ones) are used as upstream districts in the analysis. Thus, there is limited concern that specific characteristics of upstream districts are driving the upstream--downstream results.}

The definition of upstream districts, that is, how far upstream one should search for districts, matters because pollution decays as it flows downstream. 
Because the decay rates depend on the temperature and other environmental factors of rivers, I adopt a variety of distances from a given district (station) to identify the upstream districts.
Specifically, for a given district (station), the upstream districts are selected from those that fall within a range of $[X,Y]$ km of the given district (station), where $X\in \{0,50,100\}$, $Y\in\{100,150\}$, and $X<Y$. 
I use a range of $[0,150]$ km as the baseline specification; however, the results remain robust when using alternative buffer sizes or when considering all upstream districts without buffers, as shown in Appendix Table \ref{tab:updown_change_buffers}.

The upstream--downstream analysis adopts regressions \ref{iv_st_updown} and \ref{iv_ft_updown} modified from the baseline IV specification. 
I change the independent variable to the upstream number of latrines per square kilometer, and the instrument to the upstream AWC.\footnote{~If there are multiple upstream districts, I construct the independent variable by dividing the total number of latrines by the total area of these districts. Additionally, I construct the instrument by taking the average of the AWC values from these districts.} 
I also control for AWC in the reference district because the instrument (upstream AWC) can be spatially correlated with AWC in the reference district, which can also affect the outcomes.

\vspace*{-0.8cm}

\begin{equation}\label{iv_st_updown}
	\begin{aligned}
		Y_{i,d,t} =  \delta_i + \theta_t + \beta_{IV}^{U} Upstream\_Latrine_{d,t} + \gamma_1 Precip_{d,t} + \gamma_2 AWC_{d} \cdot Post_{t} + \varepsilon_{i, t} 
	\end{aligned}
\end{equation}

\vspace*{-0.6cm}

\begin{equation}\label{iv_ft_updown}
	\begin{aligned}
	Upstream\_Latrine_{d,t} &= \delta_i + \theta_t + \pi_1 Upstream\_AWC_{d} \cdot Post_{t} + \pi_2 Precip_{d,t} \\ &+ \pi_3 AWC_{d} \cdot Post_{t} +  \nu_{d, t}
	\end{aligned}
\end{equation}

\vspace*{0.2cm}

In the health analysis, the outcome variable ($Y_{d,t}$) is defined as the district-level diarrheal child mortality rate.
I focus specifically on the post-neonatal mortality rate because it is the closest available measure to the infant mortality rate, which is known to be significantly impacted by poor sanitation and water pollution \citep{do2018can,geruso2018neighborhood}.\footnote{~Post-neonatal and infant mortality rates refer to the probabilities of a child dying between 28 days after birth and the age of one year and dying between the birth and the age of one year, respectively.}
As the health analysis uses district-level panel data, district fixed effects are used instead of monitoring station fixed effects.
The standard errors are clustered similarly at the district level.

The coefficient of interest $\beta_{IV}^{U}$ captures the effect of upstream latrine construction, which comprises two underlying channels: (i) the direct effect of upstream latrine construction on outcomes and (ii) the indirect effect of upstream latrine construction on outcomes via correlated latrine construction in the reference district, as illustrated in Appendix Figure \ref{fig:updown_analysis}.
The correlation between latrine construction in upstream and reference districts in the second channel arises because this upstream--downstream specification does not explicitly control for latrine construction in the reference district.
The presence of both channels indicates that this analysis captures the combined effects of both the spillover effect (first channel) and the local effect (second channel).\footnote{~One potential approach to isolate the spillover effect from the local effect would be to control for latrine construction in the reference district by using AWC in that district as an additional instrument, resulting in two endogenous variables and two instruments. However, I do not adopt this approach owing to the associated weak instrument issue. For the same reason, this approach was not used as the main specification in the analogous upstream--downstream analysis by \cite{dias2023down}.}
Although this analysis does not disentangle the two effects, the total effects of water pollution externalities from latrine construction should remain a key policy interest.

I expect that, in the first channel, upstream latrine construction leads to water pollution that flows downstream, subsequently causing a negative externality to health in the reference district.\footnote{~This upstream--downstream interpretation assumes river-mediated transport. However, if fecal sludge is trucked and directly dumped at downstream locations, spillovers could arise without river transport. This is unlikely to be a major driver of the results, as dumping plausibly occurs close to collection sites, often within the same district, and I focus on spillovers to downstream districts up to 150 km away along rivers.}
In the second channel, latrine construction in the reference district, which is positively correlated with upstream latrine construction, is expected to contribute to increased water pollution in the same district.\footnote{~The positive correlation of latrine construction is expected because upstream and reference districts are usually subject to similar levels of SBM policy implementation by the same state government, which is discussed in Section \ref{sec:result_health}.}
The sign of the health effect in the second channel depends on the relative magnitude of the direct positive health effects and water pollution externalities resulting from latrine construction (reduced open defecation) in the reference district.
Thus, $\beta_{IV}^{U}$ is expected to be positive for the water quality outcome because an increase in water pollution is expected in both channels.
However, the sign of the overall health effect is ambiguous because it depends on the sign and relative magnitude of the health effect in each channel.
For example, if the net health effect becomes positive in the second channel and the magnitude of this net positive health effect surpasses that of the water pollution externalities in the first channel, the overall estimated health effect could be positive.

\subsection{Validity of Exclusion Restriction} \label{sec:validity_exclusion_restriction}

Two tests are conducted to check the validity of the exclusion restriction: parallel pre-trends and falsification tests.

First, I check parallel pre-trends in the reduced-form regressions of water quality and health outcomes on the interaction of AWC with the year dummies.
The exclusion restriction implies that AWC should not affect outcomes through other channels before the implementation of the SBM policy.
During the pre-SBM period, AWC is unlikely to affect latrine construction because the official technical guideline, which stipulated requirements based on soil infiltration rates, was not published until 2013, just before the start of the SBM.\footnote{~The non-differential effect of AWC on latrine construction during the pre-SBM period is illustrated in the first-stage event study plots presented in Appendix Figure \ref{fig:eventstudy_awc_latrine_wqreg}. The differential effect of AWC becomes statistically significant from 2014 onward.}
Thus, the association between AWC and outcomes during the pre-SBM period captures causal pathways other than the latrine construction.
Conversely, after the SBM started to incentivize latrine construction in 2014, AWC is expected to have a significant relationship with outcomes by affecting latrine construction.
By extending the upstream--downstream specification, I test whether upstream AWC has differential effects on outcomes in the reference district during both the pre-SBM and post-SBM periods.
The reference year in this event study analysis is set to 2013, one year before the SBM started.

The reduced-form event study results show that upstream AWC did not have a differential effect on either water quality or health prior to the SBM policy (up to 2013), supporting the validity of the exclusion restriction in the upstream--downstream specification (Panels B and C of Figure \ref{fig:event_result_awc}).
Furthermore, AWC did not have a differential effect on water quality before the SBM policy in the baseline IV specification (Panel A).
In contrast to water quality, the baseline IV specification shows a differential pre-trend for health outcomes prior to the SBM policy. 
Therefore, for health outcomes, I treat the upstream--downstream specification as the preferred design and present only these results in the subsequent sections.

While differential pre-trends are not observed in the event study analyses, one may still be concerned that other district-level factors correlated with AWC and changing after 2014, such as post-2014 agricultural policies or weather shocks, could confound the results.\footnote{~However, to the best of my knowledge, major agricultural reforms were limited to the 2020 attempt (beyond my sample period) to introduce three new farm acts, following unsuccessful efforts to induce state-level agricultural market reforms through two acts in 2017 and 2018 \citep{chand2020}.}
These agriculture-related confounders could increase agricultural wastewater (e.g., fertilizer and pesticide pollution) or affect agricultural income, which, in turn, could influence health outcomes.
To test for the presence of these confounders indirectly, I examine the effects on a water quality indicator related to agricultural wastewater and a health outcome influenced by income in the following falsification tests.

As a second test of the validity of the exclusion restriction, I conduct falsification tests to examine the effects on water quality and health outcomes that are unrelated to fecal contamination but may be related to my instrument.
Specifically, I examine the effects of latrine construction on other water quality indicators from the NWMP dataset, as well as the prevalence of overweight in children aged 0--5 years from \cite{ihme2020overweight}.
Reassuringly, I find no effect on nitrate-nitrite levels, which primarily reflect fertilizer contamination in agricultural wastewater, in both the baseline and upstream--downstream specifications (Columns 1--2 of Appendix Table \ref{tab:falsification}).
This null result suggests that my health results are not driven by agriculture-related confounders affecting the volume of agricultural wastewater.
I also find a largely insignificant effect on water temperature, which is also unrelated to fecal contamination (Columns 3--4).\footnote{~In further analyses, I find insignificant effects on biochemical oxygen demand (BOD) and dissolved oxygen (DO), both of which measure water contamination from various pollution sources, including agricultural and industrial wastewater (Columns 5--8 of Appendix Table \ref{tab:falsification}). While BOD and DO can also be affected by fecal sludge from latrines, its partial contribution to overall BOD and DO levels may explain the null effects. These results, along with the null effect on nitrate-nitrite, also suggest no major sources of water pollution other than fecal contamination from latrine construction.}
Moreover, I find no effect on overweight prevalence, which could be influenced by agricultural income, in the upstream--downstream specification (Column 9).
This null result reinforces my previous argument that, in this specification, my instrument is unlikely to affect health outcomes through the income channel.

\section{Results} \label{sec:result}

\subsection{Effects on Water Quality} \label{sec:result_wq}

I find that latrine construction under the SBM degrades river water quality, and the water pollution externality spills over to downstream areas.

In the baseline specification, Table \ref{tab:result_water} shows that one additional latrine per square kilometer increases fecal coliform by 3\% on average (Column 3 of Panel A). 
This IV estimate is substantially larger than the OLS estimate, which is approximately 0.6\% (Column 1 of Panel A). 
This difference may be due to a downward bias in the OLS estimate caused by time-varying omitted variables, such as unobserved practices of open defecation that increase water pollution and slow down latrine construction.
Another possible explanation is that the IV estimate reflects the LATE in backward (complier) districts, where fecal sludge dumping may be more prevalent due to lower awareness of water pollution risks and weaker waste-management practices (as discussed in Section \ref{sec:iv_baseline}).

The estimated effect is robust to alternative inference and outcome specifications.
The standard errors remain stable when adjusting for spatial correlation in water pollution that may extend beyond district borders using the \cite{conley1999gmm} approach with a 150 km cutoff, which corresponds to the baseline and maximum distance threshold used to define upstream districts (Columns 1 and 3 of Panel A of Table \ref{tab:result_water}).
The effect remains robust when the yearly maximum value of fecal coliform is used as an outcome, showing that latrine construction also increases peak pollution levels, which may matter more for health risks (Appendix Table \ref{tab:result_water_max}).
The results are also robust when fecal coliform values are winsorized at the 99th, 95th, 90th, and 75th percentiles (Appendix Table \ref{tab:result_water_winsorized}).

The result of the first stage shows an expected positive association between AWC and the number of latrines (Column 2 of Panel A of Table \ref{tab:result_water}). 
Although the F-statistics of the first stage are not low (29.954), I compute the 95\% confidence interval of the \cite{anderson1949estimation} test, which is robust to weak instruments. 
The positive left and right ends of the 95\% confidence interval ([0.015, 0.049]) show that the results are robust to the \cite{anderson1949estimation} specification.

The total effect of the SBM (hereinafter called the ``average policy effect'') is a 72\% increase in fecal coliform, which shows a substantial water pollution externality (Column 3 of Panel A). 
The average policy effect is calculated as a back-of-the-envelope estimate by multiplying the estimated coefficient (3\%) by the difference in the mean number of latrines per square kilometer between the pre-SBM period (2012--2013) and the post-SBM period (2014--2019) in the regression sample (24.19).\footnote{~This back-of-the-envelope calculation relies on simplifying assumptions, including a constant per-unit effect of latrine construction. It also abstracts from the natural decay of fecal coliform over time. However, this is unlikely to be a major concern because the outcome is measured annually and fecal coliform levels typically decay on shorter time scales than a year, so within-year decay is reflected in the annual fecal coliform levels. The same approach is used to calculate the average policy effects reported in subsequent tables.}
To provide a more policy-relevant interpretation of the effect size, I also examine the effect on the probability that the average fecal coliform level exceeds the maximum permissible level for bathing water specified in the ``Primary Water Quality Criteria for Bathing Water'' (2,500 MPN/100ml).
I find that the SBM increases the probability of such violations by 15.6 percentage points, which is a sizable effect relative to the pre-policy mean of 23.3\% (Column 1 of Appendix Table \ref{tab:result_water_violation}).

My estimate of a 72\% increase in fecal contamination owing to the SBM surpasses those of most previous studies. 
This substantial increase in water pollution can be attributed to more than a doubling of latrine coverage (from 39.2\% in 2013) during the SBM period.
My estimate is considerably larger than the effect of each additional border crossing induced by a border change on water pollution levels (3\% increase) in Brazil \citep{lipscomb2016decentralization} and the effect of each additional Clean Water Act grant to wastewater treatment plants on fecal coliform (3.6 \% decrease) in the United States \citep{keiser2019consequences}. 

\begin{landscape}
\begin{figure}[p]
	\begin{center}
		
		\minipage{0.55\textwidth}
			\includegraphics[width=\linewidth]{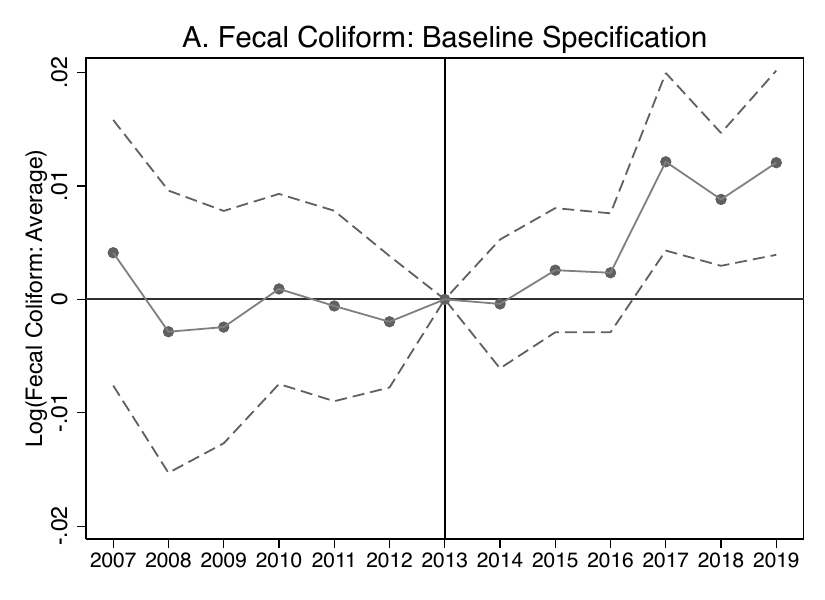}
			\endminipage\hfill \\
			\minipage{0.55\textwidth}
			\includegraphics[width=\linewidth]{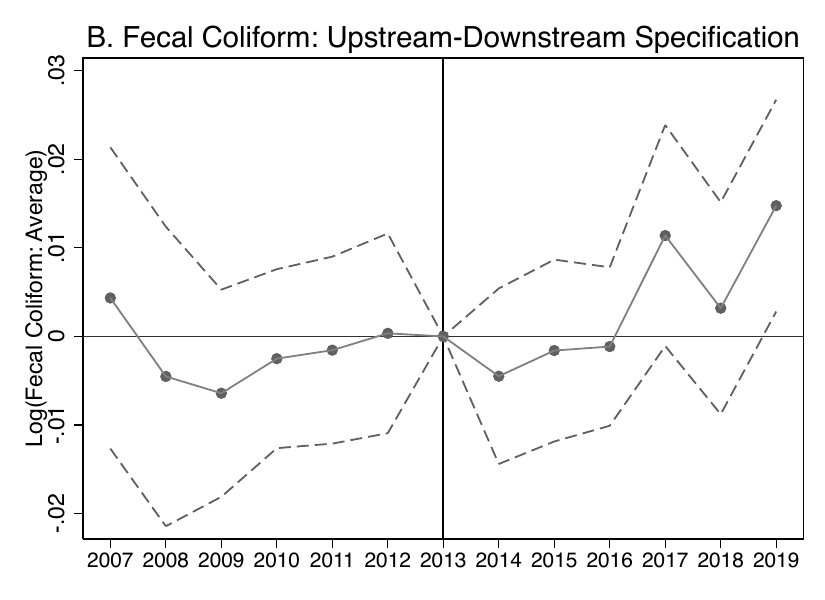}
			\endminipage
			\minipage{0.55\textwidth}%
			\includegraphics[width=\linewidth]{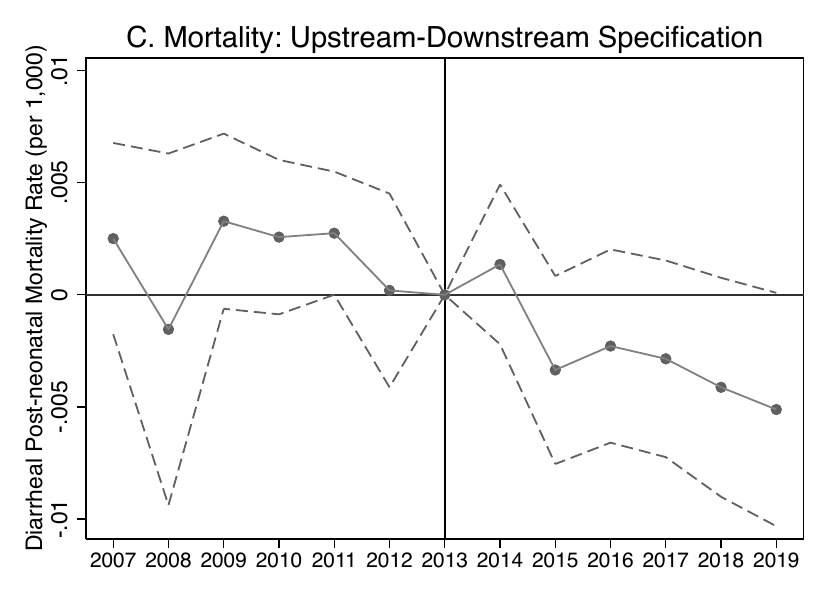}
		\endminipage
					
		\caption{\label{fig:event_result_awc} Event Study Plots of Reduced-Form Regressions}
		\medskip 
		\begin{minipage}{1.2\textwidth} 
			{\small 
				Notes: This figure shows the regression coefficients of the logarithm of fecal coliform (Panels A and B) and diarrheal post-neonatal mortality rate per 1,000 children (Panel C) on the interaction terms between Available Water Capacity in Panel A (upstream Available Water Capacity in Panels B and C) and year dummies.
				The 95\% confidence intervals are shown with dashed lines. 
				Standard errors are clustered at the district level. 
				Panel A includes monitoring station fixed effects, year fixed effects, and precipitation as a control.
				Panel B includes monitoring station fixed effects, year fixed effects, and the following controls: precipitation and the interaction of Available Water Capacity and the post-SBM indicator of a reference district, while Panel C includes district fixed effects, year fixed effects, and the same controls.
				\par}
		\end{minipage}
	\end{center}
\end{figure}

\end{landscape}

As for dynamic effects, the reduced-form event study analyses show that effects on water quality emerge with lags and intensify over time.
In the event study analyses described in Section \ref{sec:validity_exclusion_restriction}, the differential effects of AWC on fecal coliform grow over time and become statistically significant from 2017, three years after the start of the SBM, in the baseline specification (Panel A of Figure \ref{fig:event_result_awc}).
This lagged effect is consistent with the pit filling and emptying cycles.
Typically, pit latrines take between 1.5 to 3 years to fill with fecal sludge and to require emptying, as outlined in the technical guideline \citep{cpheeo2013manual}.
Thus, the onset of increased water pollution is likely to lag by a similar length of time, as water quality effects are likely to emerge once desludging begins and emptied fecal sludge is dumped.
Moreover, because the timing of SBM implementation can vary across places and the policy effort intensified over time, staggered latrine construction and subsequent desludging can also contribute to the lagged and increasing pattern in water pollution.

In the upstream--downstream specification, I find that the water pollution externality of latrine construction spills over to the downstream districts, especially several years after the SBM started.
Column 3 of Panel B of Table \ref{tab:result_water} shows that one additional upstream latrine construction per square kilometer increases fecal coliform by 1.5\% on average, which amounts to a total increase of 43\% under the SBM.
Although this average effect is imprecise, the results of the reduced-form event study indicate statistically significant water pollution externalities around the years 2017--2019, with lags similar to those in the baseline specification (Panel B of Figure \ref{fig:event_result_awc}).
Moreover, in the heterogeneous analysis in Section \ref{sec:result_het_treatment_wq}, I find statistically significant pollution spillover effects when upstream areas have low treatment capacities for fecal sludge (Columns 3 and 5 of Panel B of Table \ref{tab:result_het_water}).
The average effect shown in Column 3 of Panel B of Table \ref{tab:result_water} obscures these dynamic and heterogeneous effects, leading to the imprecise estimate.
Nonetheless, I find that the increased downstream water pollution translates into a statistically significant increase in the probability of violating the water quality criteria for bathing water by 11.5 percentage points under the SBM (Column 2 of Appendix Table \ref{tab:result_water_violation}).

\subsection{Effects on Health} \label{sec:result_health}

I find that latrine construction under the SBM improves overall health, which suggests that the direct positive health effect of reduced open defecation outweighs the negative externality on health owing to increased water pollution.
For health outcomes, I report results using only the upstream--downstream specification, which is my preferred design based on the validity checks for the exclusion restriction in Section \ref{sec:validity_exclusion_restriction} (the same applies to the heterogeneity analysis in Section \ref{sec:result_het_treatment_health}).

In the upstream--downstream specification, I find that upstream latrine construction reduced diarrheal post-neonatal mortality in the reference district overall (Column 3 of Panel A of Table \ref{tab:result_health}).
Although upstream latrine construction can negatively affect health by causing water pollution spillovers to the reference district (the direct effect discussed in Section \ref{sec:iv_updown}), it can also improve health outcomes via correlated latrine construction (reduced open defecation) in the reference district (the indirect effect discussed in Section \ref{sec:iv_updown}).
The overall positive health effect indicates that the net positive health effect from increased latrine construction in the reference district outweighs the water pollution externalities from the upstream districts.
This overall positive health effect remains robust when the diarrheal mortality rates for other age groups are used as outcomes (Appendix Table \ref{tab:result_health_multiple}).

As supporting evidence for the indirect effect, I find a positive correlation between latrine construction in the upstream and reference districts (Column 3 of Panel B of Table \ref{tab:result_health}).
This positive correlation can be attributed to the fact that these districts are usually located within the same state, given that the buffer size for identifying the upstream districts is 150 km.\footnote{~84\% of the 103 reference districts have at least one upstream district within the same state.} 
Because states play a central role in implementing the SBM policy in India, districts within the same state are likely to undertake similar levels of latrine construction.\footnote{~To test this claim, I estimate the intra-cluster correlation coefficient, which measures the proportion of the overall variance that is explained by the within-state variance in the change in the number of latrines per square kilometer from 2013 to 2019. The coefficient is estimated to be 0.704, indicating that districts within the same state behave similarly in terms of latrine construction.}
While other state-level policies, such as agricultural policies, could similarly lead to correlated agricultural investments across districts and affect health outcomes through changes in agricultural wastewater or income, the pre-trends and falsification tests in Section \ref{sec:validity_exclusion_restriction} suggest a low likelihood of these alternative channels.

Regarding the magnitude of the health effect, one additional upstream latrine construction per square kilometer reduces the diarrheal post-neonatal mortality rate per 1,000 children by 0.011, which is a 0.4\% decrease from the pre-SBM period (Column 3 of Panel A of Table \ref{tab:result_health}).\footnote{~The IV estimate is larger in magnitude than the OLS estimate (Columns 1 and 3 of Panel A of Table \ref{tab:result_health}). This difference may be due to bias in the OLS estimate caused by reverse causality, where an increase in the diarrheal mortality rate leads to more latrine construction as a response. Another possible explanation is that the LATE in the IV design is driven by backward districts, which have greater potential for reducing diarrheal mortality due to their higher baseline levels (as discussed in Section \ref{sec:iv_baseline}).}
The average policy effect of the SBM is calculated to be a 0.269 reduction in diarrheal post-neonatal mortality rate per 1,000 children, which amounts to a 10\% reduction from the pre-SBM period.
This total 10\% reduction in diarrheal mortality under the SBM policy is smaller than the effects noted in previous studies, such as that of \cite{geruso2018neighborhood}, who reported a 48\% decrease in the infant mortality rate associated with a 60 percentage point reduction in the fraction of neighbors defecating in the open, a change similar in magnitude to the SBM policy.
This discrepancy is consistent with water pollution externalities that offset the health gains from reductions in open defecation.
By studying a large-scale sanitation program, this paper highlights that scaling up latrine construction can generate negative externalities, leading to smaller health gains than those documented in earlier studies that focus on local, village-level impacts.

Turning to dynamic effects on health, the reduced-form event study analyses show that the net positive health effects become more pronounced over time.
Specifically, the differential effects of AWC on diarrheal post-neonatal mortality become increasingly negative over time (Panel C of Figure \ref{fig:event_result_awc}).
This gradual increase in health gains could reflect a lag between latrine construction and sustained use: my treatment captures latrine construction rather than actual use.
Because regular latrine use requires behavioral change, which prior studies show is challenging \citep[e.g.,][]{augsburg2022nature}, consistent use may increase only gradually, generating delayed improvements in child health.

\subsection{Robustness Checks} \label{sec:robustness}

The results are robust to an alternative DiD design, consideration of spillovers from neighboring districts and urban areas, and the adoption of a balanced panel and alternative mortality dataset.
While this section mainly tests the robustness of the baseline results, it also discusses the robustness of the heterogeneous effects examined in Section \ref{sec:result_het_treatment}.

\vspace{0.1in}


\noindent \textit{Alternative DiD Design.}---I adopt a DiD design that exploits the differential increase in latrine coverage across districts with different levels of baseline coverage. 
All districts achieved almost universal latrine coverage by the target date of 2019, regardless of their baseline latrine coverage.
Consequently, districts with lower baseline latrine coverage experienced a larger increase in latrine coverage, which may have led to a larger increase in water pollution.
The DiD design thus uses the baseline latrine non-coverage in 2013, interacted with a post-SBM indicator, as a treatment variable (see Appendix \ref{sec:did} for more details).

As shown in Appendix Table \ref{tab:did_result_waterquality}, the DiD results are similar to those of the IV design. 
I find a negative effect on water quality, although the overall effect is imprecise (Column 1).
Consistent with the heterogeneous effects shown in Section \ref{sec:result_het_treatment}, this negative effect is significant only in areas with lower treatment capacities (Columns 2--5).
The event study results show that parallel pre-trends hold, and the water pollution effect becomes more pronounced over time in states with lower treatment capacities (Panel B of Appendix Figure \ref{fig:wq_event_result_did}).

\vspace{0.15in}


\noindent \textit{Spillovers from Neighboring Districts.}---The baseline analysis assumes that the water quality at a given monitoring station is affected only by latrine construction in the district where the station is located. 
However, monitoring stations can be situated in rivers flowing along the borders of several districts. 
In this case, the water quality at these stations is likely to be affected by these neighboring districts. 
Therefore, I conduct an additional analysis that incorporates spillover effects from neighboring districts.
For monitoring stations located within 2 km of more than one district, I compute the weighted average of variables of neighboring districts using district areas as weights.
Data from other monitoring stations remain unchanged. 
I then re-run the baseline IV regression using this modified dataset.

As shown in Appendix Table \ref{tab:neighbor}, I find similar results: a negative effect on water quality (Column 1), driven by areas with lower treatment capacities (Columns 3 and 5).

\vspace{0.15in}


\noindent \textit{Influence from Urban Areas.}---While this paper focuses on the effects of latrine construction in rural India, the results could also be driven by latrine construction in urban areas. 
Therefore, I estimate the effects after excluding monitoring stations and districts close to urban areas from the sample. 
Specifically, I drop monitoring stations and districts within 50, 100, or 150 km of cities with a population of 1 million or more, according to the 2011 Census.

As shown in Appendix Table \ref{tab:influence_urban}, the results are robust to the exclusion of urban areas, regardless of distance. 
I find a negative effect on water quality, while the health effects remain positive but become less precise, possibly due to stronger offsetting by water pollution effects.

\vspace{0.15in}

\noindent \textit{Balanced Panel.}---The baseline analysis uses an unbalanced panel of water quality data to cover as many districts as possible to enhance the external validity of the results.
As a robustness check, I conduct the same analysis on a balanced panel to mitigate the concern that monitoring stations may have been endogenously installed in less-polluted locations over the sample period.
As shown in Appendix Table \ref{tab:balanced_wq_iv}, I find a negative effect on water quality, especially in areas with lower treatment capacities.

\vspace{0.15in}

\noindent \textit{Alternative Mortality Dataset.}---The baseline health analysis uses diarrheal mortality rate estimates from \cite{ihme2020} as outcomes, but these are predicted based on multiple household surveys.
Thus, I conduct the robustness check using the original infant mortality data from the National Family Health Survey 5 (NFHS-5), conducted in 2019--2021.
From the birth histories of women in households surveyed in NFHS-5, I use data concerning the year of childbirth and whether the child died within 12 months of birth, which serves as an infant mortality indicator.\footnote{~The NFHS data is limited by its mortality indicator encompassing all types of mortality rather than isolating those caused by water pollution, such as diarrheal mortality, which is why I do not adopt this dataset in the baseline specification. To match the NFHS-5 dataset with other datasets, I use the year of birth of the child and the geocoordinates of NFHS clusters (villages).}
I then conduct the same upstream--downstream analysis on this alternative outcome by focusing on children living close to rivers (within 5 or 10 km).\footnote{~I use rivers $\geq$30 m wide at mean annual discharge, available in the Global River Widths from Landsat Database \citep{allen2018global}. This dataset covers smaller rivers than the dataset of major rivers used to identify upstream districts. Including smaller rivers enables me to capture the pollution exposure of children living near these rivers, which branch out from the major rivers and are affected by their pollution.}
These children are more likely to be exposed to water pollution externalities from latrine construction that flow along rivers, as examined in the upstream--downstream analysis.
Although the outcome in this analysis is measured at the child level, the district-level variables remain the same as in the baseline health analysis, with additional controls included at the child and household levels.\footnote{~Child-level controls include indicators for being a first-born child and part of a multiple birth. Household-level controls include religion (Hindu, Muslim, others), caste (Scheduled Caste, Scheduled Tribe, Other Backward Class, others), education (primary, secondary, or higher), and wealth quintiles.}

As shown in Column 1 of Appendix Table \ref{tab:result_het_close_river_nfhs}, I find a consistent overall positive health effect, regardless of distance.
The magnitude of this effect is a reduction in the infant mortality rate by 1.3--2.0 per 1,000 children, representing a 3.4--5.5\% decrease from the pre-SBM period.\footnote{~Average policy effects are not presented in this robustness check because using the change in the number of latrines for the entire district to scale the effect only on children near rivers may not provide accurate estimates of the average policy effects. The same applies to the analysis of diarrheal mortality rates in river-adjacent areas, as discussed later in Section \ref{sec:result_het_treatment_health}.}
This effect size is larger than the baseline result for the diarrheal post-neonatal mortality rate, likely due to the inclusion of a broader age range (including neonatal) and the influence of other mortality causes correlated with diarrheal mortality.
Regarding the heterogeneous effects of the treatment capacity of fecal sludge, the positive health effects are smaller or statistically insignificant when upstream areas have lower treatment capacities (Columns 2-5), as demonstrated in the heterogeneity analysis in Section \ref{sec:result_het_treatment_health}.

\section{Heterogeneous Effects by Treatment Capacity of Fecal Sludge} \label{sec:result_het_treatment}

To identify the mechanism behind the negative externalities on water quality and health, I examine whether the effects of latrine construction on water quality and health vary by the level of complementary infrastructure for the treatment of fecal sludge.
The negative externalities are pronounced in areas with lower treatment capacities, where dumping of untreated fecal sludge is more likely to occur.
This suggests that insufficient treatment (or dumping) of fecal sludge is the primary mechanism driving these negative externalities.

In the heterogeneity analysis, I use geographical variation in the treatment capacities of STPs.
Based on the inventory of STPs compiled by the CPCB \citep{cpcb2015}, I calculate the STP capacities at both the state and district levels in 2013, one year before the SBM started.\footnote{~The district-level STP capacities are susceptible to measurement errors owing to missing observations of STPs in the CPCB inventory. Some districts may be flagged as districts with zero treatment capacity owing to missing observations, even though they may actually have STPs. Therefore, I also use state-level STP capacities, which are less susceptible to measurement errors because of broader aggregation.}
The baseline level of STP capacity is adopted to address concerns about the endogenous construction of STPs in response to water pollution caused by latrine construction.
While a potential concern of using the baseline STP capacity is that treatment effects could also be mediated by STP construction during the post-SBM period, CPCB data show that the change in STP capacity during this period was limited.\footnote{~First, consistent with the fact that planning and constructing STPs can take 5--10 years, the total STP capacity across India increased by only 52\% from 2013 to 2021, even though latrine coverage more than doubled in the same timeframe. Second, areas with lower baseline STP capacities did not experience a more substantial increase in STP construction, as indicated by the positive correlation between the baseline level of STP capacity in 2013 and the change in STP capacity from 2013 to 2021 at the state level (analysis based on \cite{cpcb2015,cpcb2021}). This second finding shows that baseline STP differences do not disappear or reverse during the post-SBM period.}
Therefore, in the baseline specification, I compare the effects in states (districts) that have higher baseline treatment capacities than the median in the sample with those in states (districts) with lower treatment capacities.\footnote{~I present heterogeneity results based on subgroup analyses rather than analyses using interaction terms between latrine construction and the high-capacity indicator because the latter approach introduces two endogenous variables and two instruments, leading to a weak instrument issue. The median value is calculated after assigning zero capacity to the states (districts) without any STP.}
In the upstream--downstream specification, I examine the heterogeneous effects of different levels of baseline treatment capacities in upstream states (districts).

This heterogeneity analysis does not explicitly consider STP treatment fees, which can also affect the amount of treated fecal sludge due to data limitations.
However, this approach is justified because the tipping fees for discharging at STPs are substantially lower than the revenues generated from emptying latrine pits for desluding truck operators.\footnote{~The tipping fees for discharge at STPs are approximately USD 1.2 (INR 100) and USD 6 (INR 500) per visit per truck in Chennai and Goa, respectively, according to case studies in \cite{cotreatment2018}. These fees are much lower than the revenues of truck operators, who typically charge between USD 6-30 (INR 500-2500) per household and visit a large number of households before discharging at STPs \citep{rao2020business}.}
This context suggests that truck operators are likely to transport fecal sludge to STPs when available.
Furthermore, variations in STP capacity implicitly account for variations in fees, as larger capacities are expected to result in lower marginal treatment costs and tipping fees.

\subsection{Effects on Water Quality} \label{sec:result_het_treatment_wq}

I find that the negative externality of water quality is concentrated in areas without adequate wastewater infrastructure. 
As shown in Table \ref{tab:result_het_water}, an additional latrine per square kilometer leads to a 5.1\% (3.7\%) increase in fecal coliform in districts (states) with lower treatment capacities, which amounts to a 134\% (98\%) increase under the SBM (Columns 3 and 5 of Panel A). 
Conversely, I find insignificant effects in states and districts with higher treatment capacities (Columns 2 and 4 of Panel A).
Similarly, in the upstream--downstream specification, I find that a negative externality from upstream latrine construction spills over downstream only when upstream areas have lower treatment capacities. The magnitude of these effects (82-114\% increase under the SBM) is comparable to those observed in the baseline IV specification (Columns 3 and 5 of Panel B).\footnote{~These insignificant effects in the case of higher treatment capacities also suggest that STP treatment fees do not substantially affect the amount of treated fecal sludge.}

These differential effects by the treatment capacity of fecal sludge suggest that the dumping of fecal sludge emptied from latrines is the primary mechanism contributing to increased river pollution.
In contrast, the insignificant effect on water quality in areas with higher treatment capacities suggests a low probability that alternative mechanisms, such as the direct seepage of fecal matter from latrines into rivers, affect water quality.
This alternative mechanism is discussed further in Section \ref{sec:alternative_mechanism}.

\subsection{Effects on Health} \label{sec:result_het_treatment_health}

I find that the overall positive health effect is eliminated when upstream areas lack adequate wastewater infrastructure.
The heterogeneity analysis allows me to explicitly investigate the negative externality on health through exposure to increased water pollution. 
A negative externality can be captured as the difference between health effects in areas with lower treatment capacities (where significant river pollution is observed) and those in areas with higher treatment capacities (where river pollution is insignificant).

Table \ref{tab:result_het_health} shows the heterogeneous effects on diarrheal child mortality rates by treatment capacities.
I find that the total effect of the SBM is a 15\% (42\%) decrease in diarrheal post-neonatal mortality rate from the pre-SBM period when upstream districts (states) have higher treatment capacities (Columns 2 and 4 of Panel A).
This corresponds to the cases in which the water pollution externalities are found to be insignificant in the water quality analysis.
However, the positive health effect is eliminated when upstream areas have lower treatment capacities and water pollution externalities are significant (Columns 3 and 5 of Panel A).
These heterogeneous health effects remain robust when examining the effects on mean diarrheal mortality rates only in areas close to rivers (within 5 or 10 km), where children are more likely to be exposed to water pollution, as demonstrated in the analysis of the alternative mortality dataset in Section \ref{sec:robustness} (Appendix Table \ref{tab:result_het_close_river_ihme}).\footnote{~A substantial portion of the population in the sample districts used for my analysis lived in close proximity to rivers. Specifically, 36\% of the population resided within 5 km of rivers, while 58\% lived within 10 km in 2011, according to the WorldPop raster data (Appendix Table \ref{tab:sumstats_appendix}). Additionally, for this robustness check, I use the same river dataset adopted in the analysis of the alternative mortality dataset.}
As another check of the role of pollution exposure, an additional heterogeneity analysis that splits districts by the population share living close to rivers, constructed from pre-SBM population raster data as detailed in Section \ref{sec:alternative_mechanism}, shows that the differential health effects by upstream district-level treatment capacity are present only in districts with high near-river population shares (Appendix Table \ref{tab:health_by_popshare}).
This pattern is consistent with negative health externalities being concentrated in areas with greater exposure to river pollution.

These findings, together with the water quality results, suggest that increased river pollution owing to the dumping of fecal sludge offsets the direct positive health effects.\footnote{~Although groundwater contamination in wells by latrines could be another mechanism for this negative health effect, it is unlikely, given the significant first-stage relationship suggesting that latrine construction generally complies with the technical guideline intended to prevent groundwater pollution.}
Although the overall health effect across India is positive, water pollution externalities diminish the effectiveness of latrine construction in improving child health.\footnote{~The heterogeneity results in this section allow me to gauge the offsetting magnitude of water pollution externalities. Specifically, these results provide a back-of-the-envelope decomposition of the two channels embedded in my upstream--downstream analysis: a direct pollution spillover and an indirect effect operating through local correlated construction. Using the district-level heterogeneity estimates, the implied spillover (negative for health) and the indirect component (positive for health) are of comparable magnitude and can offset each other (see Appendix \ref{sec:appendix_boe_0} for details).}

\subsection{Alternative Mechanism: Direct Contamination} \label{sec:alternative_mechanism}

The primary mechanism emphasized so far is pit emptying and subsequent dumping of fecal sludge into rivers.
This mechanism is supported by heterogeneity analyses showing that water pollution externalities are pronounced where fecal sludge treatment capacity is limited (Section \ref{sec:result_het_treatment_wq}), and it is also consistent with the lagged effects observed in the reduced-form event study, which align with typical pit filling and emptying cycles (Section \ref{sec:result_wq}).

However, another potential alternative mechanism is direct contamination, whereby fecal matter stored in pits leaks directly into rivers, particularly in settings where latrines are located close to rivers.
If this channel were quantitatively important, water pollution effects should be larger in districts where a greater share of the population lives near rivers, because more latrines are situated close to rivers and leaked fecal contaminants can reach river water through shorter hydrologic pathways.

To assess this channel, I examine heterogeneity in water pollution effects by the share of the population living near rivers.
Specifically, using WorldPop population raster data in 2011 (pre-SBM period), I compare districts above versus below the median share of the population living within 5 km or 10 km of rivers.

The results provide no supporting evidence for the direct contamination mechanism.
The estimated effects on fecal coliform are statistically significant in both districts with high and low near-river population shares (Appendix Table \ref{tab:wq_by_popshare}).
The estimates are, if anything, larger in districts with low near-river population shares, which is difficult to reconcile with direct contamination as the dominant mechanism.
This pattern is also consistent with the dumping mechanism, since fecal sludge emptied from pits can be transported for disposal in rivers far from households, and dumping may be easier in sparsely populated river areas due to lower visibility.
As additional evidence against direct contamination, the significant first-stage relationship between AWC and latrine construction suggests that latrine construction generally follows the technical guideline, making widespread leakage from pits into rivers unlikely.

\section{Conclusion} \label{sec:conclusion}

My analysis documents the unintended negative consequences of latrine construction when scaled up as a nationwide policy. 
Although open defecation is commonly blamed for water pollution externalities, I show that large-scale household latrine construction can generate even greater negative externalities due to insufficient infrastructure for fecal sludge treatment.

Specifically, I examine the consequences of the world's largest sanitation policy, the SBM in India, which subsidized the construction of over 100 million latrines.
I exploit the SBM's requirements for latrine construction to identify its causal effects on water quality and health. 
According to the official technical guideline, the soil infiltration rate determines the cost and difficulty of latrine construction after the SBM starts.
Therefore, the interaction of the soil infiltration rate with a post-SBM indicator is used as an instrument for latrine construction.

I find that the SBM increases fecal contamination of rivers by 72\% in rural India, which is a substantial effect. 
This water pollution externality exists only in areas with lower treatment capacities for fecal sludge, where dumping of fecal sludge is more likely to occur. 
Although the SBM reduces diarrheal mortality in children by 10\% overall, this positive health effect is eliminated when upstream areas have lower treatment capacities.
These heterogeneous effects suggest that water pollution externalities owing to the dumping of fecal sludge offset the direct positive health effects of reduced open defecation.

Back-of-the-envelope cost--benefit analyses of the SBM confirm the importance of complementing latrine construction with adequate wastewater infrastructure.
The mortality benefit alone is not worth the cost of the SBM policy under insufficient fecal sludge treatment.\footnote{~The total benefits would be larger than the estimated mortality benefit if I also included health effects for other age groups and additional benefits such as improved educational outcomes and reduced violence against women. Conversely, the total costs would be higher if I also considered the reduced recreational value of water quality, or they would be lower if some households do not use the full amount of the subsidy.}
The mortality benefit, that is, reduction in diarrheal post-neonatal mortality because of latrine construction, is calculated to be USD 5.6 million, which is about one-third of the subsidy cost for latrine construction (USD 16.9 million) at the district level.\footnote{~District-level cost--benefit analyses are conducted because the health effect sizes used are derived from district-level analyses of health and latrine data. The mortality benefit is calculated using the estimated average policy effect and the value of a statistical life in India. Meanwhile, the subsidy cost is calculated by multiplying the amount of the SBM subsidy by the increased number of latrines under the SBM. More detailed steps are described in Appendix \ref{sec:appendix_boe_1}.}
However, complementing latrine construction with adequate treatment of fecal sludge to mitigate negative externalities would substantially increase the mortality benefit at a lower cost.
The additional mortality benefit of having higher treatment capacity is calculated to be USD 7.4 million, which is larger than the additional cost of constructing and operating more STPs (USD 4.5 million) at the district level.\footnote{~The additional mortality benefit is calculated based on the difference in the estimated average policy effects between districts with higher and lower treatment capacities. The additional cost is calculated by multiplying the unit cost of sewage treatment plants by the difference in STP capacity between districts with higher and lower treatment capacities. More detailed steps are described in Appendix \ref{sec:appendix_boe_2}.}

The results have several policy implications for developing countries that promote sanitation and other developmental policies. 
The first clear implication is that policymakers should consider the possibility of the negative externalities of sanitation investments on water quality and health. 
An enabling environment that includes the effective treatment of fecal sludge by infrastructure can make sanitation policies more effective.
In the Indian context, this implies that state and local governments, supported by central government financing schemes, should align latrine expansion with investment in wastewater treatment infrastructure.
This policy direction is reflected in the second phase of the SBM, launched in 2020, which places greater emphasis on fecal sludge management, including co-treatment in STPs and the construction of FSTPs.
The need for better fecal sludge management is also a common issue in other developing countries, including Bangladesh, Nepal, and Pakistan, despite good progress in improving access to toilets \citep{wateraid2019}.
My heterogeneity results also suggest distributional implications: unequal complementary wastewater treatment infrastructure can lead to unequal health benefits from latrine construction across districts, underscoring the importance of coordinating latrine expansion with more evenly distributed investment in such infrastructure.

Second, my findings broadly highlight the unintended negative environmental consequences that can arise when a program is scaled up to a nationwide policy. 
This negative externality of a scaled-up policy aligns with a growing body of research that links conditional cash transfers to deforestation \citep{alix2013ecological} and rural road access to air pollution \citep{garg2023rural}.
By connecting a large-scale sanitation policy to water pollution, this paper further demonstrates that intensely promoting private goods without complementary infrastructure investments can cause environmental pollution, consequently reducing the effectiveness of these private goods in improving health.

\clearpage

\setlength\bibsep{0pt}
\begingroup
\fontsize{11}{11.5}\selectfont
\bibliographystyle{econ-aer}
\bibliography{sanitation_wq}
\endgroup

\clearpage


\vspace{-0.6cm}

\vspace{-0.6cm}

\clearpage

\clearpage


\begin{table}[p]\centering \caption{Summary Statistics \label{tab:sumstats}}
	\resizebox{1\textwidth}{!}{
		\begin{threeparttable}

			\begin{tabular}{@{}lccccc@{}}
	\toprule
	& Mean & SD & Min & Max & Obs. \\ 
	\midrule
	\textit{Panel A. Time-varying variables: pre-SBM (2007-2013)} &     &   &    &   &  \\
	\addlinespace
Fecal coliform: Average (million MPN/100ml) & 2.61 & 143.77 & 0 & 10000.04 & 4939\\
Diarrheal post-neonatal mortality rate (per 1,000) & 2.69 & 1.8 & 0.07 & 9.48 & 2359\\
Number of latrines (ten thousand) & 12.93 & 13.39 & 0.01 & 89.7 & 586\\
Number of latrines per sq. km & 35.55 & 41.92 & 0.03 & 283.01 & 586\\
Precipitation (thousand mm) & 1.34 & 0.78 & 0.21 & 5.59 & 1946\\
\addlinespace
	\midrule
	\textit{Panel B. Time-varying variables: post-SBM (2014-2019)} &      &    &     &     &   \\
	\addlinespace
Fecal coliform: Average (million MPN/100ml) & 0.72 & 30.39 & 0 & 1750.01 & 5553\\
Diarrheal post-neonatal mortality rate (per 1,000) & 1.46 & 1.07 & 0.05 & 5.21 & 2022\\
Number of latrines (ten thousand) & 22.52 & 18.96 & 0.01 & 146.87 & 1814\\
Number of latrines per sq. km & 59.06 & 57.05 & 1.12 & 430.09 & 1814\\
Precipitation (thousand mm) & 1.31 & 0.88 & 0.2 & 10.06 & 1814\\
\addlinespace
\midrule
\textit{Panel C. Variables not varying over time} &      &    &     &     &   \\
\addlinespace
Available water capacity (mm/m) & 128.03 & 25.91 & 19.79 & 150 & 337\\
2013 district-level sewage treatment plant capacity (MLD) & 28.17 & 105.03 & 0 & 947.5 & 337\\
2013 state-level sewage treatment plant capacity (MLD) & 389.93 & 624.77 & 0 & 2307.75 & 28\\
\bottomrule
\end{tabular}

			\begin{tablenotes}
				\setlength{\itemindent}{-2.49997pt}
				\item 
				Notes: This table shows summary statistics of time-varying variables for pre-SBM periods (2007--2013) in Panel A and post-SBM periods (2014--2019) in Panel B, and summary statistics of time-invariant variables in Panel C. 
				The latrine data are available only from 2012--2019, while data of other time-varying variables are available from 2007--2019. 
				MPN and MLD denote ``most probable number'' and ``million liters per day,'' respectively.
			\end{tablenotes}
		\end{threeparttable}
	}
\end{table}

\clearpage


\begin{table}[p]\centering \caption{The Effect on Water Quality (Log of Fecal Coliform) \label{tab:result_water}}
	\resizebox{1\textwidth}{!}{
		\begin{threeparttable}

			{
\def\sym#1{\ifmmode^{#1}\else\(^{#1}\)\fi}
\begin{tabular}{l*{3}{c}}
\toprule
                    &\multicolumn{1}{c}{OLS}&\multicolumn{1}{c}{IV: First Stage}&\multicolumn{1}{c}{IV: Second Stage}\\\cmidrule(lr){2-2}\cmidrule(lr){3-3}\cmidrule(lr){4-4}
                    &\multicolumn{1}{c}{(1)}         &\multicolumn{1}{c}{(2)}         &\multicolumn{1}{c}{(3)}         \\
                    &\multicolumn{1}{c}{Log(Fecal Coliform)}&\multicolumn{1}{c}{\# of Latrines per sq. km}&\multicolumn{1}{c}{Log(Fecal Coliform)}\\
\midrule
\multicolumn{4}{l}{\textit{Panel A. Baseline Specification}} \\ 
\addlinespace
Number of latrines  &       0.006\sym{***}&                     &       0.030\sym{***}\\
per sq. km          &     (0.002)         &                     &     (0.008)         \\
\addlinespace
AWC * Post (=1)     &                     &       0.283\sym{***}&                     \\
                    &                     &     (0.052)         &                     \\
\addlinespace
Observations        &       7,201         &       7,201         &       7,201         \\
R$^2$               &       0.020         &       0.438         &           -         \\
Number of Stations  &       1,189         &       1,189         &       1,189         \\
Number of Districts &         337         &         337         &         337         \\
KP F-Stat           &           -         &          29.954          &    -          \\
AR 95\% CI          &           -         &           -         &[.015, .049]         \\
Conley SE  &      (0.003)       &           -         &       (0.010)        \\
Average Policy Effect&       0.142         &           -         &       0.719         \\
\addlinespace
\midrule
\multicolumn{4}{l}{\textit{Panel B. Upstream-Downstream Specification}} \\ 
\addlinespace
Upstream number of  &       0.009\sym{***}&                     &       0.015         \\
latrines per sq. km &     (0.003)         &                     &     (0.011)         \\
\addlinespace
Upstream AWC  &                     &       0.322\sym{***}&                     \\
* Post (=1)                &                     &     (0.045)         &                     \\
\addlinespace
Observations        &       2,228         &       2,228         &       2,228         \\
R$^2$               &       0.057         &       0.533         &           -         \\
Number of Stations  &         365         &         365         &         365         \\
Number of Districts &         154         &         154         &         154         \\
KP F-Stat           &           -         &           50.475         &    -           \\
AR 95\% CI          &           -         &           -         &[-.008, .039]         \\
Conley SE &       (0.003)       &           -         &       (0.012)        \\
Average Policy Effect&       0.250         &           -         &       0.431         \\
\bottomrule
\end{tabular}
}

			\begin{tablenotes}
				\setlength{\itemindent}{-2.49997pt}
				\small
				\item 
				Notes: The coefficients are reported. 
				Standard errors, clustered at the district level, are in parentheses.
				***, **, and * indicate significance at the 1\%, 5\%, and 10\% levels, respectively.
				Regressions in Panel A include monitoring station fixed effects, year fixed effects, and precipitation as a control.
				Regressions in Panel B include monitoring station fixed effects, year fixed effects, and the following controls: precipitation and the interaction of Available Water Capacity and the post-SBM indicator of a reference district.
				In Panel B, the sample is limited to monitoring stations located along major rivers in India, and upstream districts are defined as those within the range of $[0,150]$ km from a reference station. 
				The KP F-Stat refers to the Wald version of the \cite{kleibergen2006generalized} rk-statistic on the excluded instrumental variables for non-i.i.d. errors. 
				The AR 95\% CI reports the 95\% confidence interval, which is robust to the weak instrument based on the \cite{anderson1949estimation} test. 
				The Conley SE refers to the standard errors that are spatially clustered with a cutoff of 150 km following the \cite{conley1999gmm} approach.
				Average policy effects are calculated by multiplying the estimated coefficients by the change in the number of latrines per square kilometer between pre-SBM and post-SBM periods.
			\end{tablenotes}
		\end{threeparttable}
	}
\end{table}

\clearpage

\begin{table}[p]\centering \caption{The Effect on Health (Diarrheal Post-neonatal Mortality Rate) \label{tab:result_health}}
	\resizebox{1\textwidth}{!}{
		\begin{threeparttable}

			{
\def\sym#1{\ifmmode^{#1}\else\(^{#1}\)\fi}
\begin{tabular}{l*{3}{c}}
\toprule
                    &\multicolumn{1}{c}{OLS}&\multicolumn{1}{c}{IV: First Stage}&\multicolumn{1}{c}{IV: Second Stage}\\\cmidrule(lr){2-2}\cmidrule(lr){3-3}\cmidrule(lr){4-4}
                    &\multicolumn{1}{c}{(1)}&\multicolumn{1}{c}{(2)}&\multicolumn{1}{c}{(3)}\\
                    &\multicolumn{1}{c}{Mortality/Latrine}&\multicolumn{1}{c}{\# of Latrines per sq. km}&\multicolumn{1}{c}{Mortality/Latrine}\\
\midrule
\multicolumn{4}{l}{\textit{Panel A. Dep. Variable: Diarrheal Post-neonatal Mortality Rate (per 1,000) (columns 1, 3)}} \\ 
\addlinespace
Upstream number of  &      -0.005\sym{**} &                     &      -0.011\sym{*}  \\
latrines per sq. km &     (0.002)         &                     &     (0.006)         \\
\addlinespace
Upstream AWC &                     &       0.301\sym{***}&                     \\
* Post (=1)                &                     &     (0.034)         &                     \\
\addlinespace
Observations        &         824         &         824         &         824         \\
R$^2$               &       0.664         &       0.573         &           -         \\
Number of Districts &         103         &         103         &         103         \\
KP F-Stat           &           -         &          78.696          &    -           \\
AR 95\% CI          &            -         &           -         &[-.023, .001]         \\
Mean of Dep. Variable&       2.576         &      23.684         &       2.576         \\
Average Policy Effect&      -0.111         &           -         &      -0.269         \\
\addlinespace
\midrule

\multicolumn{4}{l}{\textit{Panel B. Dep. Variable: \# of Latrines per sq. km in a Reference District (columns 1, 3)}} \\
\addlinespace
Upstream number of  &       0.894\sym{***}&                     &       0.726\sym{***}\\
latrines per sq. km &     (0.109)         &                     &     (0.154)         \\
\addlinespace
Upstream AWC  &                     &       0.301\sym{***}&                     \\
* Post (=1)                &                     &     (0.034)         &                     \\
\addlinespace
Observations        &         824         &         824         &         824         \\
R$^2$               &       0.796         &       0.573         &           -         \\
Number of Districts &         103         &         103         &         103         \\
KP F-Stat           &           -         &          78.696          &     -          \\
AR 95\% CI          &            -         &           -         &[.404, 1.05]         \\
Mean of Dep. Variable&      33.054         &      23.684         &      33.054         \\
\bottomrule
\end{tabular}
}

			\begin{tablenotes}
				\setlength{\itemindent}{-2.49997pt}
				
				\item 
				Notes: The coefficients are reported. 
				Standard errors, clustered at the district level, are in parentheses. 
				***, **, and * indicate significance at the 1\%, 5\%, and 10\% levels, respectively. 
				All regressions include district fixed effects, year fixed effects, and the following controls: precipitation and the interaction of Available Water Capacity and the post-SBM indicator of a reference district. 
				The sample is limited to districts that have monitoring stations used in the water quality regression along major rivers in India.
				Upstream districts are defined as those within the range of $[0,150]$ km from a reference district.
				The KP F-Stat refers to the Wald version of the \cite{kleibergen2006generalized} rk-statistic on the excluded instrumental variables for non-i.i.d. errors. 
				The AR 95\% CI reports the 95\% confidence interval, which is robust to the weak instrument based on the \cite{anderson1949estimation} test. 
				The means of the dependent variables are calculated for the pre-SBM period.
				Average policy effects are calculated by multiplying the estimated coefficients by the change in the number of latrines per square kilometer between pre-SBM and post-SBM periods.
			\end{tablenotes}
		\end{threeparttable}
	}
\end{table}

\clearpage

\begin{table}[p]\centering \caption{The Heterogeneous Effects on Water Quality by Treatment Capacity of Fecal Sludge \label{tab:result_het_water}}
	\resizebox{1\textwidth}{!}{
		\begin{threeparttable}
			
			{
\def\sym#1{\ifmmode^{#1}\else\(^{#1}\)\fi}
\begin{tabular}{l*{5}{c}}
\toprule
                    &\multicolumn{1}{c}{All}&\multicolumn{2}{c}{State-level Capacity}        &\multicolumn{2}{c}{District-level Capacity}     \\\cmidrule(lr){2-2}\cmidrule(lr){3-4}\cmidrule(lr){5-6}
                    &\multicolumn{1}{c}{(1)}&\multicolumn{1}{c}{(2)}&\multicolumn{1}{c}{(3)}&\multicolumn{1}{c}{(4)}&\multicolumn{1}{c}{(5)}\\
                    &\multicolumn{1}{c}{All}&\multicolumn{1}{c}{High}&\multicolumn{1}{c}{Low}&\multicolumn{1}{c}{High}&\multicolumn{1}{c}{Low}\\
\midrule
\multicolumn{6}{l}{\textit{Panel A. Dependent Variable: Log(Fecal Coliform) - Baseline Specification}} \\ 
\addlinespace
Number of latrines  &       0.030\sym{***}&      -0.031         &       0.037\sym{***}&       0.014         &       0.051\sym{***}\\
per sq. km          &     (0.008)         &     (0.025)         &     (0.007)         &     (0.009)         &     (0.017)         \\
\addlinespace
Observations        &       7,201         &       3,453         &       3,748         &       2,902         &       4,299         \\
Number of Stations  &       1,189         &         579         &         610         &         466         &         723         \\
Number of Districts &         337         &         182         &         155         &          96         &         241         \\
KP F-Stat           &      29.954         &       7.576         &      39.516         &      13.648         &      11.931         \\
AR 95\% CI          &[.015, .049]         &[-.123, .018]         &[.025, .054]         &[-.012,  .034]         &[.023, .105]         \\
Average Policy Effect&       0.719         &      -0.666         &       0.976         &       0.286         &       1.342         \\
\addlinespace
\midrule
\multicolumn{6}{l}{\textit{Panel B. Dependent Variable: Log(Fecal Coliform) - Upstream--Downstream Specification}} \\
\addlinespace
Upstream number of  &       0.015         &      -0.046         &       0.031\sym{***}&      -0.004         &       0.037\sym{*}  \\
latrines per sq. km &     (0.011)         &     (0.032)         &     (0.011)         &     (0.011)         &     (0.023)         \\
\addlinespace
Observations        &       2,228         &       1,107         &       1,119         &       1,097         &       1,131         \\
Number of Stations  &         365         &         171         &         194         &         180         &         185         \\
Number of Districts &         154         &          73         &          84         &          75         &          93         \\
KP F-Stat           &      50.475         &      19.767         &      41.298         &      53.262         &      15.137         \\
AR 95\% CI          &[-.008, .039]         &[-.112, .040]         &[.010, .063]         &[-.033, .018]         &[-.013, .121]         \\
Average Policy Effect&       0.431         &      -1.367         &       0.820         &      -0.092         &       1.139         \\
\bottomrule
\end{tabular}
}

			\begin{tablenotes}
				\setlength{\itemindent}{-2.49997pt}
				\small
				\item 
				Notes: The coefficients are reported. 
				Standard errors, clustered at the district level, are in parentheses. 
				***, **, and * indicate significance at the 1\%, 5\%, and 10\% levels, respectively. 
				Regressions in Panel A include monitoring station fixed effects, year fixed effects, and precipitation as a control. 
				Regressions in Panel B include monitoring station fixed effects, year fixed effects, and the following controls: precipitation and the interaction of Available Water Capacity and the post-SBM indicator of a reference district.
				In Panel B, the sample is limited to monitoring stations located along major rivers in India, and upstream districts are defined as those within the range of $[0,150]$ km from a reference station.
				In Panel A, Column 2 reports a result in states where the treatment capacities of sewage treatment plants are higher than the median, while Column 3 reports a result in states with lower treatment capacities.
				Panel B instead uses variation in treatment capacities of upstream states in Columns 2 and 3.
				Columns 4 and 5 compare results based on the different levels of treatment capacities at the district level. 
				The KP F-Stat refers to the Wald version of the \cite{kleibergen2006generalized} rk-statistic on the excluded instrumental variables for non-i.i.d. errors. 
				The AR 95\% CI reports the 95\% confidence interval, which is robust to the weak instrument based on the \cite{anderson1949estimation} test. 
				Average policy effects are calculated by multiplying the estimated coefficients by the change in the number of latrines per square kilometer between pre-SBM and post-SBM periods.
			\end{tablenotes}
		\end{threeparttable}
	}
\end{table}

\clearpage

\begin{table}[p]\centering \caption{The Heterogeneous Effects on Health by Treatment Capacity of Fecal Sludge \label{tab:result_het_health}}
	\resizebox{1\textwidth}{!}{
		\begin{threeparttable}
			
			{
\def\sym#1{\ifmmode^{#1}\else\(^{#1}\)\fi}
\begin{tabular}{l*{5}{c}}
\toprule
                    &\multicolumn{1}{c}{All}&\multicolumn{2}{c}{State-level Capacity}        &\multicolumn{2}{c}{District-level Capacity}     \\\cmidrule(lr){2-2}\cmidrule(lr){3-4}\cmidrule(lr){5-6}
                    &\multicolumn{1}{c}{(1)}&\multicolumn{1}{c}{(2)}&\multicolumn{1}{c}{(3)}&\multicolumn{1}{c}{(4)}&\multicolumn{1}{c}{(5)}\\
                    &\multicolumn{1}{c}{All}&\multicolumn{1}{c}{High}&\multicolumn{1}{c}{Low}&\multicolumn{1}{c}{High}&\multicolumn{1}{c}{Low}\\
\midrule
\multicolumn{6}{l}{\textit{Panel A. Dependent Variable: Diarrheal Post-neonatal Mortality Rate (per 1,000)}} \\
\addlinespace
Upstream number of  &      -0.011\sym{*}  &      -0.041\sym{***}&      -0.010         &      -0.014\sym{**} &      -0.000         \\
latrines per sq. km &     (0.006)         &     (0.010)         &     (0.006)         &     (0.007)         &     (0.010)         \\
\addlinespace
Observations        &         824         &         432         &         392         &         456         &         368         \\
Number of Districts &         103         &          54         &          49         &          57         &          46         \\
KP F-Stat           &      78.696         &      33.304         &      33.484         &      59.873         &      18.756         \\
AR 95\% CI          &[-.023, .001]         &[-.073, -.024]         &[-.026, .002]         &[-.030,-.000]         &[-.029, .026]         \\
Mean of Dep. Variable&       2.576         &       2.534         &       2.623         &       2.428         &       2.759         \\
Average Policy Effect&      -0.269         &      -1.058         &      -0.230         &      -0.364         &      -0.009         \\
\addlinespace
\midrule
\multicolumn{6}{l}{\textit{Panel B. Dependent Variable: Number of Latrines per sq. km in a Reference District}} \\
\addlinespace
Upstream number of  &       0.726\sym{***}&       1.327\sym{***}&       0.684\sym{***}&       0.649\sym{***}&       0.907\sym{**} \\
latrines per sq. km &     (0.154)         &     (0.214)         &     (0.160)         &     (0.170)         &     (0.358)         \\
\bottomrule
\end{tabular}
}

			\begin{tablenotes}
				\setlength{\itemindent}{-2.49997pt}
				\small
				\item 
				Notes: The coefficients are reported. 
				Standard errors, clustered at the district level, are in parentheses. 
				***, **, and * indicate significance at the 1\%, 5\%, and 10\% levels, respectively. 
				All regressions include district fixed effects, year fixed effects, and the following controls: precipitation and the interaction of Available Water Capacity and the post-SBM indicator of a reference district.
				The sample is limited to districts that have monitoring stations used in the water quality regression along major rivers in India.
				Upstream districts are defined as those within the range of $[0,150]$ km from a reference district.
				Column 2 reports results when upstream states have higher treatment capacities of sewage treatment plants than the median, while Column 3 reports results in the case of upstream states with lower treatment capacities. 
				Columns 4 and 5 compare results based on the different levels of upstream treatment capacities at the district level. 
				The KP F-Stat refers to the Wald version of the \cite{kleibergen2006generalized} rk-statistic on the excluded instrumental variables for non-i.i.d. errors. 
				The AR 95\% CI reports the 95\% confidence interval, which is robust to the weak instrument based on the \cite{anderson1949estimation} test.
				The means of the dependent variables are calculated for the pre-SBM period.
				Average policy effects are calculated by multiplying the estimated coefficients by the change in the number of latrines per square kilometer between pre-SBM and post-SBM periods.
			\end{tablenotes}
		\end{threeparttable}
	}
\end{table}

\clearpage
\appendix

\onehalfspacing

\setcounter{page}{1}

\begin{center}
	\Large{\bf{Online Appendix}}
\end{center}

\begin{center}
	\large{Unintended Consequences of Sanitation Investment: Negative Externalities on Water Quality and Health in India}
\end{center}

\begin{center}
	\normalsize{Kazuki Motohashi}
\end{center}

\listofappendices
\vspace{0.5cm}
\listofappendixfigures
\vspace{0.5cm}
\listofappendixtables

\clearpage

\begin{appendices}

\section{Conceptual Framework on Negative Externalities of Sanitation Investment} \label{sec:conceptual_framework}

\setcounter{figure}{0}
\renewcommand{\thefigure}{A\arabic{figure}}

I present a simple conceptual framework to show how latrine construction under the SBM causes water pollution externalities that offset the direct health benefits. 
A decrease in latrine price under a subsidy increases the number of constructed latrines, which increases the marginal damage (negative externalities) and offsets the marginal benefit (health benefits). 
The magnitude of these negative externalities depends on the treatment capacity of the fecal sludge.

I consider a district with $N$ households that can decide whether to construct a latrine.
I assume that a given household can build a latrine by paying a fixed price ($p_{pre}$).\footnote{~The latrine price can include both the initial construction cost of a latrine and the present value of marginal costs for emptying fecal sludge periodically.} 
The maximum number of latrines that can be built in a district is $Q^{max} = N$.

The fecal sludge emptied from latrines in this district is treated by STPs. 
The treatment capacity of the fecal sludge is given by $Q^{stp} \in [0,Q^{max}]$ where $Q^{stp}$ can be interpreted as the number of latrines whose fecal sludge can be treated by STPs. 
Thus, when the number of latrines ($Q$) exceeds $Q^{stp}$, $Q-Q^{stp}$ fecal sludge is dumped into rivers, causing negative externalities in water quality and health.
In this conceptual framework, I analyze two cases: 
(i) low treatment capacity ($Q^{stp} \leq \frac{Q^{max}}{2} $) and (ii) high treatment capacity ($Q^{stp} > \frac{Q^{max}}{2} $).

\begin{figure}[p]
	\begin{center}
		
		Panel A. Treatment Capacity Low\\
		
		\hspace{0.4cm} \minipage{0.85\textwidth}
		\includegraphics[width=\linewidth]{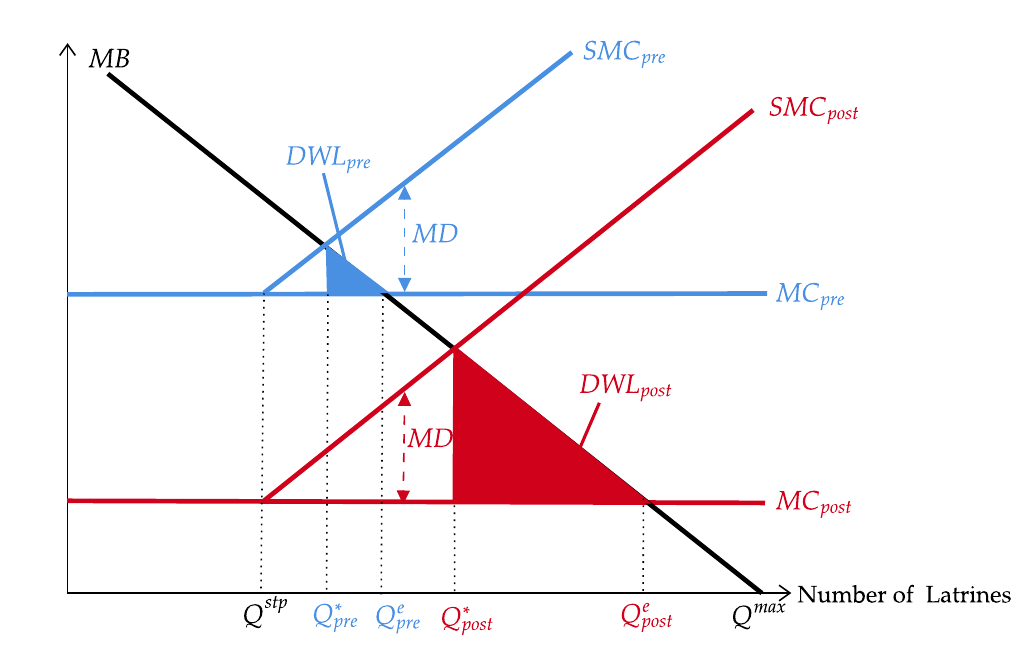}
		\endminipage \\
		
		\vspace{0.5cm}
		
		Panel B. Treatment Capacity High\\
		
		\minipage{0.85\textwidth}
		\includegraphics[width=\linewidth]{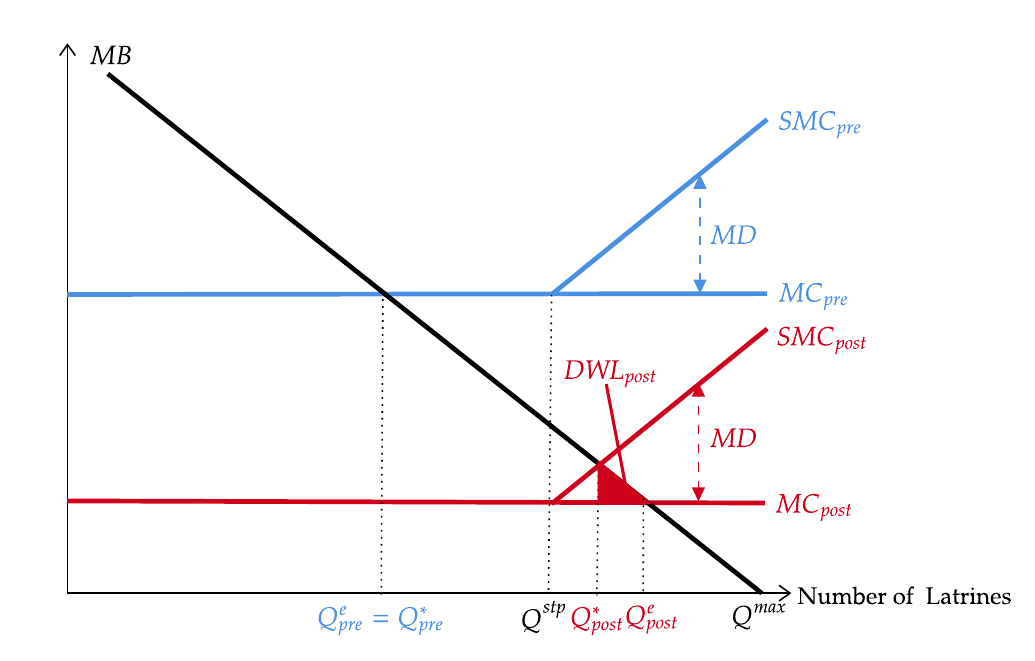}
		\endminipage 
		\caption{\label{fig:model_1} Welfare Effects of the Swachh Bharat Mission}
		\medskip 
		\begin{minipage}{1\textwidth} 
			{\footnotesize 
				Notes: This figure examines how the subsidy under the SBM changes the deadweight loss (DWL) in two cases: (A) low treatment capacity (low $Q^{stp}$) and (B) high treatment capacity (high $Q^{stp}$). 
				The subsidy shifts down the marginal cost (MC) from $MC^{pre}$ to $MC^{post}$. 
				Marginal damage (MD) represents the negative externality on health, which occurs when the number of latrines is larger than the treatment capacity level ($Q^{stp}$).
				Marginal benefit (MB) represents direct health benefits from reduced open defecation.
				This figure shows that DWL increases more substantially in the case of low treatment capacity (Panel A) than in the case of high treatment capacity (Panel B). \par}
		\end{minipage}
	\end{center}
\end{figure}

Appendix Figure \ref{fig:model_1} shows the marginal benefit ($MB$), marginal cost ($MC$), marginal damage ($MD$), and social marginal cost ($SMC$) of latrine construction for the low-treatment (Panel A) and high-treatment (Panel B) capacity cases. 

Both panels exhibit the same $MB$ and $MC$ curves.
The $MB$ curve represents the direct health benefits from reduced open defecation and exposure to fecal matter near human habitats.\footnote{~MB is assumed to only represent direct health benefits, that is, reduction in the risks of diarrheal mortality, although there could be other benefits, including an improvement in educational outcomes and reduction in violence against women.} 
This curve is downward-sloping because some households benefit more than others; for instance, if they have more infants who are vulnerable to diarrhea. 
For $MC$, the pre-SBM curves are constant at a constant price for latrines ($MC_{pre} = p_{pre}$). 
The $MC$ curves are shifted downward by the subsidy under the SBM. 
Households receive a subsidy of approximately USD 140 for latrine construction.
Therefore, the post-SBM effective price of latrines ($MC_{post} = p_{post}$) becomes substantially lower than $p_{pre}$.

The main difference between Panels A and B is $SMC$. If the treatment capacity ($Q^{stp}$) is low (Panel A), $MD$, that is, the negative externality on health through exposure to increased river pollution, becomes nonzero, starting from a lower number of latrines.
However, if the treatment capacity ($Q^{stp}$) is high (Panel B), then $MD$ occurs only for a large number of latrines.
Here, I assume a non-linear dose-response relationship: the larger the volume of dumped fecal sludge ($Q-Q^{stp}$), the larger the marginal negative externality on health ($MD$).\footnote{~The non-linear relationship is suggested by a classic epidemiological study \citep{moe1991bacterial}, which shows the evidence of threshold effects where significantly higher rates of diarrheal disease are observed once the fecal contamination level in drinking water reaches a certain threshold.}
The $SMC$ curves reflect the differences in $MD$ curves, because $SMC = MC+MD$.

Based on this conceptual framework, I examine the welfare effects of latrine construction under the SBM in Appendix Figure \ref{fig:model_1}. 
If the treatment capacity is low (Panel A), the pre-SBM market equilibrium quantity is $Q^{e}_{pre}$ at the intersection of $MB$ and $MC_{pre}$, and the pre-SBM optimal quantity is $Q^{*}_{pre}$ at the intersection of $MB$ and $SMC_{pre}$. 
The wedge between $Q^{e}_{pre}$ and $Q^{*}_{pre}$ caused by $MD$ (negative externality) generates deadweight loss ($DWL_{pre}$). 
Then, the effect of the SBM is to decrease the marginal cost from $MC_{pre}$ to $MC_{post}$ through the subsidy.
Thus, the number of latrines substantially increases from $Q^{e}_{pre}$ to $Q^{e}_{post}$. 
This increase in latrines causes a large increase in the negative externality owing to the low treatment capacity. 
The deadweight loss increased substantially from $DWL_{pre}$ to $DWL_{post}$. 
Conversely, if the treatment capacity is high (Panel B), the increase in deadweight loss owing to latrine construction is limited because the negative externality only occurs in a large number of latrines. 
A comparison of Panels A and B suggests that the subsidy under the SBM adversely impacts welfare more substantially in the case of low treatment capacity.

I further examine the effects of latrine construction under the SBM on water quality and health in Appendix Figure \ref{fig:model_2}, which is based on the welfare analysis in Appendix Figure \ref{fig:model_1}. 
In Appendix Figure \ref{fig:model_2}, the total benefit represents the total direct health effects, whereas the total damage represents the total negative externality on health owing to water pollution.\footnote{~The total benefit in Appendix Figure \ref{fig:model_2} is the area under the $MB$ curves of Appendix Figure \ref{fig:model_1}. The total damage in Appendix Figure \ref{fig:model_2} represents the area bounded by the $SMC$ and $MC$ curves shown in Appendix Figure \ref{fig:model_1}.} 
I examine the difference between the total benefit and total damage (net benefit) as a health outcome in the empirical analysis.\footnote{~I assume that the total benefit is larger than the total damage. In this case, the net benefit is positive, which means that latrines are health-improving. This is consistent with the empirical results of this paper.}
The total damage can be interpreted as the degree of water pollution, which corresponds to the water quality outcome of the empirical analysis. 

This paper estimates the effects of an increase in the number of latrines at market equilibrium from $Q^{e}_{pre}$ to $Q^{e}_{post}$ on water quality and health under the SBM. 
According to Appendix Figure \ref{fig:model_2}, there are three testable hypotheses for the empirical analyses. 
The first hypothesis is tested in the baseline analysis (Section \ref{sec:result}), and the second and third hypotheses are tested in the heterogeneity analysis (Section \ref{sec:result_het_treatment}). 

\begin{enumerate}
	\setlength{\itemsep}{2.5pt}
	\setlength{\parskip}{0pt}
	\setlength{\parsep}{0pt}
	\item The SBM improves health overall (increase in net benefit) if the total benefit increases more substantially than the total damage and increases water pollution (increase in total damage) regardless of treatment capacity.\footnote{~While theoretically, net benefit may decrease, Appendix Figure \ref{fig:model_2} demonstrates a case where the total benefit increases more substantially than the total damage (increase in net benefit), which is consistent with the empirical results of this paper.}
	\item The magnitude of positive health effects is smaller in the case of low treatment capacity.
	\item The magnitude of negative effects on water quality (increased water pollution) is larger in the case of low treatment capacity.
\end{enumerate}

\clearpage

\begin{figure}[p]
	\begin{center}
		
		Panel A. Treatment Capacity Low\\
		
		\hspace{0.4cm} \minipage{0.85\textwidth}
		\includegraphics[width=\linewidth]{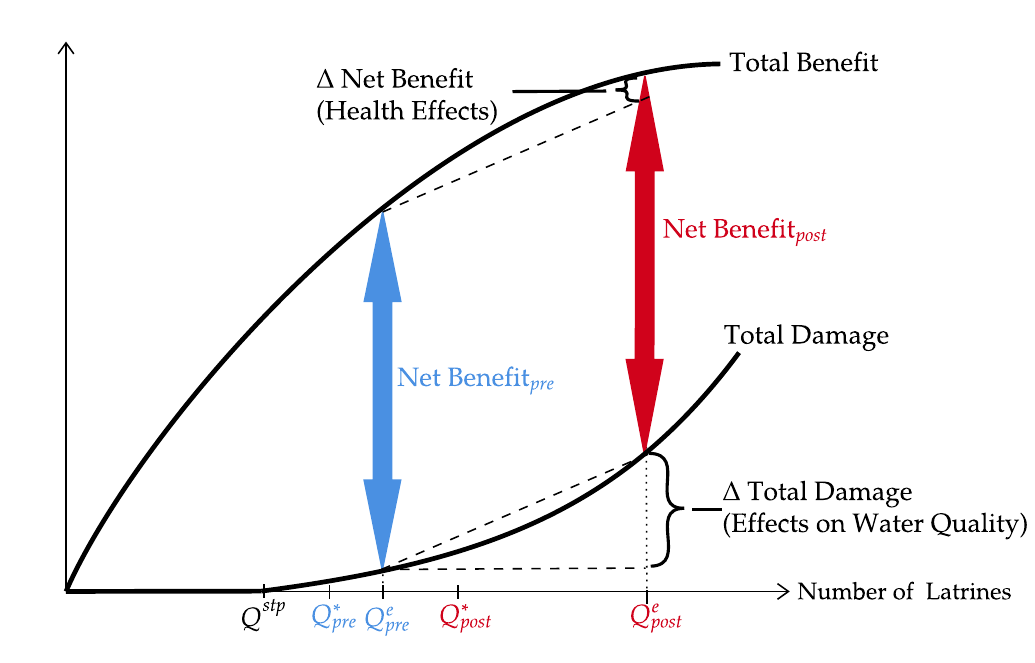}
		\endminipage \\
		
		\vspace{0.5cm}
		
		Panel B. Treatment Capacity High\\
		
		\minipage{0.85\textwidth}
		\includegraphics[width=\linewidth]{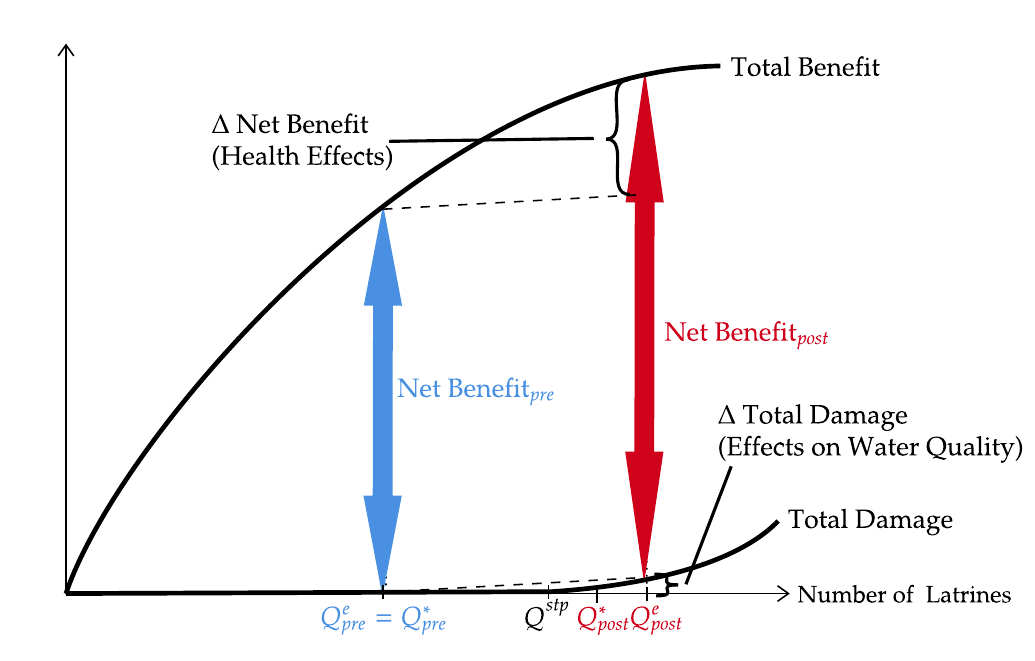}
		\endminipage 
		\caption{\label{fig:model_2} Effects of the Swachh Bharat Mission on Water Quality and Health}
		\medskip 
		\begin{minipage}{1\textwidth} 
			{\footnotesize 
				Notes: This figure examines how the SBM affects water quality and health in two cases: (A) low treatment capacity (low $Q^{stp}$) and (B) high treatment capacity (high $Q^{stp}$).
				The total benefit and total damage in this figure are based on the marginal benefit and marginal damage plotted in Appendix Figure \ref{fig:model_1}.
				Effects on health and water quality are represented by the changes in net benefit and total damage, respectively.
				This figure shows that the SBM improves health overall and increases water pollution. 
				In the case of low treatment capacity in Panel A, the magnitude of health effects is smaller, while the magnitude of water quality effects is larger.
				\par}
		\end{minipage}
	\end{center}
\end{figure}

\clearpage

\section{Data Appendix} \label{sec:appendix_data}

\subsection{Water Quality}

\begin{itemize}
	\setlength{\itemsep}{1.5pt}
	\setlength{\parskip}{0pt}
	\setlength{\parsep}{0pt}
	\item I obtain the monitoring station-level water quality data of rivers from the following data sources. 
	\begin{enumerate}
		\item 2012-2019: NWMP (National Water Quality Monitoring Programme) Data, Central Pollution Control Board
		\\ \url{https://cpcb.nic.in/nwmp-data/} (accessed January 15, 2021)
		\item 2007-2011: Water Quality Database, National Water Quality Monitoring
		\\ \url{http://www.cpcbenvis.nic.in/water_quality_data.html} (accessed January 15, 2021)
	\end{enumerate}
	
	\item These datasets are separated by water body types: (i) rivers; (ii) medium and minor rivers; (iii) canals, seawater, drains, STPs (sewage treatment plants), and WTPs (water treatment plants); (iv) lakes, ponds, and tanks; and (v) groundwater. In this paper, I use data from the first and second types, encompassing all types of rivers.
	
	\item I use the yearly average values of the following water quality indicators in the analysis.
	\begin{itemize}
		\item Fecal Coliform: Yearly maximum values are also used for the robustness check.
		\item Nitrate and Nitrite, Temperature, Dissolved Oxygen, Biochemical Oxygen Demand (for falsification tests)
	\end{itemize}
	
	\item I drop outliers of water temperature (one observation of over 100 $^\circ$C) from the sample in the falsification tests.
	
	\item I complement these datasets with the GPS location data of water quality monitoring stations in the following document.
	
	\begin{itemize}
		\item Central Pollution Control Board Website
		\\ \url{https://cpcb.nic.in/wqm/WQMN_list.pdf} (accessed January 15, 2021)
	\end{itemize}
	
	\item I use 2011 district boundary data and GPS location data of monitoring stations to identify districts where monitoring stations are located.
	
\end{itemize}

\subsection{Health}

\begin{itemize}
	\setlength{\itemsep}{1.5pt}
	\setlength{\parskip}{0pt}
	\setlength{\parsep}{0pt}
	\item I obtain the 5 km raster data of mortality and overweight prevalence estimates from 2000 to 2019 from the following data sources.
	\begin{enumerate}
		\item Diarrheal Mortality: Global Under-5 Diarrhea Incidence, Prevalence, and Mortality Geospatial Estimates 2000--2019, Institute for Health Metrics and Evaluation \citep{ihme2020}
		\\ \url{https://ghdx.healthdata.org/record/ihme-data/global-under-5-diarrhea-incidence-prevalence-mortality-geospatial-estimates-2000-2019} (accessed May 30, 2021)
		\item Overweight Prevalence: Global Under-5 Overweight Prevalence Geospatial Estimates 2000-2019 \citep{ihme2020overweight}
		\\ \url{https://ghdx.healthdata.org/record/ihme-data/global-under-5-overweight-prevalence-geospatial-estimates-2000-2019} (accessed May 29, 2022)
	\end{enumerate}
	
	\item These estimates are computed by applying the Bayesian model-based geostatistical framework to the data in the following household surveys.
	\begin{itemize}
		\item Diarrheal Mortality: India Demographic and Health Survey 2005--2006, 2015--2016, India District Level Household Survey 2002--2005, 2007--2008, 2012--2014, and India Human Development Survey 2004--2005, 2011--2013
		\item Overweight Prevalence: India Coverage Evaluation Survey 2009-2010, India Demographic and Health Survey 2005--2006, 2015--2016, India District Level Household Survey 2002--2005, 2007--2008, 2012--2014, and India Human Development Survey 2004--2005, 2011--2013
	\end{itemize}
	
	\item I use estimates of diarrheal mortality rates of five age groups, that is, early-neonatal (0--6 days), late-neonatal (7--27 days), post-neonatal (28--364 days), ages 1--4 years, and under 5 years. I also use estimates of overweight prevalence for ages 0--5 years.
	\item I use the mean estimates of mortality rates (per 1,000 children, or per child multiplied by 1,000) and overweight prevalence.
	\item For the analysis, I compute the district-level means of mortality rates and overweight prevalence estimates based on the raster data and 2011 district boundary data.
	\item As a robustness check, I adopt the following alternative mortality dataset.
	\begin{itemize}
	\item National Family Health Survey 5 (NFHS-5) 2019-2021 \\ \url{https://dhsprogram.com/methodology/survey/survey-display-541.cfm} (accessed June 21, 2023)
	\item The NFHS-5 interviews all women aged 15--49 years within the sample households and records detailed information about their birth histories.
	\item In their birth histories, I use the data concerning the year of birth and whether the child died within 12 months of birth, that is, an infant mortality indicator. This mortality indicator encompasses all types of mortality, not solely those driven by water pollution.
	\end{itemize}
\end{itemize}

\subsection{Latrines}

\begin{itemize}
	\setlength{\itemsep}{1.5pt}
	\setlength{\parskip}{0pt}
	\setlength{\parsep}{0pt}
	\item I obtain data on the number of constructed household latrines from 2012 to 2019 in rural India from the following official database of the SBM policy. 
	\begin{itemize}
		\item Format A03: Swachh Bharat Mission Target Vs Achievement On the Basis of Detail entered, Swachh Bharat Mission - Gramin (All India)
		\\ \url{https://sbm.gov.in/sbmReport/Report/Physical/ SBM_TargetVsAchievementWithout1314.aspx} (originally accessed March 28, 2020)
		\item The format name and the URL have been updated as follows:\\ 
		ER 77: Swachh Bharat Mission Target Vs Achievement On the Basis of Detail entered
		\\\url{https://sbm.gov.in/sbmphase2/Secure/Entry/UserMenu.aspx} (accessed March 5, 2023)
	\end{itemize}
	
	\item The raw tables scraped from this database record numbers of constructed latrines at the village level, so I aggregate them to the district-level data for analysis.
	\item This dataset uses district names in 2019, so I transform the data to follow the 2011 boundary by considering district splits from 2011 to 2019. For example, if District A is divided into Districts B and C between 2011 and 2019, the number of latrines in District A is computed as the total number of latrines in Districts B and C. This aggregation allows me to match the latrine data with the water quality data based on the 2011 boundary. 
	\item For the IV design, I compute the number of latrines per square kilometer by dividing the number of latrines by the district area. The district area is computed using 2011 district boundary data.
	\item For the DiD design in the robustness check, I compute the latrine coverage in 2013 by dividing the number of latrines in 2013 by the total number of recorded households in each district. 
	
\end{itemize}

\subsection{GIS (Geographic Information System) Data} \label{sec:appendix_data_gis}

\begin{itemize}
	\setlength{\itemsep}{1.5pt}
	\setlength{\parskip}{0pt}
	\setlength{\parsep}{0pt}
	\item 2011 District Boundary
	\begin{itemize}
		\item I obtain the shape files of the 2011 district boundary of the ML Infomap from the Data Lab at Tufts University. 
		\item This dataset includes 640 districts that were available in India in the 2011 Census.
		\item I use this boundary data to match all datasets used in the analysis.
	\end{itemize}
	
	\item River Basin
	\begin{itemize}
		\item I obtain the shape files of the ``Watershed Map of India'' of the ML Infomap from the Data Lab at Tufts University. 
		\item This dataset records the boundaries of 34 river basins in India.
		\item I use this basin data to identify the basin of each monitoring station.
	\end{itemize}
	
	\item River Line 1
	\begin{itemize}
		\item I obtain polygons of rivers from the following data source.
		\begin{itemize}
			\item The version 4.1.0 GIS polygons of rivers and lakes (1:10m), Natural Earth 
			\\ \url{https://www.naturalearthdata.com/downloads/10m-physical-vectors/10m-rivers-lake-centerlines/} (accessed April 15, 2021)
		\end{itemize}   
		\item This dataset covers 43 major rivers in India.
		\item I use this dataset of river lines for identifying upstream districts.
	\end{itemize}
	
	\item River Line 2
	\begin{itemize}
		\item I obtain polygons of rivers from the following data source.
		\begin{itemize}
			\item Global River Widths from Landsat (GRWL) Database \citep{allen2018global} 
			\\ \url{https://zenodo.org/record/1297434} (accessed April 16, 2021)
		\end{itemize}   
		\item This dataset covers rivers that are $\geq$30 m wide at mean annual discharge globally.
		\item I use this river data in robustness checks that employ an alternative mortality dataset, in health analyses focusing on areas close to rivers, and in analyses of an alternative mechanism related to direct contamination.
	\end{itemize}
	
	\item Digital Elevation Data
	\begin{itemize}
		\item I obtain 90 m raster elevation data from the following database.
		\begin{itemize}
			\item Shuttle Radar Topography Mission data Version 4.1, International Centre for Tropical Agriculture \citep{reuter2007evaluation} 
			\\ \url{https://cgiarcsi.community/data/srtm-90m-digital-elevation-database-v4-1/} (accessed May 3, 2021)
		\end{itemize}   
		\item I use this elevation data for identifying upstream districts.
	\end{itemize}
	
\end{itemize}

\subsection{Available Water Capacity}

\begin{itemize}
	\setlength{\itemsep}{1.5pt}
	\setlength{\parskip}{0pt}
	\setlength{\parsep}{0pt}
	\item I obtain the 30 arc-second raster data of Available Water Capacity from the following data source. 
	\begin{itemize}
		\item Harmonized World Soil Database v1.2, Food and Agriculture Organization of the United Nations 
		\\ \url{https://www.fao.org/soils-portal/soil-survey/soil-maps-and-databases/harmonized-world-soil-database-v12/ru/} (accessed July 22, 2021)
	\end{itemize}
	\item For the analysis, I compute the district-level mean of Available Water Capacity based on this raster data and 2011 district-level boundary data.
	
\end{itemize}

\subsection{Sewage Treatment Plants (STPs)}

\begin{itemize}
	\setlength{\itemsep}{1.5pt}
	\setlength{\parskip}{0pt}
	\setlength{\parsep}{0pt}
	\item I obtain an inventory of STPs from the following data source. 
	\begin{itemize}
		\item Inventorization of Sewage Treatment Plants, Central Pollution Control Board \citep{cpcb2015}
		\\ \url{https://nrcd.nic.in/writereaddata/FileUpload/NewItem_210_Inventorization_of_Sewage-Treatment_Plant.pdf} (accessed April 12, 2021)
	\end{itemize}
	
	\item This dataset includes detailed information on 816 STPs in 28 states and union territories in India in 2015.
	\item I first extract 467 STPs that were operational in 2013 and had information on the installed capacity. 
	\item Next, I manually assign state and district names to these STPs based on their city/town locations. 
	\item Lastly, I calculate the aggregated STP capacities at both state and district levels in 2013 for the heterogeneity analysis.
	
\end{itemize}

\subsection{Other District Characteristics}

\begin{itemize}
	\setlength{\itemsep}{1.5pt}
	\setlength{\parskip}{0pt}
	\setlength{\parsep}{0pt}
	\item Precipitation
	\begin{itemize}
		\item I obtain 0.25-degree raster data of precipitation from 2007 to 2019 from the following data source.
		\begin{itemize}
			\item Gridded Rainfall (0.25 x 0.25) NetCDF File, India Meteorological Department \citep{pai2014development}
			\\ \url{https://www.imdpune.gov.in/cmpg/Griddata/Rainfall_25_NetCDF.html} (accessed April 8, 2021)
		\end{itemize}   
		\item First, I aggregate daily raw data into annual data.
		\item Then, for the analysis, I compute the district-level mean of annual precipitation based on this raster data and 2011 district-level boundary data.
	\end{itemize}
	
	\item Nighttime Light
	\begin{itemize}
		\item I obtain 15 arc-second raster data of nighttime light in 2013 from the following data source.
		\begin{itemize}
			\item V.2 annual composites of Visible and Infrared Imaging Suite (VIIRS) Day Night Band (DNB), Earth Observation Group, National Oceanic and Atmospheric Administration \citep{elvidge2021annual}
			\\ \url{https://eogdata.mines.edu/products/vnl/} (accessed April 20, 2021)
		\end{itemize}
		\item Specifically, I use the values of masked average radiance, which represent stable lights from which background noises, biomass burning, and aurora are removed. 
		\item For the DiD design in the robustness check, I compute the district-level mean of nighttime luminosity in 2013 based on the annual composite of 2013 and 2011 district-level boundary data.
	\end{itemize}
	
	\item Population Raster
	\begin{itemize}
		\item I obtain 100 m resolution raster population data from the following data source.
		\begin{itemize}
			\item India 100m Population, WorldPop 
			\\\url{https://hub.worldpop.org/doi/10.5258/SOTON/WP00532} (accessed July 10, 2024)
		\end{itemize}   
		\item For the analysis of an alternative mechanism related to direct contamination, I compute the share of the population living within 5 km and 10 km of rivers in 2011 using the population raster and district-level boundary data from 2011.
	\end{itemize}
	
	\item Other Socio-demographic Characteristics
	\begin{itemize}
		\item For the DiD design in the robustness check, I obtain district-level data on population, the proportions of Scheduled Caste and Scheduled Tribe members, and literacy rates of rural India in 2011 from the 2011 Census of India.
		\begin{itemize}
			\item Basic Population Figures of India/State/District/Sub-District/Village, 2011 Census
			\\ \url{https://censusindia.gov.in/nada/index.php/catalog/42560} (accessed May 30, 2022)
		\end{itemize}
	\end{itemize}
	
\end{itemize}

\subsection{Identification of Upstream Districts}\label{identify_updown}

\begin{itemize}
	\setlength{\itemsep}{1.5pt}
	\setlength{\parskip}{0pt}
	\setlength{\parsep}{0pt}
	\item I identify upstream districts for the upstream--downstream analysis, which is discussed in Section \ref{sec:iv_updown}.
	\item I first focus on districts located along 43 major rivers in the GIS polygons of the Natural Earth. Some of the districts are further dropped if they have no further upstream districts. Then, for the water quality data, I use 365 monitoring stations that are within 4 km of the major rivers. The district-level health analysis focuses on 103 districts that have monitoring stations along the major rivers.
	\item Second, I use elevation data along the major rivers to identify the upstream--downstream relationships between monitoring stations and districts. The upstream districts of a given district (station) are selected as the districts that intersect with river segments whose elevations are higher than the elevation of the given district (station). This operation, aimed at identifying upstream districts, is repeated for each major river.
	\item If the major rivers have several branches, I divide them into smaller segments at each branching point. Ultimately, I decompose 43 major rivers into 60 segments, which I use to determine the upstream--downstream relationships.
	\item If two or more rivers flow through a given district, I do not include this district in the final sample because the upstream--downstream relationships become unclear.
	\item I adopt a variety of distances from a given district (station) for identifying upstream districts. Specifically, for a given district (station), the upstream districts are selected from districts that fall within a range of $[X,Y]$ km from the given district (station), where $X\in \{0,50,100\}$, $Y\in\{100,150\}$, and $X<Y$.
\end{itemize}

\subsection{Identification of Neighboring Districts}\label{identify_neighbor}

\begin{itemize}
	\setlength{\itemsep}{1.5pt}
	\setlength{\parskip}{0pt}
	\setlength{\parsep}{0pt}
	\item I identify neighboring districts for the robustness check of considering the spillovers from neighboring districts, which is discussed in Section \ref{sec:robustness}.
	\item First, I identify monitoring stations that are situated in more than one district. I create 2 km buffers around stations and select stations whose buffers intersect with more than one district.
	\item Out of 1,189 monitoring stations, 324, 26, and 1 monitoring station(s) are situated among two, three, and four districts, respectively.
	\item Then, for these identified monitoring stations, I compute the weighted average of variables of neighboring districts by using district areas as weights.
	\item The data of other monitoring stations remain unchanged. 
	
\end{itemize}

\subsection{Identification of Urban Areas}\label{identify_urban}

\begin{itemize}
	\setlength{\itemsep}{1.5pt}
	\setlength{\parskip}{0pt}
	\setlength{\parsep}{0pt}
	\item I identify urban areas for the robustness check of excluding the influence of urban areas, which is discussed in Section \ref{sec:robustness}.
	\item First, I focus on 53 urban agglomerations/cities that have a population of 1 million and above in 2011. These cities are identified from the following data source of the 2011 Census.
	\begin{itemize}
		\item \url{https://web.archive.org/web/20111113152754/http://www.censusindia.gov.in/2011-prov-results/paper2/data_files/India2/Table_3_PR_UA_Citiees_1Lakh_and_Above.pdf} (accessed March 7, 2022)
	\end{itemize}
	\item Second, I obtain the GPS locations of these cities by using the GeoNames geographical database (\url{http://www.geonames.org/about.html}).
	\item Finally, I drop monitoring stations and districts that are within 50, 100, or 150 km of the GPS locations of these cities in the robustness check.
\end{itemize}

\clearpage

\section{Robustness Check: Difference-in-Differences Design} \label{sec:did}

As a robustness check, I adopt an alternative DiD design that exploits the differential increase in latrine coverage across districts with different baseline coverage levels.

\setcounter{figure}{0}
\renewcommand{\thefigure}{C\arabic{figure}}
\setcounter{table}{0}
\renewcommand{\thetable}{C\arabic{table}}

\subsection{Empirical Strategy}

An alternative DiD design uses the district-level baseline latrine coverage as a treatment, which affects the number of latrines constructed under the SBM.\footnote{~This DiD design that uses variation in the baseline degree of policy implementation is in the same vein as \cite{duflo2001schooling} and \cite{bleakley2007disease}. While another potential approach could be to leverage variation in the timing of latrine construction initiation, this is challenging to implement because only one district had 0\% baseline coverage in 2013, and most districts were already treated before the implementation of SBM.}
This design exploits the fact that all districts achieved almost universal latrine coverage by the 2019 target date, regardless of their baseline latrine coverage. 
Thus, districts with lower baseline latrine coverage experienced larger increases in latrine coverage.
As shown in Appendix Figure \ref{fig:latrine_coverage}, there were substantial differences in baseline latrine coverage across districts in 2013, suggesting a differential increase in the number of latrines by 2019.
As expected, I find that lower baseline coverage is positively correlated with the number of latrines constructed under the SBM (Appendix Figure \ref{fig:did_fs}).
I expect that districts with higher latrine non-coverage (lower latrine coverage) in 2013 experienced a greater increase in water pollution owing to a larger increase in latrine coverage.

In this alternative DiD design, I adopt the following regression:

\vspace*{-0.5cm}

\begin{equation}\label{did_eq}
	\begin{split}
		Y_{i,d,t} = \delta_i + \theta_{b,t} + \beta_{DID} (1 - Latrine_{d}^{pre}) \cdot Post_{t} + \gamma \mathbf{X_{d, t}} + \varepsilon_{i, t} 
	\end{split}
\end{equation}

\noindent where $Y_{i,d,t}$ is a water quality indicator, represented by the logarithm of fecal coliform, at monitoring station $i$ located in district $d$ in year $t$.
$Latrine_{d}^{pre}$ is the latrine coverage in district $d$ in 2013, one year before the SBM started. 
$Post_{t}$ is an indicator that takes the value of one after 2014, when the SBM started.
$\mathbf{X_{d, t}}$ is a set of control variables, which are time-varying precipitation and time-invariant district characteristics, including VIIRS nighttime luminosity in 2013, population, proportions of Scheduled Caste and Scheduled Tribe members, and literacy rates in 2011.
Time-invariant variables are added as control variables after interacting with the year dummies.
Monitoring station fixed effects ($\delta_{i}$) are included as per the IV design.
Basin-year fixed effects ($\theta_{b,t}$) are also included to account for secular trends in water quality across years, which may vary across river basins.
Standard errors are clustered at the district level because the baseline latrine coverage varies across districts.
The coefficient of interest is $\beta_{DID}$ and is expected to be positive; that is, a higher increase in water pollution.

To examine pre-trends and the dynamic evolution of the treatment effects, I also adopt the following event-study specification: 

\vspace*{-0.6cm}

\begin{equation}\label{did_event_eq}
	\begin{split}
		Y_{i,d,t} = \delta_i + \theta_{b,t} + \sum_{l=2007}^{2019} \beta_l (1 - Latrine_{d}^{pre}) \cdot T_{l} + \gamma \mathbf{X_{i, t}} + \varepsilon_{i, t} 
	\end{split}
\end{equation}

\noindent where the reference year is set to 2013, and $T_{l}$ is a year dummy variable. 
The standard errors are clustered similarly at the district level.
The coefficient of interest is $\beta_l$, which measures the treatment effect on water quality for each year relative to 2013. 
The $\beta_l$'s for 2007--2012 are examined to test the assumption of parallel pre-trends, whereas the $\beta_l$'s for 2014--2019 capture the dynamic evolution of the treatment effects. 
Based on the tests of parallel pre-trends, I show the results of the DiD design only for water quality outcomes.

\subsection{Data}

The DiD design uses the same datasets for water quality and latrines introduced in Section \ref{sec:data}. 
This design uses longer panel water quality data from 2007 to 2019.
I additionally identify the basin of each monitoring station using the GPS coordinates of the monitoring stations and the ``Watershed Map of India'' of the ML Infomap.
It also uses latrine coverage in 2013 as a treatment, which is computed by dividing the number of household latrines in 2013 by the total number of households recorded in each district.

This design uses two additional datasets to account for other district characteristics that affect latrine construction and water quality, and achieve a better balance between the treatment and control groups.

First, I use 15 arc-second ($<$500 m at the equator) raster data of nighttime lights to account for the size of the economy at the district level. 
Specifically, I use the V.2 annual composites of the Visible and Infrared Imaging Suite Day Night Band (VIIRS DNB) \citep{elvidge2021annual}.\footnote{~I use the values of masked average radiance, which represent stable lights from which background noises, biomass burning, and aurora are removed.}
The district-level mean nighttime luminosity in the pre-SBM period is computed based on the 2013 annual composite.

Second, I use data on district-level socio-demographic characteristics, including population, the proportions of Scheduled Caste and Scheduled Tribe members, and literacy rates in rural India from the 2011 Census of India.

\subsection{Results}

As in the IV design, I find that latrine construction under the SBM increases river pollution, especially in areas with lower treatment capacities. 
The positive coefficient in Column 1 of Appendix Table \ref{tab:did_result_waterquality} suggests that latrine construction increases water pollution, although the effect becomes imprecise in the DiD design.
Heterogeneity analysis by fecal sludge treatment capacity confirms that the negative externality on water quality is concentrated in areas with lower treatment capacities (Columns 2--5).
The coefficients of $(1 - Latrine_{d}^{pre}) \cdot Post_{t}$ show that a district with a baseline latrine coverage of 50\% would experience an increase in fecal coliform of about 75-90\%, relative to a district with 100\% baseline latrine coverage, in areas with lower treatment capacities (Columns 3 and 5). 
Considering that the baseline latrine coverage was 39.2\% in 2013, the total effects of the SBM in states with lower treatment capacities can be calculated as $(1-0.392)\times1.790 = 1.088$, which is relatively close to the average policy effect (0.976) in Column 3 of Panel A of Table \ref{tab:result_het_water} in the IV design. 
Conversely, consistent with the results of the IV design, no negative externality is found in areas with higher treatment capacities (Columns 2 and 4).

The event study results show that the negative externality on water quality in areas with lower treatment capacities has become significant two years after the start of the SBM, and this effect has increased over time.
The estimated coefficients of the event-study specification are reported in Appendix Figure \ref{fig:wq_event_result_did}. 
First, Appendix Figure \ref{fig:wq_event_result_did} shows no differential pre-trends for most panels (except Panel D), which enhances the validity of the parallel pre-trends assumption. 
Second, Appendix Figure \ref{fig:wq_event_result_did} highlights that the negative externality in states with lower treatment capacities has become significant since 2016, two years after the start of the SBM, and this effect has become larger from 2016 to 2019 (Panel B).\footnote{~Appendix Figure \ref{fig:wq_event_result_did} also shows event study plots that compare districts with higher and lower treatment capacities. Because I find differential pre-trends in the case of districts with lower treatment capacities (Panel D), I focus on the results based on state-level variations in treatment capacities.}
The lagged effect is consistent with the fact that the differential increase in the number of latrines among districts with different levels of baseline coverage began around 2016, as shown in Appendix Figure \ref{fig:did_fs}.

\clearpage


\begin{figure}
	\begin{center}
		\includegraphics[width=0.55\textwidth]{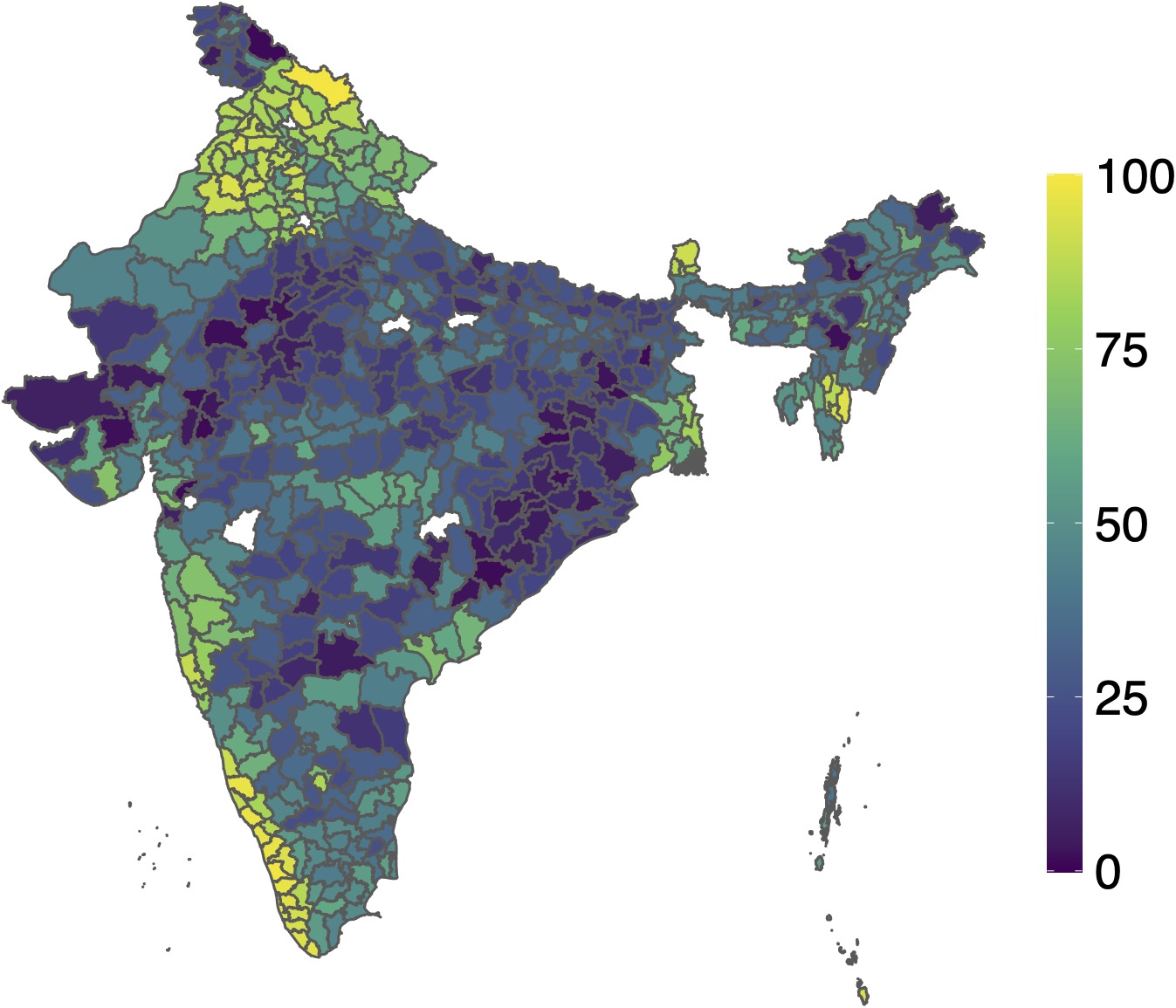}
		\caption{\label{fig:latrine_coverage} Latrine Coverage (\%) in 2013 across Districts}
		\medskip 
		\begin{minipage}{1\textwidth} 
			{\small 
				Notes: Districts with no data on latrine coverage are displayed as blank. 
				These districts correspond to urban areas where latrine data are not recorded under the SBM. \par}
		\end{minipage}
	\end{center}
\end{figure}

\begin{figure}[p]
	\begin{center}
		\includegraphics[width=0.55\textwidth]{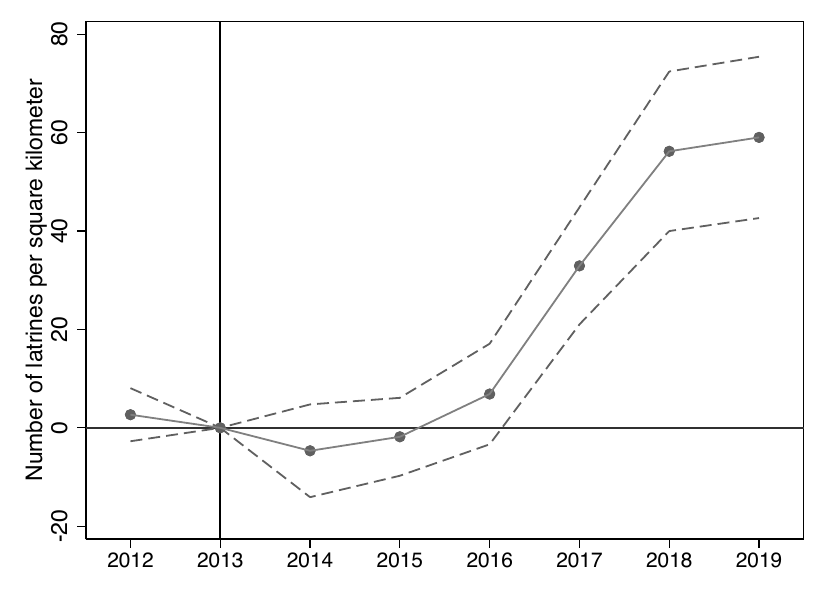}
		\caption{\label{fig:did_fs} Differential Change in the Number of Latrines between Districts with Lower Baseline Coverage and Districts with Higher Baseline Coverage}
		\medskip 
		\begin{minipage}{1\textwidth} 
			{\small
				Notes: This figure shows the district-level regression coefficients of the number of latrines per square kilometer on the interaction terms between (1- baseline latrine coverage in 2013) and year dummies. 
				The regression includes district fixed effects, year fixed effects, and the following controls: precipitation, VIIRS nighttime luminosity, population, the proportions of Scheduled Caste and Scheduled Tribe members, and literacy rates.
				The 95\% confidence intervals are shown with dashed lines. 
				Standard errors are clustered at the district level. 
				\par}
		\end{minipage}
	\end{center}
\end{figure}

\clearpage

\begin{landscape}
	
	\begin{figure}[p]
		\begin{center}
			
			\minipage{0.45\textwidth}
			\includegraphics[width=\linewidth]{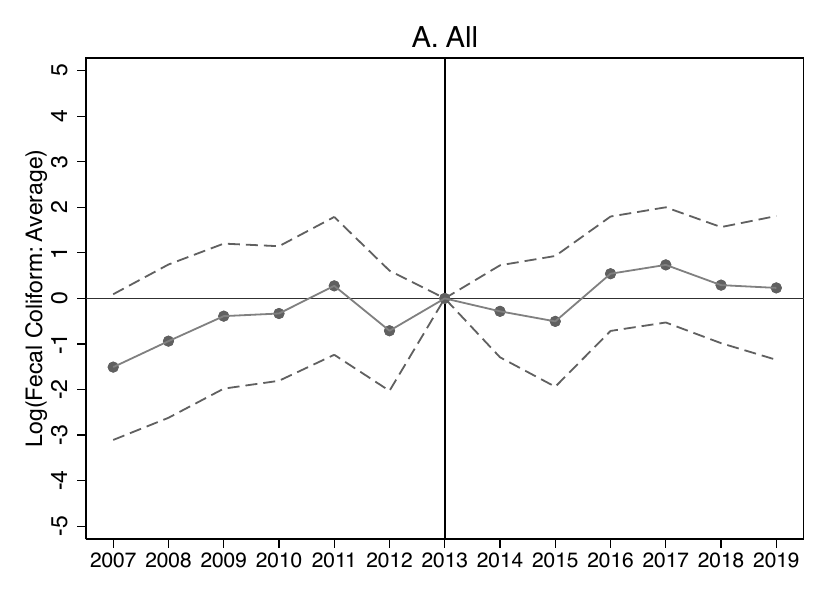}
			\endminipage
			\minipage{0.45\textwidth}
			\includegraphics[width=\linewidth]{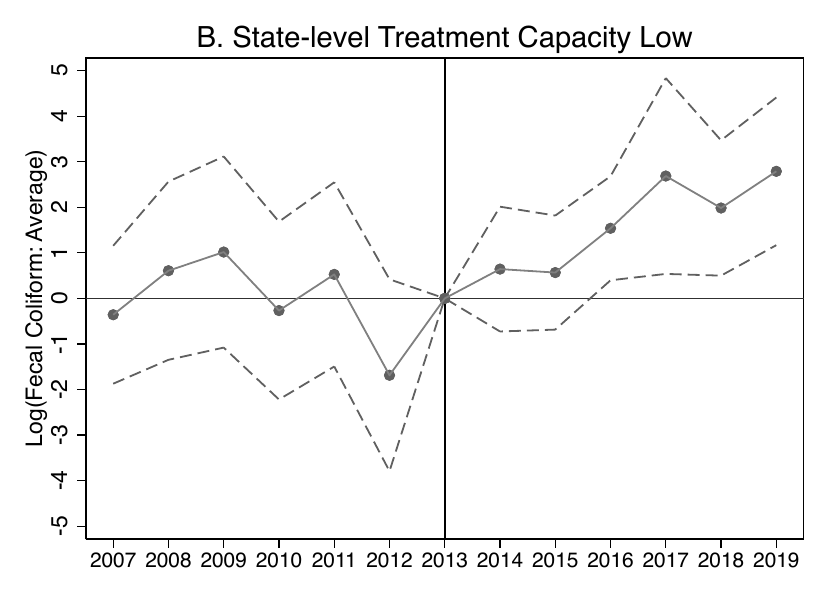}
			\endminipage
			\minipage{0.45\textwidth}%
			\includegraphics[width=\linewidth]{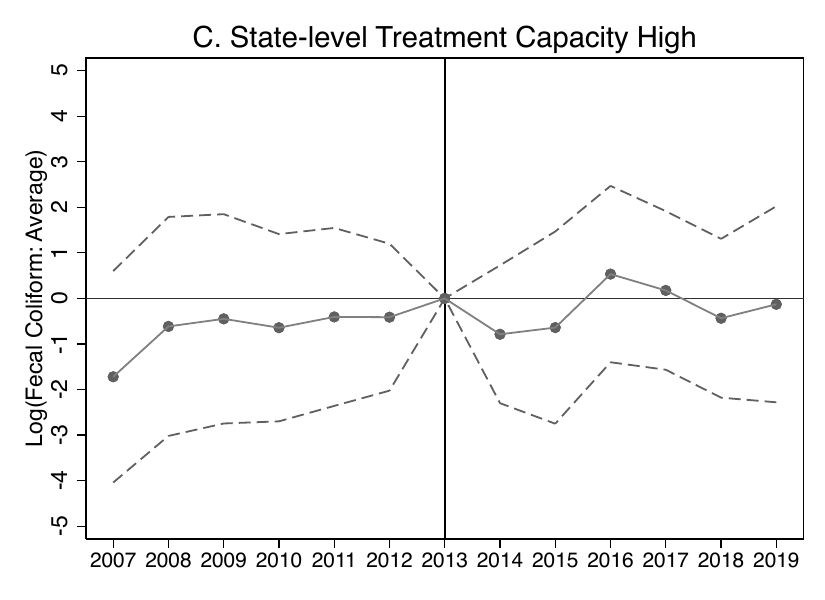}
			\endminipage \\
			\minipage{0.45\textwidth}
			\includegraphics[width=\linewidth]{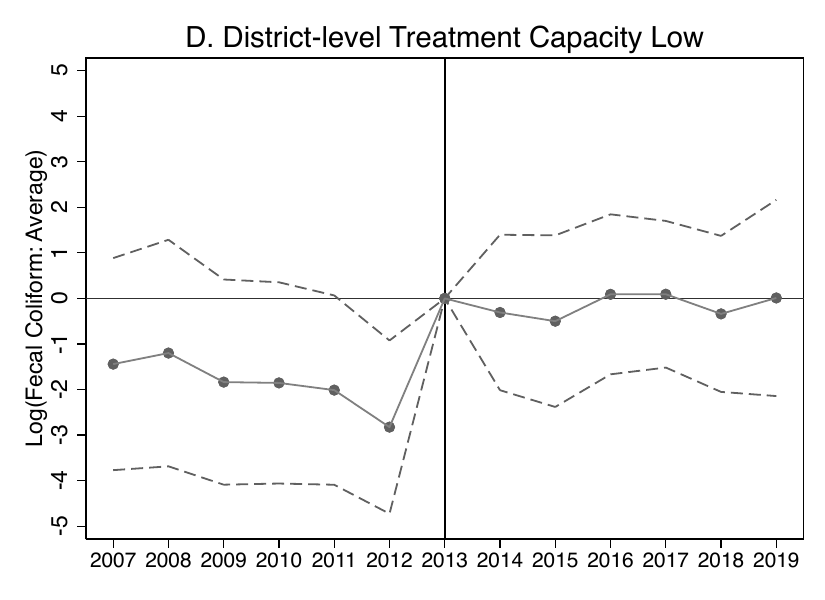}
			\endminipage
			\minipage{0.45\textwidth}%
			\includegraphics[width=\linewidth]{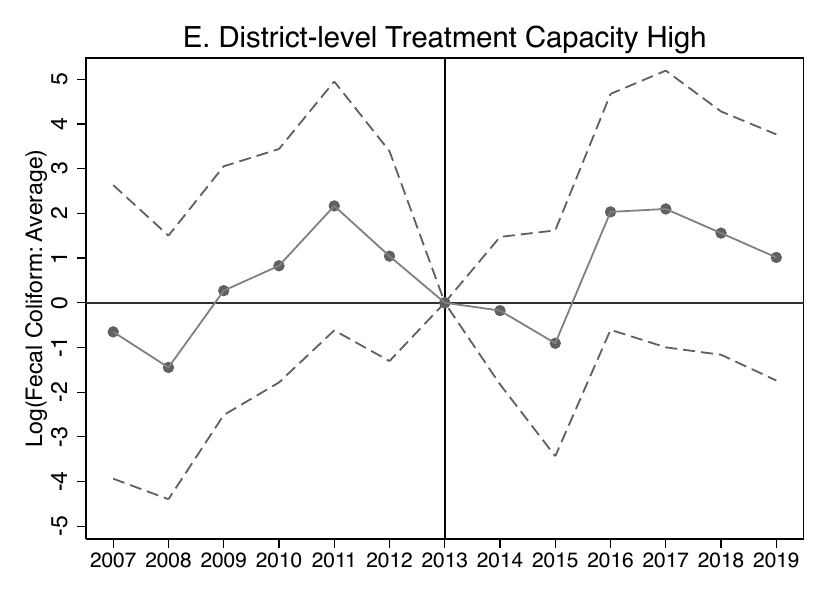}
			\endminipage
			
			\caption{\label{fig:wq_event_result_did} The Dynamic Effects on Water Pollution (Log of Fecal Coliform)}
			\medskip 
			\begin{minipage}{1.2\textwidth} 
				{\small
					Notes: This figure shows the regression coefficients of the logarithm of fecal coliform in regression \ref{did_event_eq}. 
					The 95\% confidence intervals are shown with dashed lines. 
					Standard errors are clustered at the district level. 
					All regressions include monitoring station fixed effects, basin-year fixed effects, and the following controls: precipitation, VIIRS nighttime luminosity, population, the proportions of Scheduled Caste and Scheduled Tribe members, and literacy rates.
					Panel B shows a result in states where the treatment capacities of sewage treatment plants are lower than the median, while Panel C shows a result in states with higher treatment capacities.
					Panel D shows a result in districts where the treatment capacities are lower than the median, while Panel E shows a result in districts with higher treatment capacities. \par}
			\end{minipage}
		\end{center}
	\end{figure}
	
\end{landscape}

\clearpage

\begin{table}[p]\centering \caption{DiD Results: The Effect on Water Quality (Log of Fecal Coliform) \label{tab:did_result_waterquality}}
	\resizebox{0.9\textwidth}{!}{
		\begin{threeparttable}

			{
\def\sym#1{\ifmmode^{#1}\else\(^{#1}\)\fi}
\begin{tabular}{l*{5}{c}}
\toprule
                    &\multicolumn{1}{c}{All}&\multicolumn{2}{c}{State-level Capacity}   &\multicolumn{2}{c}{District-level Capacity}\\\cmidrule(lr){2-2}\cmidrule(lr){3-4}\cmidrule(lr){5-6}
                    &\multicolumn{1}{c}{(1)}&\multicolumn{1}{c}{(2)}&\multicolumn{1}{c}{(3)}&\multicolumn{1}{c}{(4)}&\multicolumn{1}{c}{(5)}\\
                    &\multicolumn{1}{c}{All}&\multicolumn{1}{c}{High}&\multicolumn{1}{c}{Low}&\multicolumn{1}{c}{High}&\multicolumn{1}{c}{Low}\\
\midrule
(1 - 2013 Latrine   &       0.647         &       0.372         &       1.790\sym{***}&       0.496         &       1.496\sym{**} \\
Coverage) * Post (= 1)&     (0.527)         &     (0.775)         &     (0.660)         &     (0.911)         &     (0.654)         \\
\midrule
Observations        &      10,385         &       5,075         &       5,281         &       4,240         &       6,110         \\
R$^2$               &       0.860         &       0.869         &       0.883         &       0.881         &       0.879         \\
Number of Stations  &       1,187         &         577         &         606         &         465         &         719         \\
Number of Districts &         335         &         182         &         151         &          95         &         238         \\
\bottomrule
\end{tabular}
}

			\begin{tablenotes}
				\setlength{\itemindent}{-2.49997pt}
				\item 
				Notes: The coefficients are reported. 
				Standard errors, clustered at the district level, are in parentheses. 
				***, **, and * indicate significance at the 1\%, 5\%, and 10\% levels, respectively. 
				All regressions include monitoring station fixed effects, basin-year fixed effects, and the following controls: precipitation, VIIRS nighttime luminosity, population, the proportions of Scheduled Caste and Scheduled Tribe members, and literacy rates. 
				Column 2 reports a result in states where the treatment capacities of sewage treatment plants are higher than the median, while Column 3 reports a result in states with lower treatment capacities. 
				Columns 4 and 5 compare results based on the different levels of treatment capacities at the district level.
			\end{tablenotes}
		\end{threeparttable}
	}
\end{table}

\clearpage

\section{Back-of-the-Envelope Analyses} 

\subsection{Decomposition of Upstream--Downstream Effect} \label{sec:appendix_boe_0}

This appendix provides a back-of-the-envelope decomposition to gauge the relative magnitudes of two channels embedded in the upstream--downstream specification, although it relies on strong assumptions.
As discussed in Section \ref{sec:iv_updown}, the upstream--downstream coefficient combines (i) a direct pollution spillover from upstream latrine construction and (ii) an indirect effect operating through local correlated latrine construction. 
Because the specification does not separately identify these two channels, I use heterogeneity by upstream district-level treatment capacity to provide an illustrative magnitude check.

The key assumption is that the direct spillover channel is limited when upstream districts have high treatment capacity, where fecal sludge is more likely to be treated rather than dumped into rivers.
This assumption is consistent with the water quality results, which show insignificant water pollution externalities when upstream areas have higher treatment capacity.
Under this assumption, the estimated upstream--downstream effect in the high-capacity subsample can be interpreted as primarily reflecting the indirect component operating through local correlated construction. 
This estimate is -0.014 (Column 4 of Panel A of Table \ref{tab:result_het_health}), showing a net positive health effect.

In contrast, when upstream districts have low treatment capacity, the estimated upstream--downstream effect is close to zero (Column 5 of Panel A of Table \ref{tab:result_het_health}).
If the indirect component is of similar magnitude in the low-capacity subsample, then the spillover component can be inferred as the residual needed to reconcile the near-zero net estimate with the indirect component. 
This implies a spillover magnitude of about 0.014, corresponding to a negative health effect. 
Therefore, in this district-level heterogeneity analysis, the negative health spillover and the positive indirect component through local correlated construction are of comparable order of magnitude and can offset each other.

\subsection{Cost--Benefit Analysis of Swachh Bharat Mission} \label{sec:appendix_boe_1}

The district-level mortality benefit (USD 5.6 million) is calculated by multiplying the total number of mortalities reduced under the SBM (10.4) by the estimated value of a statistical life in India, which amounts to USD 0.54 million or INR 44.69 million \citep{majumder2018value}.\footnote{~In the cost--benefit analysis, an exchange rate of USD 1 = 83.2 INR (or INR 1 = USD 0.012), as of November 9, 2023, is adopted.}
The reduction in the total number of mortalities is calculated by multiplying the estimated average policy effect (0.269 per 1,000 children) by the estimated district-level mean population aged 0-1 (0.039 million people). 
This population estimate is derived from the district-level mean population (1.66 million people) and the percentage of the population aged 0-4 (9.32\%), according to the 2011 Census.
It is assumed that the population aged 0--1 constitutes one-fourth of the population aged 0--4. 

The district-level subsidy cost (USD 16.9 million) is calculated by multiplying the amount of the SBM subsidy (USD 144.1 or INR 12,000) by the increased number of latrines at the district level between pre-SBM and post-SBM periods (0.12 million).

\subsection{Cost--Benefit Analysis of Having Higher Treatment Capacity} \label{sec:appendix_boe_2}

The district-level additional benefit (USD 7.4 million) is calculated by multiplying the difference in the estimated average policy effects between districts with higher and lower treatment capacities (0.364-0.009=0.355 per 1,000 children) by the same value of a statistical life in India. 

The district-level additional cost (USD 4.5 million) is calculated by multiplying the unit cost of sewage treatment plants (0.10 million USD/million liters per day), which consists of the capital cost and the operation and maintenance cost, by the district-level difference in STP capacity between districts with higher and lower treatment capacities (45.0 million liters per day).
To construct the unit cost of sewage treatment plants, I refer to the estimates of the capital cost for secondary treatment (USD 0.03 million per million liters per day) and the operation and maintenance cost (USD 0.07 million per million liters per day) over a period of five years for the most commonly used technology, the Upflow Anaerobic Sludge Blanket, provided by the Central Pollution Control Board \citep{cpcb2013}.
These five years correspond to the duration of the post-SBM period in the analysis. 
Assuming a 15-year lifespan for STPs, the five-year capital cost is set to one-third of the total capital cost.

\clearpage

\section{Additional Figures} \label{sec:appendix_figure}
\setcounter{figure}{0}
\renewcommand{\thefigure}{E\arabic{figure}}

\begin{figure}[htbp]
	\begin{center}
		\includegraphics[width=0.8\textwidth]{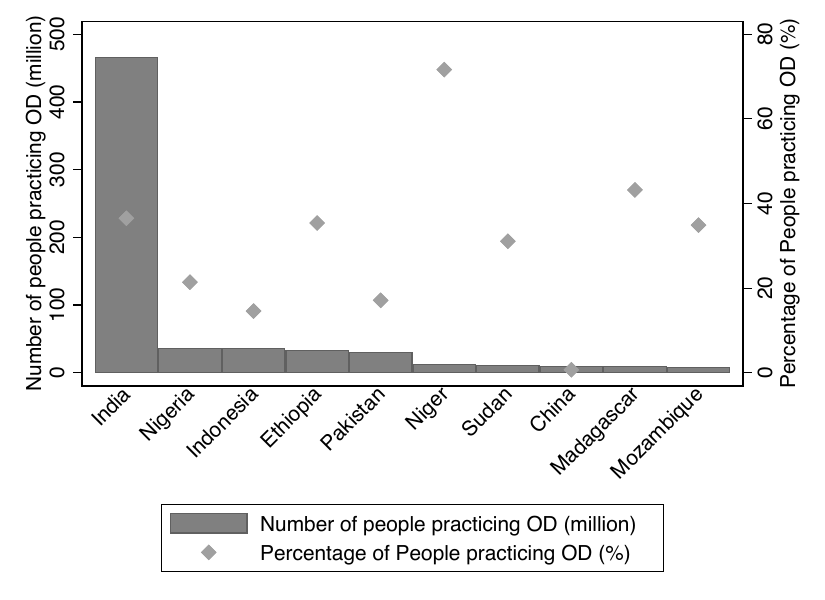}
		\caption{\label{fig:od_india} Top 10 Countries by the Number of People Practicing Open Defecation in 2013}
		\medskip 
		\begin{minipage}{1\textwidth} 
			{\small 
				Notes: This figure documents the top 10 countries by the number of people practicing open defecation (OD). 
				It plots both the number and percentage of people practicing open defecation for these 10 countries. 
				The data source is the WHO/UNICEF Joint Monitoring Programme for Water Supply, Sanitation, and Hygiene. \par}
		\end{minipage}
	\end{center}
\end{figure}

\clearpage

\begin{figure}[p]
	\begin{center}
		\includegraphics[width=0.7\textwidth]{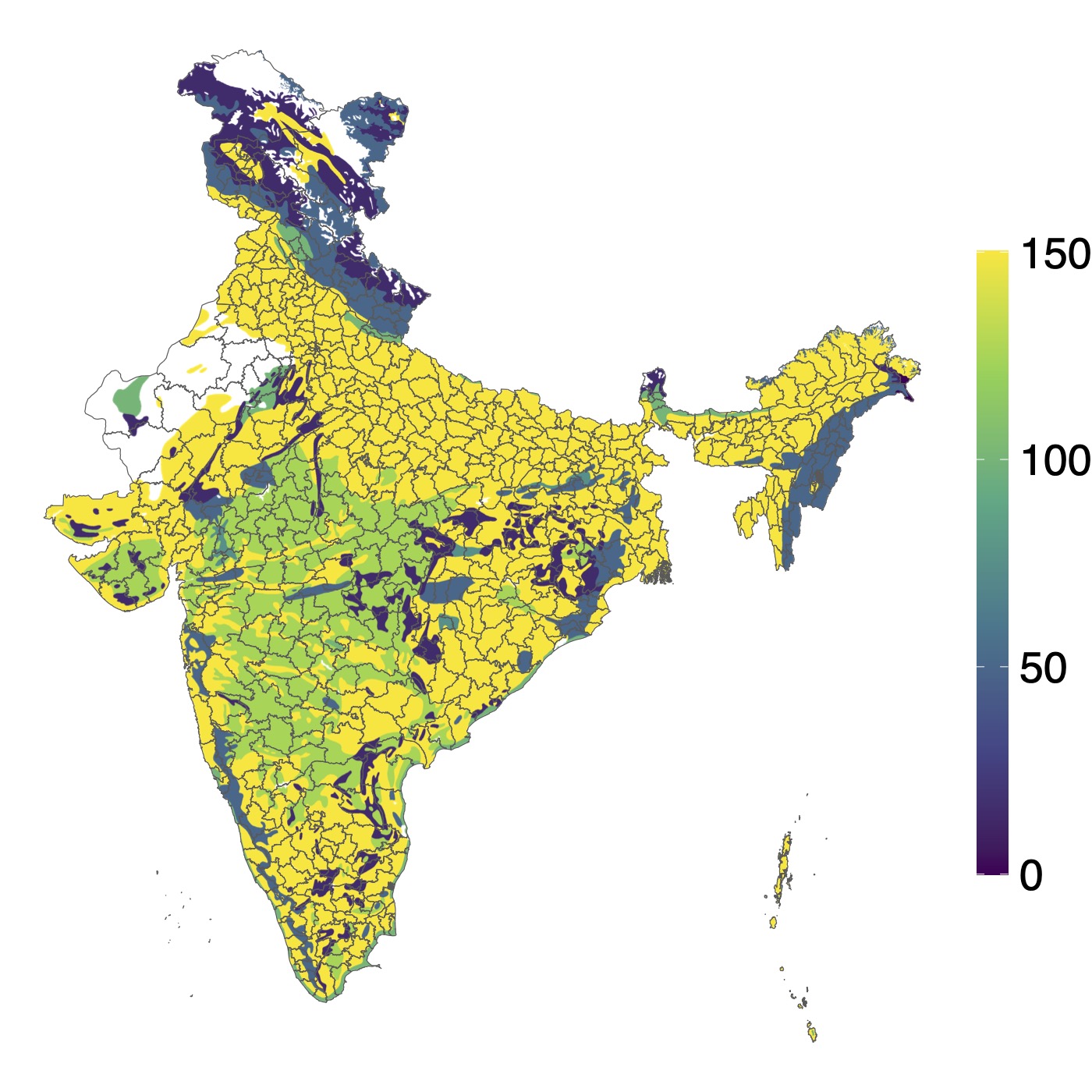}
		\caption{\label{fig:soil_awc_raw} Distribution of Available Water Capacity (mm/m)}
		\medskip 
		\begin{minipage}{1\textwidth} 
			{\small
				Notes: White areas indicate missing data on Available Water Capacity. These areas are not included in the final sample of the analysis due to the absence of water quality monitoring stations, as shown in Figure \ref{fig:stations}. District boundaries are depicted with black lines.
				\par}
		\end{minipage}
	\end{center}
\end{figure}

\clearpage

\begin{figure}[p]
	\begin{center}
		\includegraphics[width=0.9\textwidth]{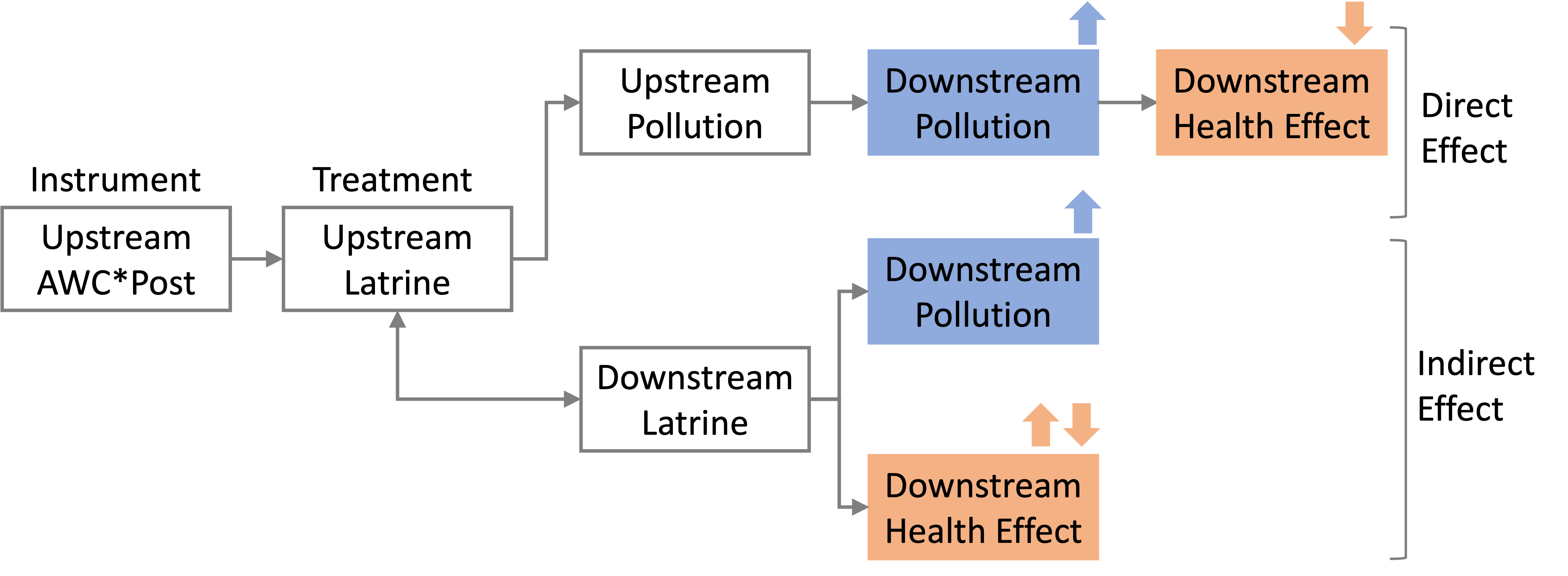}
		\caption{\label{fig:updown_analysis} Two Underlying Channels in Upstream--Downstream Analysis}
		\medskip 
		\begin{minipage}{1\textwidth} 
			{\small
				Notes: This figure illustrates two underlying channels in the upstream--downstream analysis.
				The $\beta_{IV}^{U}$ in regression \ref{iv_st_updown} represents the composite of the two underlying effects.
				In this figure, a reference district is referred to as a downstream area.
				The first underlying channel is a direct effect where upstream latrine construction leads to water pollution that flows downstream, subsequently causing a negative externality on health in downstream areas.
				The second underlying channel is an indirect effect where downstream latrine construction, which is correlated with upstream latrine construction, contributes to increased water pollution in downstream areas. 
				The sign of the health effect in the second channel depends on the relative magnitude of direct positive health effects and water pollution externalities resulting from latrine construction (reduced open defecation) in downstream areas.
				\par}
		\end{minipage}
	\end{center}
\end{figure}

\clearpage

\begin{figure}[p]
	\begin{center}
		
		\minipage{0.8\textwidth}
		\includegraphics[width=\linewidth]{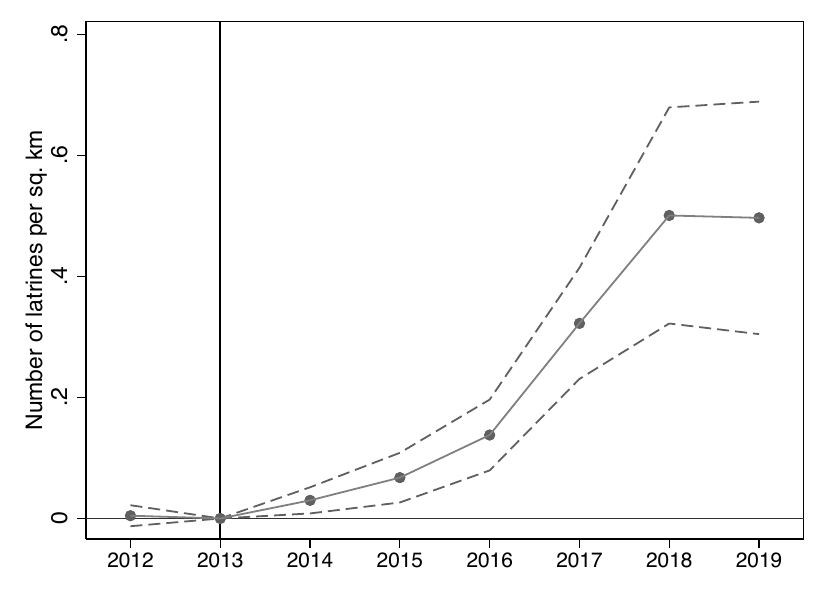}
		\endminipage
		
		\caption{\label{fig:eventstudy_awc_latrine_wqreg} Event Study Plot of First-Stage Regression of Number of Latrines}
		\medskip 
		\begin{minipage}{1\textwidth} 
			{\footnotesize
				Notes: This figure shows the regression coefficients of the number of latrines per square kilometer on the interaction terms between Available Water Capacity and year dummies in the water quality regression.
				The 95\% confidence intervals are shown with dashed lines. 
				Standard errors are clustered at the district level. 
				The regression includes monitoring station fixed effects, year fixed effects, and precipitation as a control. 
				\par}
		\end{minipage}
	\end{center}
\end{figure}

\clearpage

%
%
%

\section{Additional Tables} \label{sec:appendix_table}
\setcounter{table}{0}
\renewcommand{\thetable}{F\arabic{table}}

\begin{table}[htbp]\centering \caption{Summary Statistics of Variables for Robustness Checks \label{tab:sumstats_appendix}}
	\resizebox{0.8\textwidth}{!}{
		\begin{threeparttable}

			\begin{tabular}{@{}lccccc@{}}
	\toprule
	& Mean & SD & Min & Max & Obs. \\ 
	\midrule
	\textit{Panel A. Time-varying variables: pre-SBM (2007-2013)} &     &   &    &   &  \\
	\addlinespace
Fecal coliform: Maximum (million MPN/100ml) & 5.21 & 287.55 & 0 & 20000 & 4939\\
Nitrate/Nitrite: Average (mg/L) & 1.8 & 2.77 & 0 & 24.5 & 2436\\
Temperature: Average ($^\circ$C) & 24.79 & 4.79 & 2.9 & 43.5 & 5206\\
Biochemical Oxygen Demand: Average (mg/L) & 5.4 & 17.17 & 0 & 534.5 & 5219\\
Dissolved Oxygen: Average (mg/L) & 6.77 & 1.75 & 0 & 38.85 & 5204\\
Violation of Bathing Water Quality Criteria (=1) & 0.21 & 0.41 & 0 & 1 & 4939\\
Diarrheal early-neonatal mortality rate (per 1,000) & 20.2 & 13.65 & 0.5 & 71.86 & 2359\\
Diarrheal late-neonatal mortality rate (per 1,000) & 9.22 & 6.19 & 0.23 & 32.41 & 2359\\
Diarrheal age 1-4 mortality rate (per 1,000) & 0.48 & 0.34 & 0.01 & 1.83 & 2359\\
Diarrheal under 5 mortality rate (per 1,000) & 1.07 & 0.72 & 0.03 & 3.87 & 2359\\
Overweight prevalence age 0-5 (\%) & 6.77 & 2.33 & 0.77 & 17.57 & 2359\\
Infant mortality rate (per 1,000) $<$ 5 km of rivers & 36.44 & 187.37 & 0 & 1000 & 63125\\
Infant mortality rate (per 1,000) $<$ 10 km of rivers & 36.9 & 188.53 & 0 & 1000 & 101045\\
Latrine coverage (\%) & 43 & 25.48 & 0.08 & 100 & 586\\
\addlinespace
	\midrule
	\textit{Panel B. Time-varying variables: post-SBM (2014-2019)} &      &    &     &     &   \\
	\addlinespace
Fecal coliform: Maximum (million MPN/100ml) & 1.44 & 60.76 & 0 & 3500 & 5555\\
Nitrate/Nitrite: Average (mg/L) & 2.19 & 19.84 & 0 & 1150.02 & 5521\\
Temperature: Average ($^\circ$C) & 24.85 & 4.36 & 0 & 35 & 5765\\
Biochemical Oxygen Demand: Average (mg/L) & 5.5 & 18.52 & 0 & 719.5 & 5739\\
Dissolved Oxygen: Average (mg/L) & 6.61 & 1.81 & 0 & 51.1 & 5766\\
Violation of Bathing Water Quality Criteria (=1) & 0.26 & 0.44 & 0 & 1 & 5553\\
Diarrheal early-neonatal mortality rate (per 1,000) & 9.79 & 7.28 & 0.29 & 35.57 & 2022\\
Diarrheal late-neonatal mortality rate (per 1,000) & 4.64 & 3.44 & 0.14 & 16.76 & 2022\\
Diarrheal age 1-4 mortality rate (per 1,000) & 0.2 & 0.15 & 0.01 & 0.73 & 2022\\
Diarrheal under 5 mortality rate (per 1,000) & 0.51 & 0.38 & 0.02 & 1.86 & 2022\\
Overweight prevalence age 0-5 (\%) & 7.35 & 2.73 & 0.57 & 25.49 & 2022\\
Infant mortality rate (per 1,000) $<$ 5 km of rivers & 34.45 & 182.38 & 0 & 1000 & 41946\\
Infant mortality rate (per 1,000) $<$ 10 km of rivers & 35.31 & 184.56 & 0 & 1000 & 67633\\
Latrine coverage (\%) & 76.78 & 27.43 & 3.58 & 100 & 1814\\
\addlinespace
\midrule
\textit{Panel C. Variables not varying over time} &      &    &     &     &   \\
\addlinespace
2011 Population (thousand) & 1572.08 & 1077 & 28.99 & 6074.19 & 337\\
2011 \% of Scheduled caste population & 16.75 & 9.69 & 0 & 53.39 & 337\\
2011 \% of Scheduled tribe population & 16.98 & 25.24 & 0 & 98.10 & 337\\
2011 \% of Literate population & 61.16 & 10.44 & 28.66 & 88.7 & 337\\
2013 VIIRS nighttime luminosity (nW/cm2/sr) & 0.71 & 1.57 & 0.01 & 17.98 & 337\\
2011 \% of Population Living within 5 km of Rivers & 36.44 & 19.18 & 0.3 & 93.88 & 316\\
2011 \% of Population Living within 10 km of Rivers & 57.81 & 25.04 & 0 & 100 & 320\\
\bottomrule
\end{tabular}
			
			\begin{tablenotes}
				\setlength{\itemindent}{-2.49997pt}
				\item 
				Notes: This table shows summary statistics of time-varying variables for pre-SBM periods (2007--2013) in Panel A and post-SBM periods (2014--2019) in Panel B, and summary statistics of time-invariant variables in Panel C. 
				The latrine data are available only from 2012 to 2019, while data of other time-varying variables are available from 2007 to 2019.
			\end{tablenotes}
		\end{threeparttable}
	}
\end{table}

\begin{landscape}
\begin{table}[p]\centering \caption{Balance Tests between Sample Districts and Remaining Districts in India \label{tab:balance_samples}}
	\resizebox{1.2\textwidth}{!}{
		\begin{threeparttable}

			\begin{tabular}{l*{7}c}
\toprule
 & \multicolumn{4}{c}{Means: pre-SBM} & \multicolumn{3}{c}{Differences} \\ \cmidrule(lr){2-5} \cmidrule(lr){6-8}
 &\multicolumn{1}{c}{(1)}&\multicolumn{1}{c}{(2)}&\multicolumn{1}{c}{(3)}&\multicolumn{1}{c}{(4)}&\multicolumn{1}{c}{(5)}&\multicolumn{1}{c}{(6)} &\multicolumn{1}{c}{(7)}\\
 Variable & All  & Baseline & U--D (Water) & U--D (Health) &(2) vs not (2)& (3) vs not (3)&(4) vs not (4)  \\
                     
\midrule
Latrine coverage (\%)&40.42&42.11&39.72&39.42&3.75*&-0.93&-1.19\\
&(25.14)&(25.58)&(23.32)&(21.77)&(2.02)&(2.23)&(2.42)\\
Diarrheal post-neonatal&2.32&2.19&2.52&2.50&-0.27*&0.26*&0.22\\
mortality rate (per 1,000) &(1.93)&(1.52)&(1.56)&(1.61)&(0.16)&(0.16)&(0.18)\\
Population (thousand)&1302.73&1572.08&1723.64&1657.13&568.91***&554.28***&422.37***\\
&(1018.81)&(1077.00)&(1029.35)&(996.05)&(76.60)&(93.98)&(107.10)\\
VIIRS nighttime&1.36&0.71&0.67&0.66&-1.38***&-0.91***&-0.84***\\
 luminosity (nW/cm2/sr)  & (5.39)&(1.57)&(0.86)&(0.91)&(0.44)&(0.29)&(0.27)\\
\bottomrule
\end{tabular}

			\begin{tablenotes}
				\setlength{\itemindent}{-2.49997pt}
				\item 
				Notes: Columns 1-4 report the means of pre-SBM variables (in 2013, except for population in 2011) for the baseline sample, the upstream--downstream sample for water quality analysis, and the upstream--downstream sample for health analysis, respectively. 
				Columns 5-7 test the differences between each sample and the remaining districts in India, with ***, **, and * indicating significance at the 1\%, 5\%, and 10\% levels, respectively.
			\end{tablenotes}
		\end{threeparttable}
	}
\end{table}
\end{landscape}

\clearpage

\begin{table}[p]\centering \caption{Upstream--Downstream Analysis: Alternative Buffer Sizes \label{tab:updown_change_buffers}}
	\resizebox{1\textwidth}{!}{
		\begin{threeparttable}

			{
\def\sym#1{\ifmmode^{#1}\else\(^{#1}\)\fi}
\begin{tabular}{l*{6}{c}}
\toprule
                    &\multicolumn{6}{c}{Buffer Distances from Reference Stations/Districts}                                                                       \\\cmidrule(lr){2-7}
                    &\multicolumn{1}{c}{(1)}&\multicolumn{1}{c}{(2)}&\multicolumn{1}{c}{(3)}&\multicolumn{1}{c}{(4)}&\multicolumn{1}{c}{(5)}&\multicolumn{1}{c}{(6)}\\
                    &\multicolumn{1}{c}{0--50 km}&\multicolumn{1}{c}{0--100 km}&\multicolumn{1}{c}{0--150 km}&\multicolumn{1}{c}{50--150 km}&\multicolumn{1}{c}{100--150 km}&\multicolumn{1}{c}{Full}\\
\midrule
\multicolumn{7}{l}{\textit{Panel A. Dependent Variable: Log(Fecal Coliform)}} \\ 
\addlinespace
Upstream number of &       0.014         &       0.017         &       0.015         &       0.003         &       0.001         &       0.037\sym{**} \\
latrines per sq. km &     (0.012)         &     (0.013)         &     (0.011)         &     (0.009)         &     (0.007)         &     (0.016)         \\
\addlinespace
Observations        &       1,758         &       2,152         &       2,228         &       2,008         &       1,488         &       2,235         \\
Number of Stations  &         287         &         352         &         365         &         325         &         238         &         367         \\
Number of Districts &         133         &         151         &         154         &         140         &         112         &         155         \\
KP F-Stat           &      23.148         &      36.766         &      50.475         &      38.427         &      49.767         &      73.913         \\
AR 95\% CI          &[-.011, .049]         &[-.010, .048]         &[-.008, .039]         &[-.018, .021]         &[-.019, .014]         &[.005, .074]         \\
Average Policy Effect&       0.427         &       0.481         &       0.431         &       0.098         &       0.021         &       0.754         \\
\addlinespace
\midrule
\multicolumn{7}{l}{\textit{Panel B. Dependent Variable: Diarrheal Post-neonatal Mortality Rate (per 1,000)}} \\
\addlinespace
Upstream number of  &      -0.011\sym{*}  &      -0.012\sym{*}  &      -0.011\sym{*}  &      -0.012\sym{*}  &      -0.017\sym{***}&       0.003         \\
latrines per sq. km &     (0.006)         &     (0.006)         &     (0.006)         &     (0.006)         &     (0.006)         &     (0.010)         \\
\addlinespace
Observations        &         688         &         808         &         824         &         704         &         488         &         840         \\
Number of Districts &          86         &         101         &         103         &          88         &          61         &         105         \\
KP F-Stat           &      58.692         &      61.264         &      78.696         &      78.481         &      77.325         &      83.728         \\
AR 95\% CI          &[-.026, .002]         &[-.025, .001]         &[-.023, .001]         &[-.026, .001]         &[-.030, -.004]         &[-.014, .027]         \\
Mean of Dep. Variable&       2.695         &       2.571         &       2.576         &       2.763         &       3.078         &       2.570         \\
Average Policy Effect&      -0.309         &      -0.294         &      -0.269         &      -0.303         &      -0.458         &       0.049         \\
\bottomrule
\end{tabular}
}

			\begin{tablenotes}
				\setlength{\itemindent}{-2.49997pt}
				\item  \small
				Notes: The coefficients are reported. 
				Standard errors, clustered at the district level, are in parentheses. ***, **, and * indicate significance at the 1\%, 5\%, and 10\% levels, respectively. 
				The sample is limited to monitoring stations (Panel A) and districts (Panel B) located along major rivers in India. 
				In Columns 1--5, buffer sizes are changed to identify upstream districts. 
				In Column 6, all upstream districts are included without the restriction on buffer sizes. 
				Regressions in Panel A include monitoring station fixed effects, year fixed effects, and the following controls: precipitation and the interaction of Available Water Capacity and the post-SBM indicator of a reference district.
				Regressions in Panel B include district fixed effects, year fixed effects, and the same controls as in Panel A. 
				The KP F-Stat refers to the Wald version of the \cite{kleibergen2006generalized} rk-statistic on the excluded instrumental variables for non-i.i.d. errors. 
				The AR 95\% CI reports the 95\% confidence interval, which is robust to the weak instrument based on the \cite{anderson1949estimation} test.
				The means of the dependent variables are calculated for the pre-SBM period.
				Average policy effects are calculated by multiplying the estimated coefficients by the change in the number of latrines per square kilometer between pre-SBM and post-SBM periods.
			\end{tablenotes}
		\end{threeparttable}
	}
\end{table}

\clearpage

\begin{landscape}
	\begin{table}[p]\centering \caption{Falsification Tests \label{tab:falsification}}
		\resizebox{1.3\textwidth}{!}{
			\begin{threeparttable}

				{
\def\sym#1{\ifmmode^{#1}\else\(^{#1}\)\fi}
\begin{tabular}{l*{9}{c}}
\toprule
                    &\multicolumn{2}{c}{Log(Nitrate/Nitrite)}   &\multicolumn{2}{c}{Log(Temperature)}       &\multicolumn{2}{c}{Log(BOD)}               &\multicolumn{2}{c}{Log(DO)}                &\multicolumn{1}{c}{Overweight Prevalence}\\\cmidrule(lr){2-3}\cmidrule(lr){4-5}\cmidrule(lr){6-7}\cmidrule(lr){8-9}\cmidrule(lr){10-10}
                    &\multicolumn{1}{c}{(1)}         &\multicolumn{1}{c}{(2)}         &\multicolumn{1}{c}{(3)}         &\multicolumn{1}{c}{(4)}         &\multicolumn{1}{c}{(5)}         &\multicolumn{1}{c}{(6)}         &\multicolumn{1}{c}{(7)}         &\multicolumn{1}{c}{(8)}         &\multicolumn{1}{c}{(9)}         \\
\midrule
Number of latrines  &      0.0003         &                     &     -0.0021         &                     &     -0.0034         &                     &     -0.0008         &                     &                     \\
per sq. km          &    (0.0162)         &                     &    (0.0014)         &                     &    (0.0037)         &                     &    (0.0010)         &                     &                     \\
\addlinespace
Upstream number of  &                     &      0.0205         &                     &     -0.0022\sym{*}  &                     &      0.0048         &                     &     -0.0006         &     -0.0131         \\
latrines per sq. km &                     &    (0.0147)         &                     &    (0.0012)         &                     &    (0.0058)         &                     &    (0.0010)         &    (0.0184)         \\
\midrule
Observations        &       6,379         &       1,848         &       7,102         &       2,171         &       7,084         &       2,191         &       7,094         &       2,205         &         824         \\
Number of Stations  &       1,142         &         341         &       1,179         &         359         &       1,184         &         364         &       1,181         &         364         &           -         \\
Number of Districts &         319         &         142         &         334         &         151         &         336         &         153         &         336         &         153         &         103         \\
KP F-Stat           &       7.991         &      44.278         &      28.449         &      47.154         &      28.067         &      47.218         &      29.955         &      50.664         &      78.696         \\
AR 95\% CI          &[... , .033]         &[-.017, .052]         &[-.005, .001]         &[-.006, .000]         &[-.011, .005]         &[-.007, .018]         &[-.003, .001]         &[-.003, .002]         &[-.050, .027]         \\
Mean of Dep. Variable&       2.110         &       1.633         &      24.775         &      24.543         &       5.157         &       3.361         &       6.653         &       7.205         &       6.781         \\
\bottomrule
\end{tabular}
}

				\begin{tablenotes}
					\setlength{\itemindent}{-2.49997pt}
					\item 
					Notes: The coefficients are reported. 
					Standard errors, clustered at the district level, are in parentheses. ***, **, and * indicate significance at the 1\%, 5\%, and 10\% levels, respectively. 
					Regressions in Columns 1, 3, 5, and 7 include monitoring station fixed effects, year fixed effects, and precipitation as a control.
					Regressions in Columns 2, 4, 6, 8, and 9 include monitoring station (or district) fixed effects, year fixed effects, and the following controls: precipitation and the interaction of Available Water Capacity and the post-SBM indicator of a reference district, and upstream districts are defined as those within the range of $[0,150]$ km from a reference district.
					Column 9 uses overweight prevalence (\%) for ages 0--5 as an outcome.
					The KP F-Stat refers to the Wald version of the \cite{kleibergen2006generalized} rk-statistic on the excluded instrumental variables for non-i.i.d. errors. 
					The AR 95\% CI reports the 95\% confidence interval, which is robust to the weak instrument based on the \cite{anderson1949estimation} test.
					The open-ended confidence interval shows that the searched grids do not extend far enough to capture the point where the rejection probability crosses above 95\%.
					The means of the dependent variables are calculated for the pre-SBM period. 
				\end{tablenotes}
			\end{threeparttable}
		}
	\end{table}
\end{landscape}

\clearpage

\begin{table}[p]\centering \caption{The Effect on Log of Maximum Values of Fecal Coliform \label{tab:result_water_max}}
	\resizebox{0.95\textwidth}{!}{
		\begin{threeparttable}

			{
\def\sym#1{\ifmmode^{#1}\else\(^{#1}\)\fi}
\begin{tabular}{l*{5}{c}}
\toprule
                    &\multicolumn{1}{c}{All}&\multicolumn{2}{c}{State-level Capacity}   &\multicolumn{2}{c}{District-level Capacity}\\\cmidrule(lr){2-2}\cmidrule(lr){3-4}\cmidrule(lr){5-6}
                    &\multicolumn{1}{c}{(1)}&\multicolumn{1}{c}{(2)}&\multicolumn{1}{c}{(3)}&\multicolumn{1}{c}{(4)}&\multicolumn{1}{c}{(5)}\\
                    &\multicolumn{1}{c}{All}&\multicolumn{1}{c}{High}&\multicolumn{1}{c}{Low}&\multicolumn{1}{c}{High}&\multicolumn{1}{c}{Low}\\
\midrule
Number of latrines  &       0.033\sym{***}&      -0.033         &       0.040\sym{***}&       0.016\sym{*}  &       0.055\sym{***}\\
per sq. km          &     (0.009)         &     (0.026)         &     (0.007)         &     (0.010)         &     (0.018)         \\
\midrule
Observations        &       7,201         &       3,453         &       3,748         &       2,902         &       4,299         \\
Number of Stations  &       1,189         &         579         &         610         &         466         &         723         \\
Number of Districts &         337         &         182         &         155         &          96         &         241         \\
KP F-Stat           &      29.954         &       7.576         &      39.516         &      13.648         &      11.931         \\
AR 95\% CI          &[.017, .054]         &[-.130, .018]         &[.028, .059]         &[-.010, .038]         &[.026, .114]         \\
Average Policy Effect&       0.801         &      -0.715         &       1.069         &       0.340         &       1.463         \\
\bottomrule
\end{tabular}
}

			\begin{tablenotes}
				\setlength{\itemindent}{-2.49997pt}
				\item Notes: The coefficients are reported. 
				Standard errors, clustered at the district level, are in parentheses. 
				***, **, and * indicate significance at the 1\%, 5\%, and 10\% levels, respectively. 
				All regressions include monitoring station fixed effects, year fixed effects, and precipitation as a control. 
				Column 2 reports a result in states where the treatment capacities of sewage treatment plants are higher than the median, while Column 3 reports a result in states with lower treatment capacities. 
				Similarly, Columns 4 and 5 compare results based on the different levels of treatment capacities at the district level. 
				The KP F-Stat refers to the Wald version of the \cite{kleibergen2006generalized} rk-statistic on the excluded instrumental variables for non-i.i.d. errors. 
				The AR 95\% CI reports the 95\% confidence interval, which is robust to the weak instrument based on the \cite{anderson1949estimation} test.
				Average policy effects are calculated by multiplying the estimated coefficients by the change in the number of latrines per square kilometer between pre-SBM and post-SBM periods.
				
			\end{tablenotes}
		\end{threeparttable}
	}
\end{table}

\clearpage

\begin{table}[p]\centering \caption{The Effect on Log of Fecal Coliform After Winsorizing \label{tab:result_water_winsorized}}
	\resizebox{0.9\textwidth}{!}{
		\begin{threeparttable}

			{
\def\sym#1{\ifmmode^{#1}\else\(^{#1}\)\fi}
\begin{tabular}{l*{4}{c}}
\toprule
                    &\multicolumn{4}{c}{Cutoff Percentiles for Winsorizing}                                 \\\cmidrule(lr){2-5}
                    &\multicolumn{1}{c}{(1)}&\multicolumn{1}{c}{(2)}&\multicolumn{1}{c}{(3)}&\multicolumn{1}{c}{(4)}\\
                    &\multicolumn{1}{c}{99 Percentile}&\multicolumn{1}{c}{95 Percentile}&\multicolumn{1}{c}{90 Percentile}&\multicolumn{1}{c}{75 Percentile}\\
\midrule
Number of latrines  &       0.031\sym{***}&       0.031\sym{***}&       0.031\sym{***}&       0.024\sym{***}\\
per sq. km          &     (0.008)         &     (0.008)         &     (0.008)         &     (0.007)         \\
\midrule
Observations        &       7,201         &       7,201         &       7,201         &       7,201         \\
Number of Stations  &       1,189         &       1,189         &       1,189         &       1,189         \\
Number of Districts &         337         &         337         &         337         &         337         \\
KP F-Stat           &      29.954         &      29.954         &      29.954         &      29.954         \\
AR 95\% CI          &[.016, .050]         &[.016, .050]         &[.017, .050]         &[.012, .040]         \\
Average Policy Effect&       0.741         &       0.748         &       0.752         &       0.590         \\
\bottomrule
\end{tabular}
}

			\begin{tablenotes}
				\setlength{\itemindent}{-2.49997pt}
				\item Notes: The coefficients are reported. 
				Standard errors, clustered at the district level, are in parentheses. 
				***, **, and * indicate significance at the 1\%, 5\%, and 10\% levels, respectively. 
				All regressions include monitoring station fixed effects, year fixed effects, and precipitation as a control. 
				Columns 1--4 report results after winsorizing fecal coliform values at the 99th, 95th, 90th, and 75th percentiles, respectively.
				The KP F-Stat refers to the Wald version of the \cite{kleibergen2006generalized} rk-statistic on the excluded instrumental variables for non-i.i.d. errors. 
				The AR 95\% CI reports the 95\% confidence interval, which is robust to the weak instrument based on the \cite{anderson1949estimation} test.
				Average policy effects are calculated by multiplying the estimated coefficients by the change in the number of latrines per square kilometer between pre-SBM and post-SBM periods.
			\end{tablenotes}
		\end{threeparttable}
	}
\end{table}

\clearpage

\begin{table}[p]\centering \caption{The Effect on Violation of Bathing Water Quality Criteria \label{tab:result_water_violation}}
	\resizebox{0.9\textwidth}{!}{
		\begin{threeparttable}

			{
\def\sym#1{\ifmmode^{#1}\else\(^{#1}\)\fi}
\begin{tabular}{l*{2}{c}}
\toprule
                    &\multicolumn{2}{c}{Violation of Bathing Water Quality Criteria (=1)}\\\cmidrule(lr){2-3}
                    &\multicolumn{1}{c}{(1)}&\multicolumn{1}{c}{(2)}\\
                    &\multicolumn{1}{c}{Baseline Specification}&\multicolumn{1}{c}{Upstream-Downstream Specification}\\
\midrule
Number of latrines  &       0.006\sym{***}&                     \\
per sq. km          &     (0.002)         &                     \\
\addlinespace
Upstream number of  &                     &       0.004\sym{*}  \\
latrines per sq. km &                     &     (0.002)         \\
\midrule
Observations        &       7,201         &       2,228         \\
Number of Stations  &       1,189         &         365         \\
Number of Districts &         337         &         154         \\
KP F-Stat           &      29.954         &      50.475         \\
AR 95\% CI          &[.003, .012]         &[-.001, .009]         \\
Mean of Dep. Variable&       0.233         &       0.297         \\
Average Policy Effect&       0.156         &       0.115         \\
\bottomrule
\end{tabular}
}

			\begin{tablenotes}
				\setlength{\itemindent}{-2.49997pt}
				\item Notes: The coefficients are reported. 
				Standard errors, clustered at the district level, are in parentheses.
				***, **, and * indicate significance at the 1\%, 5\%, and 10\% levels, respectively.
				Violation is defined as an indicator equal to one if the average fecal coliform level exceeds the maximum permissible level under the Primary Water Quality Criteria for Bathing Water (2,500 MPN/100 ml).
				The regression in Column 1 includes monitoring station fixed effects, year fixed effects, and precipitation as a control.
				The regression in Column 2 includes monitoring station fixed effects, year fixed effects, and the following controls: precipitation and the interaction of Available Water Capacity and the post-SBM indicator of a reference district.
				In Column 2, the sample is limited to monitoring stations located along major rivers in India, and upstream districts are defined as those within the range of $[0,150]$ km from a reference station. 
				The KP F-Stat refers to the Wald version of the \cite{kleibergen2006generalized} rk-statistic on the excluded instrumental variables for non-i.i.d. errors. 
				The AR 95\% CI reports the 95\% confidence interval, which is robust to the weak instrument based on the \cite{anderson1949estimation} test.
				The means of the dependent variables are calculated for the pre-SBM period.
				Average policy effects are calculated by multiplying the estimated coefficients by the change in the number of latrines per square kilometer between pre-SBM and post-SBM periods.
			\end{tablenotes}
		\end{threeparttable}
	}
\end{table}

\clearpage

\begin{table}[p]\centering \caption{The Effects on Diarrheal Mortality Rates of Other Age Groups (per 1,000) \label{tab:result_health_multiple}}
	\resizebox{1\textwidth}{!}{
		\begin{threeparttable}

			{
\def\sym#1{\ifmmode^{#1}\else\(^{#1}\)\fi}
\begin{tabular}{l*{5}{c}}
\toprule
                    &\multicolumn{1}{c}{(1)}&\multicolumn{1}{c}{(2)}&\multicolumn{1}{c}{(3)}&\multicolumn{1}{c}{(4)}&\multicolumn{1}{c}{(5)}\\
                    &\multicolumn{1}{c}{Early-neonatal}&\multicolumn{1}{c}{Late-neonatal}&\multicolumn{1}{c}{Post-neonatal}&\multicolumn{1}{c}{Age 1--4}&\multicolumn{1}{c}{Under 5}\\
\midrule
Upstream number of  &      -0.092\sym{**} &      -0.042\sym{*}  &      -0.011\sym{*}  &      -0.002\sym{**} &      -0.005\sym{**} \\
latrines per sq. km &     (0.046)         &     (0.021)         &     (0.006)         &     (0.001)         &     (0.002)         \\
\midrule
Observations        &         824         &         824         &         824         &         824         &         824         \\
Number of Districts &         103         &         103         &         103         &         103         &         103         \\
KP F-Stat           &      78.696         &      78.696         &      78.696         &      78.696         &      78.696         \\
AR 95\% CI          &[-.190, .005]         &[-.086, .003]         &[-.023, .001]         &[-.005, .000]         &[-.010, .000]         \\
Mean of Dep. Variable&      18.562         &       8.656         &       2.576         &       0.411         &       0.969         \\
Average Policy Effect&      -2.221         &      -1.004         &      -0.269         &      -0.057         &      -0.116         \\
\bottomrule
\end{tabular}
}

			\begin{tablenotes}
				\setlength{\itemindent}{-2.49997pt}
				\item 
				Notes: The coefficients are reported. 
				Standard errors, clustered at the district level, are in parentheses. 
				***, **, and * indicate significance at the 1\%, 5\%, and 10\% levels, respectively. 
				All regressions include district fixed effects, year fixed effects, and the following controls: precipitation and the interaction of Available Water Capacity and the post-SBM indicator of a reference district. 
				The sample is limited to districts that have monitoring stations used in the water quality regression along major rivers in India.
				Upstream districts are defined as those within the range of $[0,150]$ km from a reference district.
				The KP F-Stat refers to the Wald version of the \cite{kleibergen2006generalized} rk-statistic on the excluded instrumental variables for non-i.i.d. errors. 
				The AR 95\% CI reports the 95\% confidence interval, which is robust to the weak instrument based on the \cite{anderson1949estimation} test.
				The means of the dependent variables are calculated for the pre-SBM period.
				Average policy effects are calculated by multiplying the estimated coefficients by the change in the number of latrines per square kilometer between pre-SBM and post-SBM periods.
			\end{tablenotes}
		\end{threeparttable}
	}
\end{table}

\clearpage

\begin{table}[p]\centering \caption{Robustness Check - Spillovers from Neighboring Districts: The Effect on Water Quality (Log of Fecal Coliform) \label{tab:neighbor}}
	\resizebox{0.95\textwidth}{!}{
		\begin{threeparttable}

			{
\def\sym#1{\ifmmode^{#1}\else\(^{#1}\)\fi}
\begin{tabular}{l*{5}{c}}
\toprule
                    &\multicolumn{1}{c}{All}&\multicolumn{2}{c}{State-level Capacity}   &\multicolumn{2}{c}{District-level Capacity}\\\cmidrule(lr){2-2}\cmidrule(lr){3-4}\cmidrule(lr){5-6}
                    &\multicolumn{1}{c}{(1)}&\multicolumn{1}{c}{(2)}&\multicolumn{1}{c}{(3)}&\multicolumn{1}{c}{(4)}&\multicolumn{1}{c}{(5)}\\
                    &\multicolumn{1}{c}{All}&\multicolumn{1}{c}{High}&\multicolumn{1}{c}{Low}&\multicolumn{1}{c}{High}&\multicolumn{1}{c}{Low}\\
\midrule
Number of latrines  &       0.027\sym{***}&      -0.027         &       0.037\sym{***}&       0.017\sym{*}  &       0.043\sym{***}\\
per sq. km          &     (0.007)         &     (0.019)         &     (0.006)         &     (0.009)         &     (0.013)         \\
\midrule
Observations        &       7,253         &       3,605         &       3,648         &       3,300         &       3,953         \\
Number of Stations  &       1,197         &         603         &         594         &         529         &         668         \\
Number of Districts &         489         &         260         &         229         &         185         &         304         \\
KP F-Stat           &      44.626         &      14.440         &      54.539         &      26.013         &      15.433         \\
AR 95\% CI          &[.013,  .042]         &[-.076,  .010]         &[.027, .050]         &[-.003, .036]         &[.021, .081]         \\
Average Policy Effect&       0.655         &      -0.599         &       0.952         &       0.362         &       1.140         \\
\bottomrule
\end{tabular}
}

			\begin{tablenotes}
				\setlength{\itemindent}{-2.49997pt}
				\item 
				Notes: The coefficients are reported. 
				Standard errors, clustered at the district level, are in parentheses. 
				***, **, and * indicate significance at the 1\%, 5\%, and 10\% levels, respectively. 
				All regressions include monitoring station fixed effects, year fixed effects, and precipitation as a control. 
				Column 2 reports a result in states where the treatment capacities of sewage treatment plants are higher than the median, while Column 3 reports a result in states with lower treatment capacities.
				Columns 4 and 5 compare results based on the different levels of treatment capacities at the district level. 
				The KP F-Stat refers to the Wald version of the \cite{kleibergen2006generalized} rk-statistic on the excluded instrumental variables for non-i.i.d. errors. 
				The AR 95\% CI reports the 95\% confidence interval, which is robust to the weak instrument based on the \cite{anderson1949estimation} test.
				Average policy effects are calculated by multiplying the estimated coefficients by the change in the number of latrines per square kilometer between pre-SBM and post-SBM periods.
			\end{tablenotes}
		\end{threeparttable}
	}
\end{table}

\clearpage

\begin{table}[p]\centering \caption{Robustness Check: Influence from Urban Areas \label{tab:influence_urban}}
	\resizebox{1\textwidth}{!}{
		\begin{threeparttable}

			{
\def\sym#1{\ifmmode^{#1}\else\(^{#1}\)\fi}
\begin{tabular}{l*{4}{c}}
\toprule
                    &\multicolumn{1}{c}{No Exclusion}&\multicolumn{1}{c}{50 km Exclusion}&\multicolumn{1}{c}{100 km Exclusion}&\multicolumn{1}{c}{150 km Exclusion}\\\cmidrule(lr){2-2}\cmidrule(lr){3-3}\cmidrule(lr){4-4}\cmidrule(lr){5-5}
                    &\multicolumn{1}{c}{(1)}         &\multicolumn{1}{c}{(2)}         &\multicolumn{1}{c}{(3)}         &\multicolumn{1}{c}{(4)}         \\
\midrule
\multicolumn{5}{l}{\textit{Panel A. Dependent Variable: Log(Fecal Coliform)}} \\ 
\addlinespace
Number of latrines  &       0.030\sym{***}&       0.039\sym{***}&       0.050\sym{***}&       0.072\sym{**} \\
per sq. km          &     (0.008)         &     (0.010)         &     (0.015)         &     (0.035)         \\
\addlinespace
Observations        &       7,201         &       5,295         &       3,716         &       2,492         \\
Number of Stations  &       1,189         &         890         &         623         &         421         \\
Number of Districts &         337         &         284         &         196         &         125         \\
KP F-Stat           &      29.954         &      25.785         &      17.574         &       5.693         \\
AR 95\% CI          &[.015, .049]         &[.021, .067]         &[.026, .099]         &[.026, ...]         \\
Average Policy Effect&       0.719         &       1.035         &       1.369         &       1.902         \\
\addlinespace
\midrule
\multicolumn{5}{l}{\textit{Panel B. Dependent Variable: Diarrheal Post-neonatal Mortality Rate (per 1,000)}} \\
\addlinespace
Upstream number of  &      -0.011\sym{*}  &      -0.010         &      -0.007         &      -0.009         \\
latrines per sq. km &     (0.006)         &     (0.008)         &     (0.011)         &     (0.015)         \\
\addlinespace
Observations        &         824         &         480         &         288         &         152         \\
Number of Districts &         103         &          60         &          36         &          19         \\
KP F-Stat           &      78.696         &      49.232         &      22.506         &      29.583         \\
AR 95\% CI          &[-.023, .001]         &[-.026, .008]         &[-.031, .025]         &[-.045, .042]         \\
Mean of Dep. Variable&       2.576         &       2.577         &       2.561         &       2.670         \\
Average Policy Effect&      -0.269         &      -0.208         &      -0.150         &      -0.137         \\
\bottomrule
\end{tabular}
}

			\begin{tablenotes}
				\setlength{\itemindent}{-2.49997pt}
				\item Notes: The coefficients are reported. 
				Standard errors, clustered at the district level, are in parentheses. 
				***, **, and * indicate significance at the 1\%, 5\%, and 10\% levels, respectively. 
				In Columns 2--4, I exclude monitoring stations (Panel A) and districts (Panel B) that are within a specified distance from cities that have a population of 1 million or more.
				Panel A includes monitoring station fixed effects, year fixed effects, and precipitation as a control.
				Panel B includes district fixed effects, year fixed effects, and the following controls: precipitation and the interaction of Available Water Capacity and the post-SBM indicator of a reference district, and upstream districts are defined as those within the range of $[0,150]$ km from a reference district.
				The KP F-Stat refers to the Wald version of the \cite{kleibergen2006generalized} rk-statistic on the excluded instrumental variables for non-i.i.d. errors. 
				The AR 95\% CI reports the 95\% confidence interval, which is robust to the weak instrument based on the \cite{anderson1949estimation} test.
				The open-ended confidence interval shows that the searched grids do not extend far enough to capture the point where the rejection probability crosses above 95\%.
				The means of the dependent variables are calculated for the pre-SBM period.
				Average policy effects are calculated by multiplying the estimated coefficients by the change in the number of latrines per square kilometer between pre-SBM and post-SBM periods.
			\end{tablenotes}
		\end{threeparttable}
	}
\end{table}

\clearpage

\begin{table}[p]\centering \caption{Robustness Check - Balanced Panel: The Effect on Water Quality (Log of Fecal Coliform) \label{tab:balanced_wq_iv}}
	\resizebox{0.95\textwidth}{!}{
		\begin{threeparttable}

			{
\def\sym#1{\ifmmode^{#1}\else\(^{#1}\)\fi}
\begin{tabular}{l*{5}{c}}
\toprule
                    &\multicolumn{1}{c}{All}&\multicolumn{2}{c}{State-level Capacity}   &\multicolumn{2}{c}{District-level Capacity}\\\cmidrule(lr){2-2}\cmidrule(lr){3-4}\cmidrule(lr){5-6}
                    &\multicolumn{1}{c}{(1)}&\multicolumn{1}{c}{(2)}&\multicolumn{1}{c}{(3)}&\multicolumn{1}{c}{(4)}&\multicolumn{1}{c}{(5)}\\
                    &\multicolumn{1}{c}{All}&\multicolumn{1}{c}{High}&\multicolumn{1}{c}{Low}&\multicolumn{1}{c}{High}&\multicolumn{1}{c}{Low}\\
\midrule
Number of latrines  &       0.024\sym{***}&      -0.010         &       0.031\sym{***}&       0.009         &       0.039\sym{***}\\
per sq. km          &     (0.009)         &     (0.022)         &     (0.008)         &     (0.012)         &     (0.014)         \\
\midrule
Observations        &       3,776         &       1,552         &       2,224         &       1,600         &       2,176         \\
Number of Stations  &         472         &         194         &         278         &         200         &         272         \\
Number of Districts &         158         &          75         &          83         &          53         &         105         \\
KP F-Stat           &      12.357         &      12.512         &      13.449         &       4.018         &       7.917         \\
AR 95\% CI          &[.009, .048]         &[-.072, .032]         &[.018, .053]         &[..., .048]         &[.018, .086]         \\
Average Policy Effect&       0.644         &      -0.209         &       0.926         &       0.210         &       1.137         \\
\bottomrule
\end{tabular}
}

			\begin{tablenotes}
				\setlength{\itemindent}{-2.49997pt}
				\item Notes: The coefficients are reported. 
				Standard errors, clustered at the district level, are in parentheses. 
				***, **, and * indicate significance at the 1\%, 5\%, and 10\% levels, respectively. 
				The sample is limited to monitoring stations that have observations every year from 2012 to 2019, which yields a balanced panel. 
				All regressions include monitoring station fixed effects, year fixed effects, and precipitation as a control. 
				Column 2 reports a result in states where the treatment capacities of sewage treatment plants are higher than the median, while Column 3 reports a result in states with lower treatment capacities. 
				Similarly, Columns 4 and 5 compare results based on the different levels of treatment capacities at the district level. 
				The KP F-Stat refers to the Wald version of the \cite{kleibergen2006generalized} rk-statistic on the excluded instrumental variables for non-i.i.d. errors. 
				The AR 95\% CI reports the 95\% confidence interval, which is robust to the weak instrument based on the \cite{anderson1949estimation} test. 
				The open-ended confidence interval shows that the searched grids do not extend far enough to capture the point where the rejection probability crosses above 95\%.
				Average policy effects are calculated by multiplying the estimated coefficients by the change in the number of latrines per square kilometer between pre-SBM and post-SBM periods.
				
			\end{tablenotes}
		\end{threeparttable}
	}
\end{table}

\clearpage

\begin{table}[p]\centering \caption{Robustness Check - Alternative Mortality Dataset: The Effects on Health of Children Living Close to Rivers (Infant Mortality Rate (per 1,000)) \label{tab:result_het_close_river_nfhs}}
	\resizebox{0.9\textwidth}{!}{
		\begin{threeparttable}
			
			{
\def\sym#1{\ifmmode^{#1}\else\(^{#1}\)\fi}
\begin{tabular}{l*{5}{c}}
\toprule
                    &\multicolumn{1}{c}{All}&\multicolumn{2}{c}{State-level Capacity}   &\multicolumn{2}{c}{District-level Capacity}\\\cmidrule(lr){2-2}\cmidrule(lr){3-4}\cmidrule(lr){5-6}
                    &\multicolumn{1}{c}{(1)}&\multicolumn{1}{c}{(2)}&\multicolumn{1}{c}{(3)}&\multicolumn{1}{c}{(4)}&\multicolumn{1}{c}{(5)}\\
                    &\multicolumn{1}{c}{All}&\multicolumn{1}{c}{High}&\multicolumn{1}{c}{Low}&\multicolumn{1}{c}{High}&\multicolumn{1}{c}{Low}\\
\midrule
\multicolumn{6}{l}{\textit{Panel A. Children Living within 5 km of Rivers}} \\ 
\addlinespace
Upstream number of  &      -1.954\sym{***}&      -4.022\sym{**} &      -1.399\sym{**} &      -2.386\sym{***}&      -1.652         \\
latrines per sq. km &     (0.579)         &     (1.966)         &     (0.551)         &     (0.685)         &     (1.066)         \\
\addlinespace
Observations        &      11,034         &       5,677         &       5,357         &       5,358         &       5,676         \\
Number of Districts &          69         &          38         &          31         &          36         &          33         \\
KP F-Stat           &      34.851         &       7.143         &      30.501         &      29.117         &      10.342         \\
Mean of Dep. Variable&      35.473         &      41.290         &       29.215         &      39.531         &      31.808         \\
\addlinespace
\midrule
\multicolumn{6}{l}{\textit{Panel B. Children Living within 10 km of Rivers}} \\ 
\addlinespace
Upstream number of  &      -1.251\sym{**} &      -3.195\sym{*}  &      -0.839\sym{**} &      -1.769\sym{***}&      -0.821         \\
latrines per sq. km &     (0.481)         &     (1.842)         &     (0.395)         &     (0.603)         &     (0.780)         \\
\addlinespace
Observations        &      18,094         &       9,417         &       8,677         &       8,902         &       9,192         \\
Number of Districts &          70         &          38         &          32         &          36         &          34         \\
KP F-Stat           &      37.937         &       7.894         &      30.501         &      32.344         &       9.160         \\
Mean of Dep. Variable&      36.321         &      37.710         &       34.805          &      40.404         &      32.579         \\
\bottomrule
\end{tabular}
}

			\begin{tablenotes}
				\setlength{\itemindent}{-2.49997pt}
				\item 
				Notes: The coefficients are reported. 
				Standard errors, clustered at the district level, are in parentheses. 
				***, **, and * indicate significance at the 1\%, 5\%, and 10\% levels, respectively. 
				All regressions include district fixed effects, year fixed effects, and the following controls: precipitation, the interaction of Available Water Capacity and the post-SBM indicator of a reference district, indicators for being a first-born child and part of a multiple birth, religion (Hindu, Muslim, others), caste (Scheduled Caste, Scheduled Tribe, Other Backward Class, others), education (primary, secondary, or higher), and wealth quintiles.
				Upstream districts are defined as those within the range of $[0,150]$ km from a reference district.
				Column 2 reports a result when upstream states have higher treatment capacities of sewage treatment plants than the median, while Column 3 reports a result in the case of upstream states with lower treatment capacities. 
				Columns 4 and 5 compare results based on the different levels of upstream treatment capacities at the district level. 
				The KP F-Stat refers to the Wald version of the \cite{kleibergen2006generalized} rk-statistic on the excluded instrumental variables for non-i.i.d. errors.
				The means of the dependent variables are calculated for the pre-SBM period.
				Average policy effects are calculated by multiplying the estimated coefficients by the change in the number of latrines per square kilometer between pre-SBM and post-SBM periods.
			\end{tablenotes}
		\end{threeparttable}
	}
\end{table}

\clearpage

\begin{table}[p]\centering \caption{The Effects on Health in Areas Close to Rivers \label{tab:result_het_close_river_ihme}}
	\resizebox{1\textwidth}{!}{
		\begin{threeparttable}
			
			{
\def\sym#1{\ifmmode^{#1}\else\(^{#1}\)\fi}
\begin{tabular}{l*{5}{c}}
\toprule
                    &\multicolumn{1}{c}{All}&\multicolumn{2}{c}{State-level Capacity}   &\multicolumn{2}{c}{District-level Capacity}\\\cmidrule(lr){2-2}\cmidrule(lr){3-4}\cmidrule(lr){5-6}
                    &\multicolumn{1}{c}{(1)}&\multicolumn{1}{c}{(2)}&\multicolumn{1}{c}{(3)}&\multicolumn{1}{c}{(4)}&\multicolumn{1}{c}{(5)}\\
                    &\multicolumn{1}{c}{All}&\multicolumn{1}{c}{High}&\multicolumn{1}{c}{Low}&\multicolumn{1}{c}{High}&\multicolumn{1}{c}{Low}\\
\midrule
\multicolumn{6}{l}{\textit{Panel A. Diarrheal Post-neonatal Mortality Rate (per 1,000) within 5 km of Rivers}} \\ 
\addlinespace
Upstream number of  &      -0.010\sym{*}  &      -0.038\sym{***}&      -0.010         &      -0.014\sym{**} &       0.000         \\
latrines per sq. km &     (0.006)         &     (0.011)         &     (0.006)         &     (0.007)         &     (0.010)         \\
\addlinespace
Observations        &         824         &         432         &         392         &         456         &         368         \\
Number of Districts &         103         &          54         &          49         &          57         &          46         \\
KP F-Stat           &      78.696         &      33.304         &      33.484         &      59.873         &      18.756         \\
AR 95\% CI          &[-.022, .002]         &[-.073, -.020]         &[-.025, .001]         &[-.029, .000]         &[-.027, .026]         \\
Mean of Dep. Variable&       2.569         &       2.530         &       2.612         &       2.423         &       2.750         \\
\addlinespace
\midrule
\multicolumn{6}{l}{\textit{Panel B. Diarrheal Post-neonatal Mortality Rate (per 1,000) within 10 km of Rivers}} \\ 
\addlinespace
Upstream number of  &      -0.011\sym{*}  &      -0.039\sym{***}&      -0.010         &      -0.014\sym{**} &       0.000         \\
latrines per sq. km &     (0.006)         &     (0.011)         &     (0.006)         &     (0.007)         &     (0.010)         \\
\addlinespace
Observations        &         824         &         432         &         392         &         456         &         368         \\
Number of Districts &         103         &          54         &          49         &          57         &          46         \\
KP F-Stat           &      78.696         &      33.304         &      33.484         &      59.873         &      18.756         \\
AR 95\% CI          &[-.023, .001]         &[-.073, -.022]         &[-.026, .002]         &[-.030, .000]         &[-.028, .027]         \\
Mean of Dep. Variable&       2.576         &       2.540         &       2.616         &       2.430         &       2.758         \\
\bottomrule
\end{tabular}
}

			\begin{tablenotes}
				\setlength{\itemindent}{-2.49997pt}
				\item 
				Notes: The coefficients are reported. 
				Standard errors, clustered at the district level, are in parentheses. 
				***, **, and * indicate significance at the 1\%, 5\%, and 10\% levels, respectively. 
				All regressions include district fixed effects, year fixed effects, and the following controls: precipitation and the interaction of Available Water Capacity and the post-SBM indicator of a reference district.
				Upstream districts are defined as those within the range of $[0,150]$ km from a reference district.
				Column 2 reports results when upstream states have higher treatment capacities of sewage treatment plants than the median, while Column 3 reports results in the case of upstream states with lower treatment capacities. 
				Columns 4 and 5 compare results based on the different levels of upstream treatment capacities at the district level. 
				The KP F-Stat refers to the Wald version of the \cite{kleibergen2006generalized} rk-statistic on the excluded instrumental variables for non-i.i.d. errors. 
				The AR 95\% CI reports the 95\% confidence interval, which is robust to the weak instrument based on the \cite{anderson1949estimation} test.
				The means of the dependent variables are calculated for the pre-SBM period.
				Average policy effects are calculated by multiplying the estimated coefficients by the change in the number of latrines per square kilometer between pre-SBM and post-SBM periods.
			\end{tablenotes}
		\end{threeparttable}
	}
\end{table}

\clearpage

\begin{table}[p]\centering \caption{The Heterogeneous Effects on Health (Diarrheal Post-neonatal Mortality Rate (per 1,000)) by the Share of the Population Living Close to Rivers \label{tab:health_by_popshare}}
	\resizebox{1\textwidth}{!}{
		\begin{threeparttable}
			
			{
\def\sym#1{\ifmmode^{#1}\else\(^{#1}\)\fi}
\begin{tabular}{l*{5}{c}}
\toprule
                    &\multicolumn{1}{c}{All}&\multicolumn{2}{c}{Near-River Pop. Share: High}&\multicolumn{2}{c}{Near-River Pop. Share: Low}\\\cmidrule(lr){2-2}\cmidrule(lr){3-4}\cmidrule(lr){5-6}
                    &\multicolumn{1}{c}{(1)}&\multicolumn{1}{c}{(2)}&\multicolumn{1}{c}{(3)}&\multicolumn{1}{c}{(4)}&\multicolumn{1}{c}{(5)}\\
                    &\multicolumn{1}{c}{All}&\multicolumn{1}{c}{STP High}&\multicolumn{1}{c}{STP Low}&\multicolumn{1}{c}{STP High}&\multicolumn{1}{c}{STP Low}\\
\midrule
\multicolumn{6}{l}{\textit{Panel A. Population Share within 5 km of Rivers}} \\ 
\addlinespace
Upstream number of  &      -0.011\sym{*}  &      -0.029\sym{**} &       0.005         &      -0.010         &      -0.042         \\
latrines per sq. km &     (0.006)         &     (0.011)         &     (0.009)         &     (0.009)         &     (0.028)         \\
\addlinespace
Observations        &         824         &         192         &         208         &         264         &         160         \\
Number of Districts &         103         &          24         &          26         &          33         &          20         \\
KP F-Stat           &      78.696         &      28.818         &      19.307         &      33.568         &       3.890         \\
AR 95\% CI          &[-.023, .001]         &[... , -.010]         &[-.020, .038]         &[-.031, .014]         &[..., -.01] U [.03, ...]         \\
Mean of Dep. Variable&       2.576         &       2.561         &       3.127         &       2.331         &       2.282         \\
Average Policy Effect&      -0.269         &      -0.921         &       0.120         &      -0.214         &      -0.887         \\
\addlinespace
\midrule
\multicolumn{6}{l}{\textit{Panel B. Population Share within 10 km of Rivers}} \\ 
\addlinespace
Upstream number of  &      -0.011\sym{*}  &      -0.025\sym{**} &       0.006         &      -0.012         &      -0.043         \\
latrines per sq. km &     (0.006)         &     (0.010)         &     (0.009)         &     (0.010)         &     (0.028)         \\
\addlinespace
Observations        &         824         &         184         &         200         &         272         &         168         \\
Number of Districts &         103         &          23         &          25         &          34         &          21         \\
KP F-Stat           &      78.696         &      28.494         &      18.542         &      32.771         &       4.244         \\
AR 95\% CI          &[-.023, .001]         &[..., -.008]         &[-.019, ...]         &[-.033, .014]         &[..., -.01] U [.04, ...]         \\
Mean of Dep. Variable&       2.576         &       2.358         &       3.134         &       2.475         &       2.314         \\
Average Policy Effect&      -0.269         &      -0.769         &       0.133         &      -0.260         &      -0.935         
\\\bottomrule
\end{tabular}
}

			\begin{tablenotes}
				\setlength{\itemindent}{-2.49997pt}
				\item Notes: The coefficients are reported. 
				Standard errors, clustered at the district level, are in parentheses. 
				***, **, and * indicate significance at the 1\%, 5\%, and 10\% levels, respectively. 
				All regressions include district fixed effects, year fixed effects, and the following controls: precipitation and the interaction of Available Water Capacity and the post-SBM indicator of a reference district.
				Upstream districts are defined as those within the range of $[0,150]$ km from a reference district.
				Columns 2 and 3 report results for districts where the share of the population living within 5 km or 10 km of rivers is above the median (5 km in Panel A and 10 km in Panel B), while Columns 4 and 5 report results for districts below the median. 
				As another dimension of heterogeneity, Columns 2 and 4 report results when upstream districts have higher treatment capacities of sewage treatment plants than the median, while Columns 3 and 5 report results in the case of upstream districts with lower treatment capacities.
				The KP F-Stat refers to the Wald version of the \cite{kleibergen2006generalized} rk-statistic on the excluded instrumental variables for non-i.i.d. errors. 
				The AR 95\% CI reports the 95\% confidence interval, which is robust to the weak instrument based on the \cite{anderson1949estimation} test. 
				The open-ended confidence interval shows that the searched grids do not extend far enough to capture the point where the rejection probability crosses above 95\%.
				The means of the dependent variables are calculated for the pre-SBM period.
				Average policy effects are calculated by multiplying the estimated coefficients by the change in the number of latrines per square kilometer between pre-SBM and post-SBM periods.
			\end{tablenotes}
		\end{threeparttable}
	}
\end{table}

\clearpage

\begin{table}[p]\centering \caption{Alternative Mechanism: The Heterogeneous Effects on Water Quality (Log of Fecal Coliform) by the Share of the Population Living Close to Rivers \label{tab:wq_by_popshare}}
	\resizebox{1\textwidth}{!}{
		\begin{threeparttable}

			{
\def\sym#1{\ifmmode^{#1}\else\(^{#1}\)\fi}
\begin{tabular}{l*{5}{c}}
\toprule
                    &\multicolumn{1}{c}{All}&\multicolumn{2}{c}{Pop. Share within 5 km of Rivers}&\multicolumn{2}{c}{Pop. Share within 10 km of Rivers}\\\cmidrule(lr){2-2}\cmidrule(lr){3-4}\cmidrule(lr){5-6}
                    &\multicolumn{1}{c}{(1)}&\multicolumn{1}{c}{(2)}&\multicolumn{1}{c}{(3)}&\multicolumn{1}{c}{(4)}&\multicolumn{1}{c}{(5)}\\
                    &\multicolumn{1}{c}{All}&\multicolumn{1}{c}{High}&\multicolumn{1}{c}{Low}&\multicolumn{1}{c}{High}&\multicolumn{1}{c}{Low}\\
\midrule
Number of latrines  &       0.030\sym{***}&       0.026\sym{***}&       0.039\sym{**} &       0.024\sym{***}&       0.049\sym{**} \\
per sq. km          &     (0.008)         &     (0.009)         &     (0.015)         &     (0.007)         &     (0.021)         \\
\midrule
Observations        &       7,201         &       3,456         &       3,480         &       3,499         &       3,541         \\
Number of Stations  &       1,189         &         543         &         604         &         553         &         608         \\
Number of Districts &         337         &         125         &         191         &         136         &         184         \\
KP F-Stat           &      29.954         &      15.649         &      17.457         &      19.741         &      10.882         \\
AR 95\% CI          &[.015, .049]         &[.009, .050]         &[.008, .080]         &[.010, .043]         &[.007, .123]         \\
Average Policy Effect&       0.719         &       0.623         &       1.036         &       0.601         &       1.235         \\
\bottomrule
\end{tabular}
}

			\begin{tablenotes}
				\setlength{\itemindent}{-2.49997pt}
				\item Notes: The coefficients are reported. 
				Standard errors, clustered at the district level, are in parentheses. 
				***, **, and * indicate significance at the 1\%, 5\%, and 10\% levels, respectively.
				All regressions include monitoring station fixed effects, year fixed effects, and precipitation as a control. 
				Column 2 reports a result for districts where the share of the population living within 5 km of rivers is above the median, while Column 3 reports a result for districts below the median. 
				Similarly, Columns 4 and 5 report results for districts with higher and lower population share within 10 km of rivers, respectively.
				The KP F-Stat refers to the Wald version of the \cite{kleibergen2006generalized} rk-statistic on the excluded instrumental variables for non-i.i.d. errors. 
				The AR 95\% CI reports the 95\% confidence interval, which is robust to the weak instrument based on the \cite{anderson1949estimation} test. 
				Average policy effects are calculated by multiplying the estimated coefficients by the change in the number of latrines per square kilometer between pre-SBM and post-SBM periods.
			\end{tablenotes}
		\end{threeparttable}
	}
\end{table}

\clearpage

\end{appendices}
\end{document}